\documentclass[10pt]{article}

\usepackage{mathrsfs}
\usepackage{amsmath,extpfeil}
\usepackage{stmaryrd}
\usepackage{mathrsfs}
\usepackage{url}
\usepackage{indentfirst}
\usepackage{enumerate}
\usepackage{amsmath,amsfonts,amssymb,amsthm}
\usepackage{amsmath,amssymb,amsthm,amscd}
\usepackage{graphicx,mathrsfs}
\usepackage{appendix}
\usepackage[numbers,sort&compress]{natbib}
\usepackage{color}
\usepackage[colorlinks, linkcolor=blue, citecolor=blue]{hyperref}

\usepackage{sidecap}
\usepackage{float}
\usepackage{extarrows}
\usepackage{booktabs}
\usepackage{verbatim}
\usepackage[usenames,dvipsnames]{xcolor}

\newtheorem{theorem}{Theorem}[section]
\newtheorem{lemma}[theorem]{Lemma}

\theoremstyle{definition}

\def\R{{\mathbb R}}
\def\C{{\mathbb C}}

\def\XXint#1#2#3{{\setbox0=\hbox{$#1{#2#3}{\int}$}
         \vcenter{\hbox{$#2#3$}}\kern-.5\wd0}}

\def\R{\mathbb{R}}

\numberwithin{equation}{section}

\allowdisplaybreaks[4]

\begin{document}

\title{Existence and Uniqueness of Constraint Minimizers for the Planar Schr\"odinger-Poisson System with Logarithmic Potentials}
\author{Yujin Guo\thanks{School of Mathematics and Statistics, and Hubei Key Laboratory of
Mathematical Sciences, Central China Normal University, P.O. Box 71010, Wuhan 430079, P. R. China. Email: \texttt{yguo@ccnu.edu.cn}. Y. J. Guo is partially supported by NSFC under Grants 12225106 and 11931012.
},
\, 	Wenning Liang\thanks{School of Mathematics and Statistics, Central China Normal University, P.O. Box 71010, Wuhan 430079,
	P. R. China. Email: \texttt{wnliang@ccnu.edu.cn}. }
\, and\, Yan Li\thanks{School of Mathematics and Statistics, Central China Normal University, P.O. Box 71010, Wuhan 430079,
		P. R. China.  Email: \texttt{yanlimath@mails.ccnu.edu.cn}.}
	}
%\thanks{}
\date{}
\maketitle

\begin{abstract}
In this paper, we study constraint minimizers $u$ of the planar Schr\"odinger-Poisson system with a logarithmic convolution potential $\ln|x|\ast u^2$ and a logarithmic external potential $V(x)=\ln(1+|x|^2)$, which can be described by the $L^2$-critical constraint minimization problem
with a subcritical perturbation. We prove that there is a threshold $\rho^*\in(0,\infty)$ such that constraint minimizers exist if and only if $0<\rho<\rho^*$.
In particular, the local uniqueness of positive constraint minimizers as $\rho\nearrow\rho^*$
is analyzed by overcoming the sign-changing property of the logarithmic convolution potential and the non-invariance under translations of the logarithmic external potential.
\end{abstract}


\section{Introduction}
In this paper, we consider the following Schr\"odinger-Poisson system
\begin{equation}\label{0.1}
\left\{
\begin{aligned}
i\psi_{t} &-\Delta\psi+V(x)\psi+\gamma w\psi=|\psi|^{p-2}\psi\ \ \hbox{in} \ \ \R^N\times\R,\\
    \Delta w&=|\psi|^2\ \ \hbox{in}\ \ \R^N\times\R,
\end{aligned}
\right.
\end{equation}
where $\psi:\R^N\times\R\rightarrow\C$ ($N\geqslant 2$) is the time-dependent wave function,
$x\mapsto V(x)$ is a real external potential, $p>2$ and $\gamma\in\R$ is a parameter. The function $w$ represents an internal potential for a nonlocal self-interaction of the wave function $\psi$, and the nonlinear term $|\psi|^{p-2}\psi$ is frequently used to model the interaction among particles \cite{BV1,Aprile,Ruiz2}. Over the past few decades, the system \eqref{0.1} has attracted considerable attentions due to its physical relevance: for instance, it arises in quantum mechanics (see \cite{Arnold,BV,BV1,Benguria,Bokanowski,Catto,Frohlich,Garnier,Lieb2,Lions,Pekar,Penrose1,Penrose2}) and particularly  semiconductor physics (see \cite{Catto1,Markowich,Sanchez1}). We refer the reader to \cite{Ando,Doniach,Fisher,Hebard} for more details on its physical aspects.

Under the standing wave ansatz $\psi(x,t)=e^{i\mu t}u(x)$, $\mu\in\R$, the Schr\"odinger-Poisson system \eqref{0.1} is reduced to the following problem
\begin{equation}\label{0.3}
    -\Delta u+\big(V(x)-\mu\big)u+\gamma \big[\Phi_{N}*u^2\big]u=|u|^{p-2}u\ \ \hbox{in}\ \
     \R^N,
\end{equation}
where
\begin{equation}\label{0:1}
\Phi_{N}(x):=\left\{
    \begin{aligned}
    &\frac{1}{N(2-N)\omega_{N}}|x|^{2-N},\ \ N\geqslant 3,\\
    &\frac{1}{2\pi}\ln|x|,\ \ N=2,
    \end{aligned}
\right.
\end{equation}
and $\omega_{N}>0$ denotes the volume of the unit ball in $\R^N$.
The case $N\geqslant3$ of (\ref{0.3}) was extensively studied in the past few years. In particular, variational methods were employed (cf. \cite{Ambrosetti,Ruiz,Chen,Azzollini1,Ruiz1,Mercuri}) to analyze the existence and multiplicity results of solutions for  (\ref{0.3}). We also refer the reader to  \cite{Bellazzini,Castella,Jeanjean,Jeanjean3,Lieb,Lions} for  analyzing  normalized  solutions $u$ of (\ref{0.3}), in the sense that $u$  satisfies $\|u\|^2_{2}=\rho$ for some given $\rho>0$.

However, the approaches mentioned above for the case $N\geqslant3$ cannot be directly adapted to the case $N=2$, due to the fact that the Newtonian potential $\Phi_{2}(x)$ defined in \eqref{0:1} is sign-changing and presents singularities as $|x|$ approaches to zero and infinity. Therefore, the existing works \cite{Silvia1,Du,Antonio,Stubbe,Silvia2,Li} show that it is more delicate to study (\ref{0.3}) for the case $N=2$. More precisely, Cingolani and Weth \cite{Silvia1} derived a strong
compactness condition for Cerami sequences, based on which they established various existence results of solutions for \eqref{0.3} under the assumptions $V(x)\in L^\infty(\R^2)$ and $p\geqslant 4$.
Further, Du and Weth \cite{Du} generalized the above results to the rest case $p\in(2,4)$ by exploring a more complicated underlying functional geometry with a different variational method.
On the other hand, by analyzing normalized solutions, Stubbe \cite{Stubbe} studied the existence and uniqueness of ground sates for \eqref{0.3} with $N=2$, where the external potential $V(x)$ and the nonlinear term  were both ignored. After Stubbe's work, Cingolani and Jeanjean \cite{Silvia2} obtained several existence and multiplicity results of normalized solutions for \eqref{0.3} with $N=2$, where $V(x)\equiv0$ and $p\in(2,4)$ were considered, see also \cite{Li} for related works.
Moreover, when $N=2$, the problem (\ref{0.3}) with  a different nonlinear term was recently analyzed in \cite{Dolbeault} and the references therein, where the authors considered the following logarithmic external potential
\begin{equation}\label{0:3}
    V(x)=\ln(1+|x|^2)\ \,\ \mbox{in}\ \ \R^2.
\end{equation}

Motivated by above facts, in this paper we focus on normalized solutions $u$ of (\ref{0.3}) with the external potential $V(x)$ satisfying \eqref{0:3}, together with $N=2$, $p=4$ and $\gamma=2\pi$, which can be obtained as  (constraint) minimizers of the following variational problem:
\begin{equation}\label{1.1}
e(\rho):=\inf_{u\in S_{\rho}}E(u),\ \ \rho>0,
\end{equation}
where the energy functional $ E(u)$ satisfies
\begin{equation}\label{1.2}
\begin{aligned}
E(u):&=\frac{1}{2}\int_{\R^2}\Big[|\nabla u(x)|^2+{\rm ln}\big(1+|x|^2\big)u^2(x)\Big]{\rm d}x\\
&\quad+\frac14\int_{\R^2}\int_{\R^2}{\rm ln}|x-y|u^2(x)u^2(y){\rm d}x{\rm d}y-\frac14\int_{\R^2}u^4(x){\rm d}x.
\end{aligned}
\end{equation}
We notice that $E(u)$ involves both a logarithmic potential term and a logarithmic convolution term, which unfortunately are not well defined in $H^1(\R^2)$. Stimulated by \cite{Silvia1,Stubbe}, we consider the space $S_\rho$ of $e(\rho)$ satisfying
\begin{equation*}
S_\rho:=\big\{u\in X: \|u\|^2_2=\rho\big\},\ \ X=\Big\{u\in H^1(\R^2):\int_{\R^2}{\rm ln}(1+|x|^2)u^2(x){\rm d}x<\infty\Big\},
\end{equation*}
which is endowed with the norm
\begin{equation*}
\|u\|_X=\left\{\int_{\R^2}\left[|\nabla u|^2+\Big(1+{\rm ln}(1+|x|^2)\Big)u^2(x)\right]{\rm d}x\right\}^{\frac{1}{2}},\ \ u\in X.
\end{equation*}
%Since the energy functional $E(u)$
%is not well-defined on the natural Sobolev space
%$H^1(\R^2)$, Stubbe \cite{Stubbe} developed a variational framework to deal with \eqref{0.3} within the smaller Hilbert space $X$, where  both the nonlinear term $|u|^{2}u$ and $V(x)$ are ignored.
One can note that the norm of $X$ and the functional $E(\cdot)$ are not invariant under translations
in $\R^2$, and the quadratic part of the energy functional \eqref{1.2} is not coercive on X. All these lead to some extra difficulties in the analysis of \eqref{1.1}, compared with the existing works \cite{guo,gwz,gzz} and the references therein. By analyzing carefully two different types of logarithmic terms involved in  $E(\cdot)$, the main purpose of this paper is to discuss the existence and local uniqueness of minimizers for $e(\rho)$.

%{\color{red}Moreover, we address a situation which is substantially different compared to the case where $\mu>0$ is fixed \cite{Silvia1,Du,Antonio}}.

If the logarithmic external potential $V(x)=\ln(1+|x|^2)$ of (\ref{0.3}) is ignored, it was proved  in \cite{Li,Silvia2} that $e(\rho)$ admits minimizers if and only if $0<\rho<\rho^*$, where $\rho^*:=\|Q\|_{2}^2$, and $Q=Q(|x|)>0$, up to translations, is the unique (cf. \cite{Gidas,Kwong}) positive solution of the following elliptic equation:
\begin{equation}\label{1.5}
-\Delta u+u-u^3=0\ \ {\rm in}\ \ \R^2,\ \ u\in H^1(\R^2).
\end{equation}
In addition, the function $Q>0$ decays exponentially as $|x|\to\infty$, in the sense that
\begin{equation}\label{1.6}
Q(x),\ \ |\nabla Q(x)|=O\Big(|x|^{-\frac{1}{2}}e^{-|x|}\Big)\ \ {\rm as}\ \ |x|\to\infty.
\end{equation}
Moreover, we have the following classical Gagliardo-Nirenberg inequality:
\begin{equation}\label{1.4}
\int_{\R^2}u^4{\rm d}x\leqslant\frac 2{\|Q\|_2^2}\int_{\R^2}|\nabla u|^2{\rm d}x\int_{\R^2}u^2{\rm d}x,\ \ u\in H^1(\R^2),
\end{equation}
where the identity is achieved at $u=Q$, and $Q=Q(x)>0$ is the unique positive solution of \eqref{1.5}.
The following theorem shows that the existence and non-existence of minimizers for $e(\rho)$ are the same as those of the case addressed in \cite{Silvia2,Li}, where the external potential $V(x)=\ln(1+|x|^2)$ is ignored.

\begin{theorem}\label{thm:1.1}
Let $Q=Q(|x|)$ be the unique positive solution of \eqref{1.5}. We have
\begin{enumerate}
\item If $0<\rho<\rho^*:=\|Q\|_2^2$, then there exists at least one minimizer of $e(\rho)$;

\item  If $\rho\geqslant\rho^*$, then there is no minimizer of $e(\rho)$.
\end{enumerate}
Moreover, we have  $\lim\limits_{\rho\nearrow\rho^*}e(\rho)=-\infty$.
\end{theorem}

\noindent The proof of Theorem \ref{thm:1.1} is similar to \cite{guo,Silvia2,Li}, which is given  in Section 2 for completion. Our analysis shows that the proof of Theorem \ref{thm:1.1} needs the Gagliardo-Nirenberg inequality \eqref{1.4} and the following Hardy-Littlewood-Sobolev inequality (cf. \cite{Lieb1}):
\begin{equation}\label{0.8}
\int_{\R^2}\int_{\R^2}\frac{1}{|x-y|}|u(x)||v(y)|{\rm d}x{\rm d}y\leqslant C\|u\|_{\frac{4}{3}}\|v\|_{\frac{4}{3}},\ \ \hbox{where}\ \ u,\ v\in L^{\frac{4}{3}}(\R^2).
\end{equation}
Actually, the inequality \eqref{0.8} is often used throughout this paper to handle with the logarithmic convolution term of $E(u)$.

Suppose $u_\rho$ is a minimizer of $e(\rho)$ for any $\rho\in(0,\rho^*)$, then the variational theory gives that $u_{\rho}$ satisfies the following Euler-Lagrange equation:
\begin{equation}\label{1.8}
-\Delta u_\rho+{\rm ln}\big(1+|x|^2\big)u_\rho+\Big(\int_{\R^2}{\rm ln}|x-y|u_\rho^2(y){\rm d}y\Big)u_\rho=\mu_\rho u_\rho+|u_\rho|^2u_{\rho}\ \ {\rm in}\ \ \R^2,
\end{equation}
where $\mu_\rho\in\R$ is the associated Lagrange multiplier. We obtain from \cite[Theorem 6.17]{Lieb1} that $E(u_{\rho})\geqslant E(|u_{\rho}|)$, which implies that $|u_{\rho}|$ is also a minimizer of $e(\rho)$. %By  one may restrict the minimizers of $e(\rho)$ to nonnegative functions.
By the strong maximum principle, it further follows from \eqref{1.8} that $|u_{\rho}|>0$ in $\R^2$.  Therefore, $u_{\rho}$ must be either strictly positive or strictly negative. Without loss of generality, we next focus on positive minimizers $u_{\rho}>0$ of $e(\rho)$.

The uniqueness of positive minimizers for $e(\rho)$ is an interesting problem in physics. Stimulated by Refs. \cite{glp,glw} and the references therein, the main result of this paper is concerned with the following local uniqueness of positive minimizers for $e(\rho)$.

\begin{theorem}\label{thm:1.3}
There exists a unique positive minimizer of $e(\rho)$ as $\rho $ is sufficiently close to $\rho^*$ from below.
\end{theorem}

It is our conjecture that $e(\rho)$ admits a unique positive minimizer for all $\rho\in(0,\rho^*)$, which however seems a more challenging problem. We now sketch the main idea of proving Theorem \ref{thm:1.3} as follows: To prove Theorem \ref{thm:1.3}, by contradiction we assume that there exist two different minimizers $u_\rho\not \equiv v_\rho$. We then define the following difference function:
\[
 \eta_\rho(x):=\frac{u_\rho - v_\rho}{\|u_\rho - v_\rho\|_{L^\infty(\R^2)}}\quad   \mbox{in}\ \ \R^2.
\]
On one hand, since $\|\eta_\rho(x)\|_{L^\infty(\R^2)}\equiv 1$, we shall show that $\eta_\rho(x)$ cannot vanish as $\rho\nearrow\rho^*$. On the other hand, we shall prove that under suitable scaling, $\eta_\rho(x)$ converges to zero in $L^\infty(\R^2)$ as  $\rho\nearrow\rho^*$, a contradiction, by constructing various types of Pohozaev identities (cf. \cite{Cao,Deng,glp,glw}). We comment that our argument is different from that of \cite{Stubbe}, where the uniqueness of minimizers was essentially obtained by analyzing the corresponding elliptic system.

However, the above procedure depends on the refined estimates of $\eta_\rho(x)$  as  $\rho\nearrow\rho^*$, for which  it is necessary to establish the following limiting behavior of positive minimizers for $e(\rho)$ as $\rho\nearrow\rho^*$.

\begin{theorem}\label{thm:1.2}
Assume $u_{\rho}$ is a positive minimizer of $e(\rho)$ for $\rho\in(0,\rho^*)$, then we have
\begin{equation*}
\lim\limits_{\rho\nearrow\rho^*}\frac{2(\rho^*-\rho )^{\frac{1}{2}}}{\rho^*}u_{\rho }\Big(\frac{2(\rho^*-\rho )^{\frac{1}{2}}}{\rho^*}x+x_{\rho}\Big)=Q(x)\,\ \ in\ \ X\cap L^\infty(\R^2),
\end{equation*}
where $x_{\rho}$ is the unique maximal point of $u_{\rho}$ as $\rho\nearrow\rho^*$ and satisfies
\begin{equation*}
\lim\limits_{\rho\nearrow\rho^*}\frac{x_{\rho}}{(\rho^*-\rho)^{\frac{1}{2}}}=0.
\end{equation*}
\end{theorem}

We remark that even though the limiting behavior of ground states for Bose-Einstein condensates was widely studied recently in \cite{guo,gwz,gzz} and the references therein, as far as we know, there exist almost no similar results for the planar Schr\"{o}dinger-Poisson system. Moreover, comparing with the existing works mentioned above, there exist some extra difficulties in the proof of Theorem \ref{thm:1.2}. Firstly, the energy functional of $e(\rho)$ is involved with the challenging logarithmic convolution term
\begin{equation}\label{0.9}
    \int_{\R^2}\int_{\R^2}{\rm ln}|x-y|u^2(x)u^2(y){\rm d}x{\rm d}y.
\end{equation}
To overcome the difficulties arising from (\ref{0.9}), we shall derive in \eqref{3.14} the following important estimate:  for any given $x\in\R^2$, there exists a constant $C>0$ such that for any $\rho\in (0,\rho^*)$,
\begin{equation*}
\int_{\R^2}{\rm ln} \Big(1+\frac1{|x-y|^2}\Big)w_\rho^2(y){\rm d}y
\leqslant C,
\end{equation*}
where $w_{\rho}>0$ is a suitably scaled function of the minimizer $u_\rho$.
Secondly, since the norm of $X$ is not invariant under translations, different from \cite{guo,gwz,gzz}, we cannot directly obtain the uniform boundedness of the scaled minimizers $w_{\rho}$ as $\rho\nearrow\rho^*$. To overcome this difficulty,  we shall borrow the idea from \cite{GLY}.
Finally, because $\lim_{\rho\nearrow\rho^*}e(\rho)=-\infty$ and the external potential term $\int_{\R^2}\ln(1+|x|^2)u_{\rho}^2{\rm d}x$ is not invariant under translations, in Section 3 we need to investigate new  techniques to derive the refined information  of the maximal point $x_{\rho}$  as $\rho\nearrow\rho^*$.

This paper is organized as follows. In Section 2, we first give some preliminary results, after which we then prove Theorem \ref{thm:1.1} on the existence and non-existence of minimizers for $e(\rho)$. In Section 3, we shall prove Theorem \ref{thm:1.2} on the refined limiting profiles of positive minimizers for $e(\rho)$ as $\rho\nearrow\rho^*$. Theorem \ref{thm:1.3} is then proved in Section 4 for the local uniqueness of positive minimizers. We finally give in Appendix A  the detailed proofs of some results used in the proof of Theorems \ref{thm:1.3} and \ref{thm:1.2}.

\section{Preliminaries}
In this section, we first establish various preliminary results which are often used throughout the whole paper. We then follow them to finish the proof of Theorem \ref{thm:1.1} in Subsection 2.1.

As performed in \cite{Silvia1}, we first define the following symmetric bilinear forms
\begin{equation*}
B_1(u,v)=\int_{\R^2}\int_{\R^2}{\rm ln}\big(1+|x-y|^2\big)u(x)v(y){\rm d}x{\rm d}y,
\end{equation*}
\begin{equation*}
B_2(u,v)=\int_{\R^2}\int_{\R^2}{\rm ln}\Big(1+\frac1{|x-y|^2}\Big)u(x)v(y){\rm d}x{\rm d}y-\frac12\int_{\R^2}\int_{\R^2}{\rm ln}\Big(1+\frac1{|x-y|^2}\Big)u(x)v(y){\rm d}x{\rm d}y,
\end{equation*}
and
\begin{equation*}
B_0(u,v)=\frac12\Big[B_1(u,v)-B_2(u,v)\Big]=\int_{\R^2}\int_{\R^2}{\rm ln}|x-y|u(x)v(y){\rm d}x{\rm d}y,\ \ u,v\in X.
\end{equation*}
Since
\begin{equation*}
{\rm ln}\big(1+|x-y|^2\big)\leqslant{\rm ln}\big(1+2|x|^2+2|y|^2\big)\leqslant2{\rm ln}\big(1+|x|^2\big)+2{\rm ln}\big(1+|y|^2\big),\ \ x,y\in\R^2,
\end{equation*}
we have the estimate
\begin{equation}
\begin{aligned}\label{2.1}
\left|B_1(uv,wz)\right|\leqslant&2\int_{\R^2}\int_{\R^2}\Big[{\rm ln}(1+|x|^2)+{\rm ln}(1+|y|^2)\Big]|u(x)v(x)w(y)z(y)|{\rm d}x{\rm d}y\\
\leqslant&2\|u\|_*\|v\|_*\|w\|_2\|z\|_2+2\|u\|_2\|v\|_2\|w\|_*\|z\|_*,\ \ u,v,w,z\in X,
\end{aligned}
\end{equation}
where
\begin{equation*}
\|u\|_*:=\left\{\int_{\R^2}{\rm ln}(1+|x|^2)u^2(x){\rm d}x\right\}^{\frac{1}{2}},\ \ u\in X.
\end{equation*}
Since $0<{\rm ln}(1+r)< r$ holds for all $r>0$, it follows from the Hardy-Littlewood-Sobolev inequality \eqref{0.8} that there exists a constant $C>0$ such that
\begin{equation}
\begin{aligned}\label{2.2}
\frac12|B_2(u,v)|\leqslant&\int_{\R^2}\int_{\R^2}{\rm ln}\Big(1+\frac1{|x-y|}\Big)|u(x)v(y)|{\rm d}x{\rm d}y\\
\leqslant&\int_{\R^2}\int_{\R^2}\frac1{|x-y|}|u(x)v(y)|{\rm d}x{\rm d}y\\
\leqslant&C\|u\|_{\frac{4}{3}}\|v\|_{\frac{4}{3}},\ \ u,v\in L^{\frac{4}{3}}(\R^2).
\end{aligned}
\end{equation}

By the continuous embedding from $X$ into $L^q(\R^2)$ for $q\in[2,\infty)$,  we derive from  \eqref{2.1} and \eqref{2.2} that the functionals $B_i(u^2,v^2)$ are well defined on $X\times X$ for $i=0,1,2$. Moreover, denote functionals on $X$ as follows:
\begin{equation*}
V_1(u)=B_1(u^2,u^2)=\int_{\R^2}\int_{\R^2}{\rm ln}(1+|x-y|^2)u^2(x)u^2(y){\rm d}x{\rm d}y,
\end{equation*}
 \begin{equation*}
V_2(u)=B_2(u^2,u^2)=\int_{\R^2}\int_{\R^2}{\rm ln}\Big(1+\frac1{|x-y|^2}\Big)u^2(x)u^2(y){\rm d}x{\rm d}y,
\end{equation*}
and
\begin{equation*}
V_0(u)=B_0(u^2,u^2)=\int_{\R^2}\int_{\R^2}{\rm ln}|x-y|u^2(x)u^2(y){\rm d}x{\rm d}y.
\end{equation*}
Using above notations, we have the following crucial properties of $V_i(\cdot)$ for $i=0,1,2$.

\begin{lemma}\label{lem:2.1} The space $X$ and the functionals $V_i(\cdot)$ $(i=0,1,2)$ have the following properties:

\begin{enumerate}
		\item  The space $X$ is compactly embedded into $L^s(\R^2)$ for all $s\in[2,\infty)$;

\item The functionals $V_0(\cdot), V_1 (\cdot), V_2(\cdot)$ and $E(\cdot)$ defined in (\ref{1.2}) are of class $C^1$ on $X$, and
 $$\langle V_i'(u), v\rangle=4B_i(u^2, uv)\ \  \mbox{for}\ \ u, v\in X, \ \
  i=0, 1, 2\ ;$$

\item $V_2(\cdot)$ is continuously differentiable on $L^{\frac{8}{3}}(\R^2)$;

\item $V_1(\cdot)$ is weakly lower semicontinuous on $H^1(\R^2)$;

\item $V_0(\cdot)$ is weakly lower semicontinuous on $X$.
\end{enumerate}
\end{lemma}

Since the proof of Lemma \ref{lem:2.1} is similar to that of  \cite[Lemma 2.2]{Silvia1}, we omit the details for simplicity.

%Based on Lemma \ref{lem:2.1}, we shall present the existence and non-existence of minimizers for $e(\rho)$ in the coming subsection.

\subsection{The existence and non-existence of minimizers for $e(\rho)$}

Applying above preliminary results, the main purpose of this subsection is to address the proof of Theorem \ref{thm:1.1} on the existence and non-existence of minimizers for $e(\rho)$. For the unique positive solution $Q=Q(|x|)>0$ of \eqref{1.5}, we note from \cite[Lemma 8.12]{Cazenave} that $Q$ satisfies the following identities
\begin{equation}\label{1.7}
\int_{\R^2}|\nabla Q|^2{\rm d}x=\int_{\R^2}Q^2{\rm d}x=\frac12\int_{\R^2}Q^4{\rm d}x.
\end{equation}
{\bf Proof of Theorem \ref{thm:1.1}.}
1. We first prove the existence of minimizers for $e(\rho)$ with $0<\rho<\rho^*$. If $u\in X$ and $\|u\|_{2}^2=\rho$ with $\rho\in (0,\rho^*)$, then we deduce from the Gagliardo-Nirenberg inequality (\cite{Weinstein}) that for any $p\geqslant 2$,
\begin{equation*}
\|u\|_p\leqslant\Big(\frac{p\rho}{2\|Q\|_2^{p-2}}\Big)^{\frac{1}{p}} \Big(\int_{\R^2}|\nabla u|^2{\rm d}x\Big)^{\frac{p-2}{2p}}.
\end{equation*}
By Hardy-Littlewood-Sobolev inequality \eqref{0.8}, we derive from above that there exists a constant $C>0$ such that
\begin{equation}\label{2.5}
\int_{\R^2}\int_{\R^2}\frac{u^2(x)u^2(y)}{|x-y|}{\rm d}x{\rm d}y\leqslant C\Big(\int_{\R^2}|u|^{\frac{8}{3}}{\rm d}x\Big)^{\frac{3}{2}}\leqslant C\rho^{\frac{3}{2}}\Big(\int_{\R^2}|\nabla u|^2{\rm d}x\Big)^{\frac{1}{2}}.
\end{equation}
It then follows from \eqref{2.2} and \eqref{2.5} that
\begin{eqnarray}\label{2.6}
\begin{aligned}
&\frac12\int_{\R^2}\int_{\R^2}{\rm ln}\Big(1+\frac1{|x-y|^2}\Big)u^2(x)u^2(y){\rm d}x{\rm d}y\\
\leqslant&\int_{\R^2}\int_{\R^2}\frac1{|x-y|}u^2(x)u^2(y){\rm d}x{\rm d}y\\
\leqslant&C\rho^{\frac{3}{2}}\Big(\int_{\R^2}|\nabla u|^2{\rm d}x\Big)^{\frac{1}{2}}.
\end{aligned}
\end{eqnarray}
Together with  \eqref{1.4}, we get from \eqref{2.6} that
\begin{equation}\label{2.7}
E(u)\geqslant\frac{\rho^*-\rho}{2\rho^*}\int_{\R^2}|\nabla u|^2{\rm d}x+\frac12\int_{\R^2}{\rm ln}(1+|x|^2)u^2{\rm d}x-\frac14C\rho^{\frac{3}{2}}\Big(\int_{\R^2}|\nabla u|^2{\rm d}x\Big)^{\frac{1}{2}},
\end{equation}
which implies that $E(u)$ is bounded from below.

Letting $\{u_n\}\subset S_\rho$ be a minimizing sequence of $e(\rho)$ for $\rho\in(0,\rho^*)$, we see from \eqref{2.7} that $\int_{\R^2}|\nabla u_{n}|^2{\rm d}x$ and $\int_{\R^2}{\rm ln}(1+|x|^2)u_n^2{\rm d}x$ are bounded uniformly in $n$. Since $\|u_{n}\|_{2}^2=\rho$, we know that $\{u_n\}$ is bounded uniformly in $X$. According to Lemma \ref{lem:2.1} (1), there exists a function $u\in X$ such that
\begin{equation*}
   u_n\rightharpoonup u\ \ \hbox{in}\ \ X\ \ \hbox{and}\ \ u_{n}\rightarrow u\ \ \hbox{strongly in}\ \ L^s(\R^2),\ \ s\in [2,\infty),
\end{equation*}
which implies that $\|u\|_{2}^2=\rho$. We hence obtain that $u\in S_{\rho}$. Moreover, we note from (1) and (5) of Lemma \ref{lem:2.1} that
\begin{equation*}
\int_{\R^2}\int_{\R^2}{\rm ln}|x-y|u^2(x)u^2(y){\rm d}x{\rm d}y\leqslant\liminf_{n\to\infty}\int_{\R^2}\int_{\R^2}{\rm ln}|x-y|u_n^2(x)u_n^2(y){\rm d}x{\rm d}y,
\end{equation*}
and
\begin{equation*}
\lim_{n\to\infty}\int_{\R^2}u_n^4{\rm d}x=\int_{\R^2}u^4{\rm d}x.
\end{equation*}
By the weakly lower semicontinuity, we then obtain from above that
\begin{equation*}
e(\rho)\leqslant E(u)\leqslant\liminf_{n\to\infty}E(u_n)=e(\rho).
\end{equation*}
Thus, $E(u)=e(\rho)$, that is, $u$ is a minimizer of $e(\rho)$ for $0<\rho<\rho^*$.

2. We next prove the non-existence of minimizers of $e(\rho)$ for $\rho\geqslant \rho^*$. In this case,  consider the test function
\begin{equation*}
u_\tau(x)=\frac{\tau\rho^{\frac{1}{2}}}{\|Q\|_2}Q(\tau x),\ \ \tau>0,
\end{equation*}
so that $u_\tau\in S_\rho$ holds for all $\tau>0$. We then deduce from \eqref{1.7} that
\begin{eqnarray}\label{2.9}
\begin{aligned}
e(\rho)\leqslant E(u_\tau)=&\frac{\rho(\rho^*-\rho)}{2\rho^*}\tau^2+\frac\rho{2\rho^*}\int_{\R^2}{\rm ln}\Big(1+\big|\frac{x}{\tau}\big|^2\Big)Q^2(x){\rm d}x\\
&+\frac{\rho^2}{4(\rho^*)^2}\int_{\R^2}\int_{\R^2}{\rm ln}|x-y|Q^2(x)Q^2(y){\rm d}x{\rm d}y-\frac14\rho^2{\rm ln}\tau,\ \ \tau>0.
\end{aligned}
\end{eqnarray}
Together with \eqref{1.6}, \eqref{2.1} and \eqref{2.6}, we derive from (\ref{2.9}) that $e(\rho)\leqslant\lim\limits_{\tau\to\infty}E(u_\tau)=-\infty$ holds for $\rho\geqslant\rho^*$. Thus, there is no minimizer of $e(\rho)$ for $\rho\geqslant\rho^*$.

Moreover, for $\rho\in (0,\rho^*)$, taking $\tau=(\rho^*-\rho)^{-\frac{1}{2}}$ into (\ref{2.9}) yields that
\begin{equation*}
    \lim\limits_{\rho\nearrow \rho^*}e(\rho)\leqslant \lim\limits_{\rho\nearrow \rho^*}E(u_{\tau})=-\infty,
\end{equation*}
and hence $\lim\limits_{\rho\nearrow \rho^*}e(\rho)=-\infty$.
This completes the proof of Theorem \ref{thm:1.1}.
\qed
\medskip

\section{Limiting Behavior of Minimizers for $e(\rho)$ as $\rho\nearrow\rho^*$}
In this section, we shall establish Theorem \ref{thm:1.2} on the limiting behavior of minimizers for $e(\rho)$ as $\rho\nearrow \rho^*$.
We first derive some estimates of the minimizers for $e(\rho)$ as $\rho\nearrow \rho^*$.

\begin{lemma}\label{lem:2.4}
Let $u_\rho$ be a positive minimizer of $e(\rho)$ for $\rho\in(0,\rho^*)$, and define
\begin{equation}\label{125}
\varepsilon_\rho=\Big(\int_{\R^2}\left|\nabla u_\rho\right|^2{\rm d}x\Big)^{-\frac{1}{2}}>0.
\end{equation}
Then we have
\begin{enumerate}
\item $\varepsilon_\rho>0$ satisfies
\begin{equation*}
\varepsilon_\rho\to0\ \ \hbox{and}\ \ \mu_{\rho}\varepsilon_{\rho}^2\rightarrow-\frac{1}{\rho^*}\ \
\hbox{as}\ \ \rho\nearrow\rho^*.
\end{equation*}

  \item Denote
\begin{equation}\label{3.2}
w_\rho(x):=\varepsilon_\rho u_\rho(\varepsilon_\rho x+x_\rho)>0\ \ in\ \ \R^2,
\end{equation} where $x_\rho$ is a global maximal point of $u_\rho$. Then there exists a constant $\eta>0$, independent of  $0 <\rho<\rho^*$, such that
\begin{equation}\label{3.3}
\int_{B_2(0)}w_\rho^2{\rm d}x\geqslant\eta>0\ \ as\ \ \rho\nearrow\rho^*.
\end{equation}

  %There exists a subsequence of $\{u_\rho\}$, still denoted by $\{u_\rho\}$, such that $u_\rho$ has a unique maximal point $x_\rho$ satisfying $x_\rho\to0$ and
\item $w_{\rho}$ satisfies
\begin{equation}\label{3.4}
w_\rho(x)\to\frac{1}{\sqrt{\rho^*}} Q\big(\frac{1}{\sqrt{\rho^*}}|x|\big)\ \ in\ \ H^1(\R^2)\ \ as\ \ \rho\nearrow\rho^*.
\end{equation}
\end{enumerate}
\end{lemma}

\noindent{\bf Proof.}
1. We first prove that $\varepsilon_{\rho}\rightarrow 0$ as $\rho\nearrow\rho^*$. Following \eqref{1.4} and \eqref{2.6}, we derive that for $\rho\in (0,\rho^*)$,
\begin{equation}\label{3.5}
\begin{aligned}
e(\rho)=&E(u_\rho)\\
\geqslant&\frac{\rho^*-\rho}{2\rho^*}\int_{\R^2}|\nabla u_\rho|^2{\rm d}x+\frac12\int_{\R^2}{\rm ln}(1+|x|^2)u_\rho^2(x){\rm d}x\\
&+\frac18\int_{\R^2}\int_{\R^2}{\rm ln}(1+|x-y|^2) u_\rho^2(x)u_\rho^2(y){\rm d}x{\rm d}y-\frac14C\rho^{\frac{3}{2}}\Big(\int_{\R^2}|\nabla u_\rho|^2{\rm d}x\Big)^{\frac{1}{2}}\\
\geqslant&\frac{\rho^*-\rho}{2\rho^*}\varepsilon_\rho^{-2}-\frac14C\rho^{\frac{3}{2}}\varepsilon_\rho^{-1}\geqslant -\frac14C\rho^{\frac{3}{2}}\varepsilon_\rho^{-1}.
\end{aligned}
\end{equation}
Moreover, we have $\lim\limits_{\rho\nearrow\rho^*}e(\rho)=-\infty$ in view of Theorem \ref{thm:1.1}, together with \eqref{3.5}, which yields that $\varepsilon_\rho\to0$ as $\rho\nearrow\rho^*$.

We next prove that $\mu_{\rho}\varepsilon_{\rho}^2\rightarrow-\frac{1}{\rho^*}$ as $\rho\nearrow\rho^*$. By the definition of $\varepsilon_{\rho}$, we have
\begin{eqnarray}\label{3.8}
\begin{aligned}
\varepsilon_\rho^2 e(\rho)=& \frac12+\frac{\varepsilon_\rho^2}{2}\int_{\R^2}\ln\big(1+|x|^2\big)u_{\rho}^2(x){\rm d}x+\frac{\varepsilon_\rho^2}{4}\int_{\R^2}\int_{\R^2}{\rm ln}|x-y|u^2_\rho(x)u^2_\rho(y){\rm d}x{\rm d}y
-\frac{\varepsilon_\rho^2}{4}\int_{\R^2}u_\rho^4(x){\rm d}x\\
=&\frac12+\frac{\varepsilon_\rho^2}{2}\int_{\R^2}\ln\big(1+|x|^2\big)u_{\rho}^2(x){\rm d}x+\frac{\varepsilon_\rho^2}{8}\int_{\R^2}\int_{\R^2}\ln\big(1+|x-y|^2\big)u^2_\rho(x)u^2_\rho(y){\rm d}x{\rm d}y\\
&-\frac{\varepsilon_\rho^2}{8}\int_{\R^2}\int_{\R^2}\ln\Big(1+\frac{1}{|x-y|^2}\Big)u^2_\rho(x)u^2_\rho(y){\rm d}x{\rm d}y
-\frac{\varepsilon_\rho^2}{4}\int_{\R^2}u_\rho^4(x){\rm d}x.
\end{aligned}
\end{eqnarray}
Since $\varepsilon_{\rho}\to 0$ as $\rho\nearrow\rho^*$, we also derive from \eqref{2.6} and   that
\begin{equation}\label{3.9}
\frac12\varepsilon_\rho^2\int_{\R^2}\int_{\R^2}{\rm ln}\Big(1+\frac1{|x-y|^2}\Big) u_\rho^2(x)u_\rho^2(y){\rm d}x{\rm d}y\leqslant
\varepsilon_\rho^2\int_{\R^2}\int_{\R^2}\frac{u_\rho^2(x) u_\rho^2(y)}{|x-y|}{\rm d}x{\rm d}y\leqslant C\rho^{\frac{3}{2}}\varepsilon_\rho\to0\ \ {\rm as}\ \ \rho\nearrow\rho^*.
\end{equation}
By the fact that $\lim\limits_{\rho\nearrow\rho^*}e(\rho)=-\infty$, we deduce from  \eqref{1.4}, \eqref{3.8} and \eqref{3.9} that
%We further deduce from \eqref{2.16}-\eqref{2.18} that
\begin{equation*}
\begin{aligned}
0\geqslant\limsup_{\rho\nearrow\rho^*}\varepsilon_\rho^2e(\rho)
%=&\limsup_{\rho\nearrow\rho^*}\frac12\left(1-\frac{\varepsilon_\rho^2}2\int_{\R^2}u_\rho^4(x){\rm d}x\right)+\limsup_{\rho\nearrow\rho^*}\frac{\varepsilon_\rho^2}{2}\int_{\R^2}\ln\big(1+|x|^2\big)u_{\rho}^2(x){\rm d}x\\
%&+\limsup_{\rho\nearrow\rho^*}\frac{\varepsilon_\rho^2}{8}\int_{\R^2}\int_{\R^2}\ln\big(1+|x-y|^2\big)u^2_\rho(x)u^2_\rho(y){\rm d}x{\rm d}y\\
%&-\lim_{\rho\nearrow\rho^*}\frac{\varepsilon_\rho^2}{8}\int_{\R^2}\int_{\R^2}\ln\Big(1+\frac{1}{|x-y|^2}\Big)u^2_\rho(x)u^2_\rho(y){\rm d}x{\rm d}y\\
=&\limsup_{\rho\nearrow\rho^*}\frac12\left(1-\frac{\varepsilon_\rho^2}2\int_{\R^2}u_\rho^4(x){\rm d}x\right)+\limsup_{\rho\nearrow\rho^*}\frac{\varepsilon_\rho^2}{2}\int_{\R^2}\ln\big(1+|x|^2\big)u_{\rho}^2(x){\rm d}x\\
&+\limsup_{\rho\nearrow\rho^*}\frac{\varepsilon_\rho^2}{8}\int_{\R^2}\int_{\R^2}\ln\big(1+|x-y|^2\big)u^2_\rho(x)u^2_\rho(y){\rm d}x{\rm d}y\geqslant0,
\end{aligned}
\end{equation*}
which implies that
\begin{equation}\label{3.10}
\begin{aligned}
&\lim\limits_{\rho\nearrow \rho^*}\varepsilon_\rho^2\int_{\R^2}u_\rho^4{\rm d}x=2,
\ \ \lim\limits_{\rho\nearrow \rho^*}\varepsilon_\rho^2\int_{\R^2}\ln(1+|x|^2)u_{\rho}^2dx=0,\\
&\lim\limits_{\rho\nearrow \rho^*}\varepsilon_\rho^2\int_{\R^2}\int_{\R^2}\ln(1+|x-y|^2)u^2_\rho(x)u^2_\rho(y){\rm d}x{\rm d}y=0,\ \ \lim\limits_{\rho\nearrow\rho^*}\varepsilon_{\rho}^2e(\rho)=0.
\end{aligned}
\end{equation}

Because $u_\rho$ is a positive minimizer of $e(\rho)$, it satisfies the Euler-Lagrange equation \eqref{1.8},
%\begin{equation*}
%-u_\rho+{\rm ln}(1+x^2)u_\rho+\Big(\int_{\R^2}{\rm ln}|x-y| u_\rho^2(y){\rm d}y\Big)u_\rho=\mu_\rho u_\rho+u_\rho^3\ {\rm in}\ \R^2,
%\end{equation*}
where the Lagrange multiplier $\mu_\rho\in\R$ satisfies
\begin{equation}\label{3.11}
\mu_\rho\rho=2e(\rho)+\frac12\int_{\R^2}\int_{\R^2}{\rm ln}|x-y| u^2_\rho(x)u^2_\rho(y){\rm d}x{\rm d}y-\frac12\int_{\R^2} u_\rho^4(x){\rm d}x.
\end{equation}
We thus obtain from \eqref{3.9}--\eqref{3.11} that
\begin{equation}\label{3.12}
\mu_\rho\varepsilon_\rho^2\to-\frac1{\rho^*}\ \  {\rm as}\ \ \rho\nearrow\rho^*.
\end{equation}

2. By the definition of $w_{\rho}$, we reduce from \eqref{1.8} that $w_\rho$ satisfies
\begin{equation}\label{1.12}
\begin{aligned}
&-\Delta w_\rho+\varepsilon_\rho^2{\rm ln}\big[1+\left|\varepsilon_\rho x+x_\rho\right|^2\big]w_\rho+\varepsilon_\rho^2\Big(\int_{\R^2}{\rm ln}|x-y|w_\rho^2(y){\rm d}y\Big)w_\rho+\rho\varepsilon_\rho^2{\rm ln}\varepsilon_\rho w_\rho\\
=&\mu_\rho\varepsilon_\rho^2w_\rho+w_\rho^3\ \ {\rm in}\ \ \R^2.
\end{aligned}
\end{equation}
Since $w_{\rho}$ is bounded uniformly in $H^1(\R^2)$, a straightfoward calculation shows that
there exists a constant $C>0$ such that
for any $x\in\R^2$ and $\rho\in (0,\rho^*)$,
\begin{equation}\label{3.14}
\begin{aligned}
&\frac12\int_{\R^2}{\rm ln} \Big(1+\frac1{|x-y|^2}\Big)w_\rho^2(y){\rm d}y\\
\leqslant&\int_{\R^2}\frac1{|x-y|}w_\rho^2(y){\rm d}y\\
=&\int_{|x-y|<1}\frac1{|x-y|}w_\rho^2(y){\rm d}y+\int_{|x-y|\geqslant 1}\frac1{|x-y|}w_\rho^2(y){\rm d}y\\
\leqslant&\Big(\int_{|x-y|<1}\frac1{|x-y|^{\frac{3}{2}}}{\rm d}y\Big)^{\frac{2}{3}}\Big(\int_{|x-y|<1}w_\rho^6{\rm d}y\Big)^{\frac{1}{3}}+\int_{|x-y|\geqslant 1}w_\rho^2(y){\rm d}y\\
\leqslant&C\|w_\rho\|^2\leqslant C,
\end{aligned}
\end{equation}
where and below we denote
\begin{equation}\label{1:1}
    \|u\|:=\|u\|_{H^1(\R^2)}=\Big\{\int_{\R^2}\big(|\nabla u|^2+u^2\big){\rm d}x\Big\}^{\frac{1}{2}}, \ \ u\in H^1(\R^2).
\end{equation}
%Since
%\begin{equation*}
%    \int_{\R^2}{\rm ln}|x-y|w_\rho^2(y){\rm d}y=\frac12\int_{\R^2}{\rm ln}(1+|x-y|^2)w_\rho^2(y){\rm d}y-\frac12\int_{\R^2}{\rm ln}\Big(1+\frac{1}{|x-y|^2}\Big)w_\rho^2(y){\rm d}y,
%\end{equation*}
We derive from \eqref{3.12}--\eqref{3.14} that
\begin{equation}\label{3.15}
-\Delta w_\rho+\frac1{2\rho^*}w_\rho-w_\rho^3\leqslant0\ \ {\rm in}\ \ \R^2\ \ {\rm as}\ \ \rho\nearrow\rho^*.
\end{equation}

Since $x_\rho$ is a global maximal point of $u_\rho$, the origin is a maximal point of $w_\rho$ for all $\rho\in(0,\rho^*)$. This implies that $-\Delta w_\rho(0)\geqslant0$ for $\rho\in(0,\rho^*)$. We then infer from \eqref{3.15} that there exists some constant $\alpha>0$, independent of $\rho$, such that $ w_\rho(0)\geqslant\alpha>0$ as $\rho\nearrow\rho^*$. Applying the De Giorgi-Nash-Moser theory \cite[Theorem 4.1]{Han}, we deduce from \eqref{3.15} that there exists a constant $C>0$ such that
\begin{equation*}
\int_{B_2(0)}w_\rho^2{\rm d}x\geqslant C\max_{x\in B_1(0)} w_\rho(x)\geqslant C(\alpha):=\eta>0\ \ {\rm as}\ \ \rho\nearrow\rho^*,
\end{equation*}
which thus completes the proof of \eqref{3.3}.

3. Since the sequence $\{w_\rho\}$ is bounded uniformly in $H^1(\R^2)$, up to a subsequence if necessary, there exists a function $w_0\in H^1(\R^2)$ such that $w_\rho\rightharpoonup w_0$ in $H^1(\R^2)$,
$w_\rho\to w_0$ in $L_{loc}^q(\R^2)$ for $q\in[2,\infty)$, and $ w_\rho\to w_0$ almost everywhere in $\R^2$ as $\rho\nearrow\rho^*$.
Moreover, we have $w_{0}\not\equiv 0$ in view of \eqref{3.3}.
%According to Lemma \ref{lem:A.1} in Appendix, by passing to a weak limit of \eqref{1.12}, the function $w_0$ satisfies
%\begin{equation}\label{3.17}
%-\Delta w_0+\frac1{\rho^*}w_0-w_0^3=0\ \ {\rm in}\ \ \R^2.
%\end{equation}
%In addition, since $w_0\neq0$, we further derive that $w_0>0$ by strong maximum principle. Because \eqref{1.5} admits a unique (up to translations) positive radially symmetric solution $Q$, by a simple rescaling, it follows from \eqref{3.17} that
%\begin{equation*}
%w_0(x)=\frac1{\sqrt{\rho^*}}Q(\frac1{\sqrt{\rho^*}}|x-y_0|)\ \ {\rm in}\ \  \R^2.
%\end{equation*}
%Moreover, since the origin is a global maximal point of $w_\rho$ for $\rho\in(0,\rho^*)$, it
%is also a global maximal point of $w_0$. We therefore obtain that $y_0=0$ and hence
%\begin{equation*}
%w_0(x)=\frac1{\sqrt{\rho^*}}Q(\frac1{\sqrt{\rho^*}}|x|)\ \ {\rm in}\ \ \R^2,
%\end{equation*}
%from which it holds that $\|w_0\|_2^2=\rho^*$ and
%\begin{equation}\label{2.26}
%\int_{\R^2}|\nabla w_0|^2{\rm d}x=\frac12\int_{\R^2}w_0^4{\rm d}x.
%\end{equation}
%we thus have that
%\begin{equation}\label{3.18}
%\int_{\R^2}w_\rho^2{\rm d}x\to\int_{\R^2}w_0^2{\rm d}x\ \ {\rm as}\ \  \rho\nearrow\rho^*.
%\end{equation}
Note from the Br$\acute{e}$zis-Lieb lemma (\cite{Willem}) that
\begin{equation*}
\|w_\rho\|_2^2=\|w_0\|_2^2+\|w_\rho-w_0\|_2^2+o(1)\ \ {\rm as}\ \ \rho\nearrow\rho^*,
\end{equation*}
\begin{equation*}
\|w_\rho\|_4^4=\|w_0\|_4^4+\|w_\rho-w_0\|_4^4+o(1)\ \ {\rm as}\ \ \rho\nearrow\rho^*,
\end{equation*}
and
\begin{equation*}
\|\nabla w_\rho\|_2^2=\|\nabla w_0\|_2^2+\|\nabla w_\rho-\nabla w_0\|_2^2+o(1)\ \ {\rm as}\ \ \rho\nearrow\rho^*.
\end{equation*}
Combining \eqref{3.10}, it yields from \eqref{1.4} that
\begin{equation}\label{10}
\begin{aligned}
0=&\lim_{\rho\nearrow\rho^*}\Big(\int_{\R^2}|\nabla w_\rho|^2{\rm d}x-\frac12\int_{\R^2}w_\rho^4{\rm d}x\Big)\\
=&\int_{\R^2}|\nabla w_0|^2{\rm d}x-\frac12\int_{\R^2} w_0^4{\rm d}x+\lim_{\rho\nearrow\rho^*}\Big(\int_{\R^2}|\nabla w_\rho-\nabla w_0|^2{\rm d}x-\frac12\int_{\R^2}|w_\rho-w_0|^4{\rm d}x\Big)\\
\geqslant&\frac{1}{2}\Big(\rho^*\|w_{0}\|_{2}^{-2}-1\Big)\int_{\R^2}\left| w_0\right|^4{\rm d}x+\lim_{\rho\nearrow\rho^*}\Big(\int_{\R^2}|\nabla w_\rho-\nabla w_0|^2{\rm d}x-\frac12\int_{\R^2}|w_\rho-w_0|^4{\rm d}x\Big)\\
\geqslant&\lim_{\rho\nearrow\rho^*}\Big(1-\frac1{\rho^*}\int_{\R^2}|w_\rho-w_0|^2{\rm d}x\Big)\int_{\R^2}|\nabla w_\rho-\nabla w_0|^2{\rm d}x\geqslant0,\\
%=&\lim_{\rho\nearrow\rho^*}\int_{\R^2}|\nabla w_\rho-\nabla w_0|^2{\rm d}x\geqslant0,
\end{aligned}
\end{equation}
from which we obtain that $\|w_0\|_2^2=\rho^*$ and
\begin{equation*}
\int_{\R^2}\big|\nabla w_\rho-\nabla w_0\big|^2{\rm d}x\to0\ \ {\rm as}\ \ \rho\nearrow\rho^*,
\end{equation*}
and hence
\begin{equation*}
    w_{\rho}\rightarrow w_{0}\ \ \mbox{strongly in}\ \ H^1(\R^2)\ \ \mbox{as}\ \ \rho\nearrow\rho^*.
\end{equation*}
%Moreover, because the convergence \eqref{3:1} is independent of the subsequence that we choose, we conclude that \eqref{3:1} holds for the whole sequence.

On the other hand,  the second equality of \eqref{10} yields that
\begin{equation*}
\int_{\R^2}|\nabla w_0|^2{\rm d}x=\frac12\int_{\R^2} w_0^4{\rm d}x,
\end{equation*}
together with the fact $\|w_{0}\|^2_{2}=\rho^*$, which implies that the identity of the inequality \eqref{1.4} is achieved by $w_0$. Thus we derive that $w_0$ satisfies
\begin{equation*}
-\Delta w_0+\frac1{\rho^*}w_0-w_0^3=0\ \ {\rm in}\ \ \R^2,
\end{equation*}
and furthermore there exists $y_0\in\R^2$ such that, up to a subsequence if necessary,
\begin{equation*}
w_\rho(x)\to w_0(x)=\frac{1}{\sqrt{\rho^*}} Q\Big(\frac{1}{\sqrt{\rho^*}}|x+y_0|\Big)\ \ {\rm in}\ \ H^1(\R^2)\ \ {\rm as}\ \ \rho\nearrow\rho^*.
\end{equation*}
Moreover, since the origin is a global maximal point of $w_\rho$ for $\rho\in(0,\rho^*)$, it
is also a global maximal point of $w_0$. This implies that $y_0=0$, and hence
\begin{equation}\label{11}
w_\rho(x)\to w_0(x)=\frac1{\sqrt{\rho^*}}Q\Big(\frac1{\sqrt{\rho^*}}|x|\Big)\ \ {\rm in}\ \ H^1(\R^2)\ \ \hbox{as}\ \ \rho\nearrow\rho^*.
\end{equation}
Since the convergence
 \eqref{11} is independent of the choice of the subsequence, we deduce that \eqref{11}
holds essentially true for the whole sequence.
This therefore completes the proof of Lemma \ref{lem:2.4}.
\qed
\medskip
%It follows from \eqref{2.26} that \eqref{1.3} can be attained by $w_0$, then we derive that there exists a $\beta\in\R$ and $y_0\in\R$ such that
%\begin{equation}\label{2.29}
%w_0(x)=\beta Q(\beta|x+y_0|)\ {\rm in}\ \R^2.
%\end{equation}

The following lemma is concerned with the exponential decay and $L^\infty$-convergence of $w_\rho$ as $\rho\nearrow\rho^*$.

\begin{lemma}\label{lem:3.2}
Assume that $u_{\rho}$ is a positive minimizer of $e(\rho)$ for $\rho\in (0,\rho^*)$,  and $w_{\rho}>0$ is defined by \eqref{3.2}. Then we have
\begin{enumerate}
  \item There exists a large constant $R>0$ such that
  \begin{equation}\label{3.20}
    |w_{\rho}(x)|\leqslant Ce^{-\frac{2|x|}{3\sqrt{\rho^*}}}\ \ and\ \
|\nabla w_\rho(x)|\leqslant Ce^{-\frac{|x|}{2\sqrt{\rho^*}}}\ \ \hbox{for}\ \ |x|\geqslant R\ \ \hbox{as}\ \ \rho\nearrow\rho^*,
\end{equation}
where the constant $C>0$ is independent of $\rho\in(0,\rho^*)$.
  \item The function $w_{\rho}$ satisfies
      \begin{equation}\label{3.21}
w_\rho(x)\to\frac1{\sqrt{\rho^*}}Q\Big(\frac1{\sqrt{\rho^*}}|x|\Big)\ \ in\ \ L^\infty(\R^2)\ \ as\ \ \rho\nearrow\rho^*.
      \end{equation}
      Moreover, the global maximal point $x_{\rho}$ of $u_{\rho}$ is unique and satisfies $x_{\rho}\to 0$ as $\rho\nearrow\rho^*$.
\end{enumerate}
\end{lemma}

\noindent{\bf Proof.}
1. Since $\varepsilon_\rho{\rm ln}\varepsilon_\rho\to0$ as $\rho\nearrow\rho^*$, inserting \eqref{3.12} and \eqref{3.14} into \eqref{1.12} yields that
\begin{equation}\label{3.01}
-\Delta w_\rho+\frac{5}{9\rho^*}w_\rho-w_\rho^3\leqslant0\ \ {\rm in}\ \ \R^2\ \ {\rm as}\ \ \rho\nearrow\rho^*.
 \end{equation}
Applying De Giorgi-Nash-Moser theory \cite[Theorem 4.1]{Han}, we then derive from \eqref{3.01} that there exists a large constant $R>0$ such that
%\begin{equation}\label{3.22}
%\max_{B_1(\xi)}w_\rho(x)\leqslant C\Big(\int_{B_2(\xi)}w_\rho^2{\rm d}x\Big)^{1/2},
%\end{equation}
%where $\xi$ is an arbitrary point in $\R^2$. Since $w_\rho\to w_0$ in $L^2(\R^2)$ as $\rho\nearrow\rho^*$, we know that
%\begin{equation*}
%\lim_{r\to\infty}\int_{|x|\geqslant r}|w_\rho|^2{\rm d}x=0\ {\rm holds\ uniformly}\ {\rm as}\ \rho\nearrow\rho^*.
%\end{equation*}
%That together with \eqref{3.22} leads to that
%\begin{equation*}
%\lim_{|x|\to\infty}w_\rho(x)=0\ {\rm holds\ uniformly}\ {\rm as}\ \rho\nearrow\rho^*,
%\end{equation*}
\begin{equation}\label{3.23}
w_\rho(x)\leqslant\frac1{3\sqrt{\rho^*}}\ \ {\rm uniformly}\ {\rm for}\ \ |x|\geqslant R\ \ {\rm as}\ \ \rho\nearrow\rho^*,
\end{equation}
due to the fact that $w_\rho\to w_0$ in $L^2(\R^2)$ as $\rho\nearrow\rho^*$.
We now derive from \eqref{3.01} and \eqref{3.23} that
\begin{equation}\label{3.24}
-\Delta w_\rho+\frac4{9\rho^*}w_\rho\leqslant0\ \ {\rm  uniformly}\ {\rm for}\ \ |x|\geqslant R\ \ {\rm as}\ \ \rho\nearrow\rho^*.
\end{equation}
Applying the comparison principle to \eqref{3.24} gives that there exists a constant $C>0$ such that
\begin{equation}\label{3:4}
\left|w_\rho(x)\right|\leqslant Ce^{-\frac{2|x|}{3\sqrt{\rho^*}}}\ \ {\rm  uniformly\ for}\ |x|\geqslant R\ \ {\rm as}\ \ \rho\nearrow\rho^*.
\end{equation}
Since the proof of exponential decay for $|\nabla w_\rho|$ is more involved, we leave it to  Lemma \ref{lem:A.2} in the Appendix. This completes the proof of \eqref{3.20}.

2. To obtain the estimate of $x_{\rho}$ as $\rho\nearrow\rho^*$, we first claim that
\begin{equation}\label{3.31}
    \int_{\R^2}{\rm ln}\big(1+|x|^2\big)u_\rho^2(x){\rm d}x\rightarrow 0\ \ \hbox{as}\ \ \rho\nearrow\rho^*.
\end{equation}
In fact, by the the exponential decays \eqref{1.6} and \eqref{3.20}, it follows from \eqref{3.4} that
\begin{equation*}
\begin{aligned}
\int_{\R^2}{\rm ln}(1+|x|^2)u_\rho^2(x+x_\rho){\rm d}x
=&\int_{\R^2}{\rm ln}(1+\varepsilon_\rho^2|x|^2)w_\rho^2(x){\rm d}x\\
\leqslant &\,\varepsilon_\rho^2\int_{\R^2}|x|^2w_\rho^2(x){\rm d}x\\
=&\big(1+o(1)\big)\rho^*\varepsilon_\rho^2\int_{\R^2}|x|^2Q^2(x){\rm d}x\ \ \ \ \hbox{as}\ \ \rho\nearrow\rho^*,
\end{aligned}
\end{equation*}
which shows that $u_{\rho}(\cdot+x_\rho)\in S_{\rho}$ as $\rho\nearrow\rho^*$. By the definition of $e(\rho)$, we have
\begin{equation*}
e(\rho)=E(u_\rho)\leqslant E\big(u_{\rho}(\cdot+x_\rho)\big),
\end{equation*}
which implies that
\begin{equation}\label{3.32}
\begin{aligned}
0\leqslant\int_{\R^2}{\rm ln}\big(1+|x|^2\big)u_\rho^2(x){\rm d}x
\leqslant&\int_{\R^2}{\rm ln}\big(1+|x|^2\big)u_\rho^2(x+x_\rho){\rm d}x\\
\leqslant&\big(1+o(1)\big)\rho^*\varepsilon_\rho^2\int_{\R^2}|x|^2Q^2(x){\rm d}x\to0\ \ \ \hbox{as}\ \ \rho\nearrow\rho^*,
\end{aligned}
\end{equation}
and hence the claim \eqref{3.31} is proved.

Based on \eqref{3.31}, we next prove that any maximal point $x_{\rho}$ of $u_{\rho}$ satisfies $x_\rho\to0$ as $\rho\nearrow\rho^*$. On the contrary, we assume that there exist a constant $\gamma>0$ and a subsequence $\{\rho_{k}\}$, where $\rho_{k}\nearrow \rho^*$ as $k\rightarrow\infty$, such that
\begin{equation}\label{122}
    |x_{\rho_{k}}|\geqslant\gamma\ \ \hbox{as}\ \ k\rightarrow\infty.
\end{equation}
By Fatou's Lemma, it then follows from \eqref{3.3} and \eqref{122} that
\begin{equation*}
\begin{aligned}
\liminf_{k\rightarrow\infty}\int_{\R^2}{\rm ln}\big(1+|x|^2\big)u_{\rho_{k}}^2(x){\rm d}x
=&\liminf_{k\rightarrow\infty}\int_{\R^2}{\rm ln}\big(1+\left|\varepsilon_{\rho_{k}}x+x_{\rho_{k}}\right|^2\big)w_{\rho_{k}}^2(x){\rm d}x\\
\geqslant&\liminf_{k\rightarrow\infty}\int_{B_2(0)}{\rm ln}\big(1+\left|\varepsilon_{\rho_{k}} x+x_{\rho_{k}}\right|^2\big)w_{\rho_{k}}^2(x){\rm d}x\\
\geqslant&\frac{1}{2}\ln(1+\gamma^2)\eta>0,
\end{aligned}
\end{equation*}
which is a contradiction in view of \eqref{3.31}. Hence, we have $x_{\rho}\rightarrow 0$ as $\rho\nearrow\rho^*$.

In the following, we prove that
\begin{equation*}
w_\rho\to w_0=\frac1{\sqrt{\rho^*}}Q\Big(\frac1{\sqrt{\rho^*}}|x|\Big)\ \ {\rm in}\ \  L^\infty(\R^2)\ \ {\rm as}\ \ \rho\nearrow\rho^*.
\end{equation*}
In view of the exponential decays \eqref{1.6} and \eqref{3.20}, it suffices to show the $L^\infty$-uniform convergence of $w_\rho$ on any compact domain of $\R^2$ as $\rho\nearrow\rho^*$. Denote
\begin{equation*}
\begin{aligned}
G_\rho(x):=&-\varepsilon_\rho^2{\rm ln}\big(1+|\varepsilon_\rho x+x_\rho|^2\big) w_\rho-\varepsilon_\rho^2\Big(\int_{\R^2}{\rm ln}|x-y|w_\rho^2(y){\rm d}y\Big) w_\rho\\
&-\rho\varepsilon_\rho^2{\rm ln}\varepsilon_\rho w_\rho+\mu_\rho\varepsilon_\rho^2 w_\rho+w_\rho^3\ \ {\rm in}\ \ \R^2.
\end{aligned}
\end{equation*}
We then get from \eqref{1.12} that
\begin{equation}\label{3.27}
-\Delta w_\rho=G_\rho\ \ {\rm in}\ \ \R^2.
\end{equation}
Because $\{w_\rho\}$ is bounded uniformly in $H^1(\R^2)$ and satisfies the exponential decay \eqref{3.20}, we obtain that $G_\rho$ is also bounded uniformly in $L^2_{loc}(\R^2)$ as $\rho\nearrow\rho^*$, in view of the fact that $x_\rho\to0$ as $\rho\nearrow\rho^*$. For any $R>0$, it follows from \cite[Theorem 8.8]{Gilbarg} that
\begin{equation*}
\|w_\rho\|_{H^2(B_R)}\leqslant C\Big(\|w_\rho\|_{H^1(B_{R+1})}+\|G_\rho\|_{L^2(B_{R+1})}\Big),
\end{equation*}
where $C>0$ is independent of $\rho$ and $R$. Therefore, $\{w_\rho\}$ is also bounded uniformly in $H_{loc}^2(\R^2)$ as $\rho\nearrow\rho^*$. Since the embedding $H^2(B_R)\hookrightarrow L^\infty(B_R)$ is compact (cf. \cite[Theorem 7.26]{Gilbarg}), we deduce from \eqref{11} that there exists a subsequence of $\{w_\rho\}$, still denoted by $\{w_\rho\}$, such that
\begin{equation*}
\lim_{\rho\nearrow\rho^*}w_\rho(x)= w_0(x)\ \ {\rm in}\ \ L^\infty(B_R).
\end{equation*}
Since $R > 0$ is arbitrary, we get that
\begin{equation}\label{3:3}
\lim_{\rho\nearrow\rho^*}w_\rho(x)=w_0(x)\ \ {\rm in}\ \ L_{loc}^\infty(\R^2).
\end{equation}
Because the above convergence is independent of the subsequence that we choose,
\eqref{3:3} holds essentially for the whole sequence.
This shows that \eqref{3.21} holds true.

We now prove the uniqueness of $x_\rho$ as $\rho\nearrow\rho^*$, where $x_\rho$ is a global maximal point of $u_\rho$. Since $\{w_{\rho}\}$ is bounded uniformly in $H^1(\R^2)$, we get that $\{ w_{\rho}\}$ is bounded uniformly in $L^q_{loc}(\R^2)$ as $\rho\nearrow\rho^*$ for $q\in[2,\infty)$. Applying the $L^p$ estimate \cite[Theorem 9.11]{Gilbarg} to \eqref{3.27}, we then derive that $\{w_{\rho}\}$ is bounded uniformly in $W^{2,q}_{loc}(\R^2)$ as $\rho\nearrow\rho^*$. By a standard Sobolev embedding, it holds that
\begin{equation}\label{3:5}
    \{w_{\rho}\}\ \ \mbox{is bounded uniformly in}\ \ C^{1,\alpha}_{loc}(\R^2)\ \ \mbox{as}\ \ \rho\nearrow\rho^*.
\end{equation}
Furthermore, since it gives from \cite[Proposition 2.3]{Silvia1} that
$$\varepsilon_\rho^2{\rm ln}\big(1+|\varepsilon_\rho x+x_\rho|^2\big)\in C^\alpha_{loc}(\R^2), \ \ \varepsilon_\rho^2\int_{\R^2}{\rm ln}|x-y|w_\rho^2(y){\rm d}y\in C^{2,\alpha}_{loc}(\R^2),$$
we have $G_{\rho}$ is bounded uniformly in $C^{\alpha}\big(B_{R}(0)\big)$, where $R>0$ is large enough. Applying the Schauder estimate \cite[Theorem 6.2]{Gilbarg}, it yields from \eqref{3.27} that $\{w_{\rho}\}$ is bounded uniformly in  $C^{2,\alpha}\big(B_{R}(0)\big)$ as $\rho\nearrow\rho^*$. Therefore, up to a subsequence if necessary, there exists a function $\tilde w_0\in C^2\big(B_{R}(0)\big)$ such that
\begin{equation*}
w_\rho\to \tilde w_0\ \ {\rm in}\ \ C^2\big(B_{R}(0)\big)\ \ {\rm as}\ \ \rho\nearrow\rho^*,
\end{equation*}
and we thus deduce from \eqref{3.21} that $\tilde w_0=w_0$, so that
\begin{equation}\label{3.33}
w_\rho\to w_0\ \ {\rm in}\ \ C^2\big(B_{R}(0)\big)\ \ {\rm as}\ \ \rho\nearrow\rho^*,
\end{equation}
where $R>0$ is a constant large enough.
Since the convergence \eqref{3.33} is independent of the subsequence that we choose, it holds essentially for the whole sequence.

Because the origin is the unique global maximal point of $Q(x)$, the convergence of \eqref{3.33} shows that all local maximal points of $w_\rho$ must approach the origin and hence stay in a small ball $B_\delta(0)$ as $\rho\nearrow\rho^*$ for some $\delta>0$. Since $Q''(0)<0$, we conclude that $Q''(r)<0$ for $r\in[0,\delta)$. It then follows from \cite[Lemma 4.2]{Ni} that as $\rho\nearrow\rho^*$, each $w_\rho$ has a unique global maximum point which is exactly the origin. This further proves the uniqueness of global maximal points for $u_\rho$ as $\rho\nearrow\rho^*$. This completes the proof of Lemma \ref{lem:3.2}.
\qed
\medskip
\vskip 0.01truein

Following the $L^\infty$-uniform convergence of positive minimizers $u_\rho$ for $e(\rho)$ as $\rho\nearrow\rho^*$, the rest part of this section is to complete the proof of Theorem \ref{thm:1.2}. In order to reach this aim, more importantly, we shall set up the refined  energy estimates, based on which we shall analyze the blow up point $x_\rho$ and the blow up rate $\varepsilon_\rho^{-1}$ of $u_\rho$  as $\rho\nearrow\rho^*$. In addition, the convergence of $u_\rho$ in $X$ can be obtained in view of the exponential decay.\vskip 0.10truein

{\noindent \bf Proof of Theorem \ref{thm:1.2}.}
To complete the proof of Theorem \ref{thm:1.2}, we note from Lemmas \ref{lem:2.4} and \ref{lem:3.2} that the rest is to prove
\begin{equation}\label{3.35M}
\bar\varepsilon_\rho:=\sqrt{\rho^*}\varepsilon_\rho=\frac2{\rho^*}(\rho^*-\rho)^{\frac{1}{2}}\big(1+o(1)\big)\ \  \hbox{as}\ \ \rho\nearrow\rho^*,
\end{equation}
and
\begin{equation}\label{3.35}
\frac{x_\rho}{\bar{\varepsilon}_\rho}\to0\ \ {\rm as}\ \ \rho\nearrow\rho^*.
\end{equation}

We first obtain the upper estimate of $e(\rho)$ as $\rho\nearrow\rho^*$.
Similar to \eqref{2.9}, one can get that for any  $\tau>0$,
\begin{equation}\label{3.6}
\begin{aligned}
e(\rho)\leqslant E(u_\tau)\leqslant&\frac{\rho(\rho^*-\rho)}{2\rho^*}\tau^2-\frac14\rho^2{\rm ln}\tau+\frac\rho{2\rho^*\tau^2}\int_{\R^2}|x|^2Q^2(x){\rm d}x\\
&+\frac{\rho^2}{4(\rho^*)^2}\int_{\R^2}\int_{\R^2}{\rm ln}|x-y|Q^2(x)Q^2(y){\rm d}x{\rm d}y.
\end{aligned}
\end{equation}
Setting
\begin{equation*}
\tau =\left[\frac{\rho\rho^*}{4(\rho^*-\rho)}\right]^{\frac{1}{2}}>0
\end{equation*}
into \eqref{3.6}, we deduce that
\begin{equation}\label{3.37}
\begin{aligned}
    e(\rho)&\leqslant \frac{(\rho^*)^2}{8}-\frac{(\rho^*)^2}{4}\ln\rho^*+\frac{(\rho^*)^2}{8}\ln\big[4(\rho^*-\rho)\big]\\
    &\quad+\frac{1}{4}\int_{\R^2}\int_{\R^2}\ln|x-y|Q^2(x)Q^2(y){\rm d}x{\rm d}y+o(1)\ \ \hbox{as}\ \ \rho\nearrow\rho^*.
\end{aligned}
\end{equation}

To derive the lower estimate of $e(\rho)$ as $\rho\nearrow\rho^*$, take $u_\rho\in X$ to be a positive minimizer of $e(\rho)$ satisfying $\|u_\rho\|_2^2=\rho<\rho^*$. Denote
\begin{equation*}
\bar w_\rho(x)=\bar\varepsilon_\rho u_\rho\big(\bar\varepsilon_\rho x+x_\rho\big)=\sqrt{\rho^*}\varepsilon_\rho u_\rho\big(\sqrt{\rho^*}\varepsilon_\rho x+x_\rho\big)>0\ \ {\rm in}\ \ \R^2,
\end{equation*}
where $x_\rho$ is the maximal point of $u_\rho$.  By the exponential decays  \eqref{1.6} and \eqref{3.20}, we obtain that for any $\epsilon>0$, there exists a constant $R>0$ such that
 \begin{equation*}
    2\int_{B_R^c}{\rm ln}(1+|x|^2)\bar w_\rho^2(x){\rm d}x+2\int_{B_R^c}{\rm ln}(1+|x|^2)Q^2(x){\rm d}x<\frac{\epsilon}{2},
 \end{equation*}
 from which and \eqref{11} we deduce that
\begin{equation*}
\begin{aligned}
&\int_{\R^2}{\rm ln}(1+|x|^2)\Big(\bar w_\rho(x)-Q(x)\Big)^2{\rm d}x\\
\leqslant&{\rm ln}(1+R^2)\int_{B_R}\Big(\bar w_\rho(x)-Q(x)\Big)^2{\rm d}x+2\int_{B_R^c}{\rm ln}(1+|x|^2)\bar w_\rho^2(x){\rm d}x+2\int_{B_R^c}{\rm ln}(1+|x|^2)Q^2(x){\rm d}x\\
<&\epsilon\ \ \hbox{as}\ \ \rho\nearrow\rho^*.
\end{aligned}
\end{equation*}
This then implies that $\bar w_\rho\to Q$ in $X$ as $\rho\nearrow\rho^*$.
Moreover, combining \eqref{3.14} with \eqref{11} yields that
\begin{eqnarray}\label{100}
\begin{aligned}
&\left|\int_{\R^2}\int_{\R^2}{\rm ln}|x-y|\bar w^2_\rho(x)\bar w^2_\rho(y){\rm d}x{\rm d}y-\int_{\R^2}\int_{\R^2}{\rm ln}|x-y|Q^2(x)Q^2(y){\rm d}x{\rm d}y\right|\\
\leqslant&\left|\int_{\R^2}\int_{\R^2}{\rm ln}|x-y|\Big(\bar w^2_\rho(x)-Q^2(x)\Big)\bar w^2_\rho(y){\rm d}x{\rm d}y\right|\\
&+\left|\int_{\R^2}\int_{\R^2}{\rm ln}|x-y|Q^2(x)\Big(\bar w^2_\rho(y)-Q^2(y)\Big){\rm d}x{\rm d}y\right|\\
\leqslant&\rho\int_{\R^2}{\rm ln}(1+|x|^2)\left|\bar w^2_\rho(x)-Q^2(x)\right|{\rm d}x+\int_{\R^2}\left|\bar w^2_\rho(x)-Q^2(x)\right|{\rm d}x\int_{\R^2}{\rm ln}(1+|y|^2)\bar w_\rho^2(y){\rm d}y\\
&+\int_{\R^2}\int_{\R^2}\frac1{|x-y|}\left|\bar w^2_\rho(x)-Q^2(x)\right|\bar w_\rho^2(y){\rm d}x{\rm d}y\\
&+\int_{\R^2}{\rm ln}(1+|x|^2)Q^2(x){\rm d}x\int_{\R^2}\left|\bar w^2_\rho(y)-Q^2(y)\right|{\rm d}y+\rho^*\int_{\R^2}{\rm ln}(1+|y|^2)\left|\bar w^2_\rho(y)-Q^2(y)\right|{\rm d}y\\
&+\int_{\R^2}\int_{\R^2}\frac1{|x-y|}Q^2(x)\left|\bar w^2_\rho(y)-Q^2(y)\right|{\rm d}x{\rm d}y\to0\ \ {\rm as}\ \ \rho\nearrow\rho^*.
\end{aligned}
\end{eqnarray}
We then deduce from \eqref{1.4}, \eqref{3.4} and \eqref{100} that
%According to the (iii) of Lemma \ref{lem:2.4}, we know that $\bar w_k\to Q$ in $H^1(\R^2)\cap L^{\infty}(\R^2)$ as $k\to\infty$. We claim that as $k\to\infty$,
%\begin{equation}\label{2.38}
%\bar\varepsilon_k\approx2\sqrt2\Big(\rho^*-\rho_k\Big)^{1/2}\left[\rho_k^2+\sqrt{\rho_k^4+64M\Big(\rho^*-\rho_k\Big)}\right]^{-1/2}.
%\end{equation}
%Indeed, we deduce from \eqref{1.3}, (iii) of Lemma \ref{lem:2.4}, and (iii) and (iv) of Lemma \ref{lem:2.1} that passing to a subsequence if necessary,
\begin{equation}
\begin{aligned}\label{3.38}
e(\rho)=&E(u_\rho)\\
=&\frac12\bar\varepsilon_\rho^{-2}\Big(\int_{\R^2}\left|\nabla\bar w_\rho\right|^2{\rm d}x-\frac{\rho^*}{2\rho}\int_{\R^2}\bar w_\rho^4{\rm d}x\big)+\frac12\int_{\R^2}{\rm ln}\big(1+\left|\bar\varepsilon_\rho x+x_\rho\right|^2\Big)\bar w_\rho^2(x){\rm d}x\\
&+\frac14\rho^2{\rm ln}\bar\varepsilon_\rho+\frac14\int_{\R^2}\int_{\R^2}{\rm ln}|x-y|\bar w^2_\rho(x)\bar w^2_\rho(y){\rm d}x{\rm d}y+\frac14\Big(\frac{\rho^*}{\rho}-1\Big)\bar{\varepsilon}_\rho^{-2}\int_{\R^2}\bar w_\rho^4{\rm d}x\\
\geqslant&\frac{\rho^*-\rho}{4\rho}\bar\varepsilon_\rho^{-2}\int_{\R^2}\bar w_\rho^4{\rm d}x+\frac14\rho^2{\rm ln}\bar\varepsilon_\rho
+\frac14\int_{\R^2}\int_{\R^2}{\rm ln}|x-y|\bar w^2_\rho(x)\bar w^2_\rho(y){\rm d}x{\rm d}y\\
=&\frac{\rho^*(\rho^*-\rho)}{2\rho}\big(1+o(1)\big)\bar\varepsilon_\rho^{-2}
%+\frac12(1+o(1))\bar\varepsilon_k^2\int_{B_{1/\sqrt{\bar\varepsilon_k}}(0)}\Big(x+\frac{x_k}{\bar\varepsilon_k}\Big)^2Q^2(x) {\rm d}x\\
+\frac14\rho^2{\rm ln}\bar\varepsilon_\rho+\frac14\big(1+o(1)\big)\int_{\R^2}\int_{\R^2}{\rm ln}|x-y|Q^2(x)Q^2(y){\rm d}x{\rm d}y\\
\geqslant& \frac{(\rho^*)^2}{8}-\frac{(\rho^*)^2}{4}\ln\rho^*+\frac{(\rho^*)^2}{8}\ln\big[4(\rho^*-\rho)\big]\\
&+\frac14\int_{\R^2}\int_{\R^2}{\rm ln}|x-y|Q^2(x)Q^2(y){\rm d}x{\rm d}y+o(1)\ \ \hbox{as}\ \ \rho\nearrow\rho^*,
\end{aligned}
\end{equation}
where the identity of the last inequality is achieved at $\bar\varepsilon_\rho>0$ satisfying (\ref{3.35M}), i.e., $\bar\varepsilon_\rho=\frac2{\rho^*}(\rho^*-\rho)^{\frac{1}{2}}\big(1+o(1)\big)$ as $\rho\nearrow\rho^*$.
Together with \eqref{3.37} and \eqref{3.38},
we now conclude that
%the energy estimate of $e(\rho_k)$ satisfies as $k\to\infty$,
\begin{equation*}
\begin{aligned}
e(\rho)&\approx\frac{(\rho^*)^2}{8}-\frac{(\rho^*)^2}{4}\ln\rho^*+\frac{(\rho^*)^2}{8} \ln\big[4(\rho^*-\rho)\big]\\
&\quad+\frac14\int_{\R^2}\int_{\R^2}{\rm ln}|x-y|Q^2(x)Q^2(y){\rm d}x{\rm d}y\ \ \hbox{as}\ \ \rho\nearrow\rho^*,
\end{aligned}
\end{equation*}
and $\bar\varepsilon_\rho>0$ satisfies (\ref{3.35M}). In addition, since $\bar w_\rho\to Q$ in $X$ as  $\rho\nearrow\rho^*$, we derive from Lemmas \ref{lem:2.4} and \ref{lem:3.2} that
%\begin{equation}\label{3.012}
%    \lim\limits_{k\rightarrow\infty}\int_{\R^2}{\rm ln}[1+\Big(\bar\varepsilon_kx+x_k\Big)^2]\bar w_k^2(x){\rm d}x=0,\ \
%\end{equation}
%and
\begin{equation*}
\lim\limits_{\rho\nearrow\rho^*}\frac{2(\rho^*-\rho)^{\frac{1}{2}}}{\rho^*}
u_{\rho}\Big(\frac{2(\rho^*-\rho)^{\frac{1}{2}}}{\rho^*}x+x_{\rho}\Big)=Q(x)\ \ \hbox{in}\ \ X\cap L^\infty(\R^2).
\end{equation*}

We finally  prove that $\bar\varepsilon_\rho^{-1}x_\rho\to0$ as $\rho\nearrow\rho^*$.
%For any sequence $\{\rho_k\}$ with $\rho_k\to\rho^*$ as $k\to\infty$,
It follows from \eqref{3.32} that
\begin{equation}\label{3.41}
\int_{\R^2}{\rm ln}(1+|x|^2)u_\rho^2(x+x_\rho){\rm d}x\leqslant\big(1+o(1)\big)\bar{\varepsilon}_\rho^2\int_{\R^2}|x|^2Q^2(x){\rm d}x\ \ \hbox{as}\ \ \rho\nearrow\rho^*.
\end{equation}
On the other hand, we note that $x_\rho\to0$ and $\bar w_\rho\to Q$ in $H^1(\R^2)$ as $\rho\nearrow\rho^*$. Since $u_\rho$ is a positive minimizer of $e(\rho)$, we now have
\begin{equation}\label{3.42}
\begin{aligned}
&\int_{\R^2}{\rm ln}(1+|x|^2)u_\rho^2(x){\rm d}x\\
%&=&\int_{\R^2}{\rm ln}\left[1+\Big(\bar\varepsilon_kx+x_k\Big)^2\right]\bar w_k^2(x){\rm d}x\\
\geqslant&\int_{B_{\frac{1}{\sqrt{\bar\varepsilon_\rho}}}(0)}{\rm ln}\big(1+\left|\bar\varepsilon_\rho x+x_\rho\right|^2\big)\bar w_\rho^2(x){\rm d}x\\
=&\big(1+o(1)\big)\bar\varepsilon_\rho^2\int_{B_{\frac{1}{\sqrt{\bar\varepsilon_\rho}}}(0)}\Big|x+\frac{x_{\rho}}{\bar{\varepsilon}_{\rho}}\Big|^2 Q^2(x){\rm d}x\ \ \hbox{as}\ \ \rho\nearrow\rho^*.\\
%&=&(1+o(1))\rho^*\varepsilon_k^2\int_{B_{1/\sqrt{\bar\varepsilon_k}}(0)}(x+\bar\varepsilon_k^{-1}x_k)^2 \bar w_k^2(x){\rm d}x
\end{aligned}
\end{equation}
Similar to \eqref{3.32}, we deduce from \eqref{3.41} and \eqref{3.42} that
\begin{equation}\label{3.43}
    \bar\varepsilon_\rho^2\int_{B_{\frac{1}{\sqrt{\bar\varepsilon_\rho}}}(0)}\Big|x+\frac{x_{\rho}}{\bar{\varepsilon}_{\rho}}\Big|^2 Q^2(x){\rm d}x\leqslant
    \big(1+o(1)\big)\bar{\varepsilon}_\rho^2\int_{\R^2}|x|^2Q^2(x){\rm d}x\ \ \hbox{as}\ \ \rho\nearrow\rho^*.
\end{equation}
On the contrary, we now assume that \eqref{3.35} is false, i.e., $\frac{x_{\rho}}{\bar{\varepsilon}_{\rho}}\nrightarrow0$ as $\rho\nearrow\rho^*$. Then there exist a constant $\tau>0$ and a subsequence $\{\rho_{k}\}$, where $\rho_{k}\nearrow \rho^*$ as $k\rightarrow\infty$, such that
\begin{equation}\label{3.44}
   \frac{ |x_{\rho_{k}}|}{\bar\varepsilon_{\rho_{k}}}\geqslant\tau>0\  \ \hbox{as}\ \ k\rightarrow\infty.
\end{equation}
By Fatou's Lemma, it then follows from \eqref{3.43} and \eqref{3.44} that
\begin{equation*}
    \int_{\R^2}|x|^2Q^2dx< \int_{\R^2}\liminf_{k\rightarrow\infty}\Big|x+\frac{ x_{\rho_{k}}}{\bar\varepsilon_{\rho_{k}}}\Big|^2Q^2(x)dx\leqslant\int_{\R^2}|x|^2Q^2dx,
\end{equation*}
which is a contradiction. This implies that \eqref{3.35} holds true, and the proof of Theorem  \ref{thm:1.2} is therefore complete.\qed
\medskip
%Moreover, it follows from \eqref{2.42}, \eqref{2.43} and \eqref{2.11} that $\{\bar\varepsilon_k^{-1}x_k\}$ is bounded. More precisely, one can verify that, passing to subsequence if necessary, there exists $y_1\in\R^2$ such that
%\begin{equation}\label{2.44}
%\frac{x_k}{\bar\varepsilon_k}\to y_1\ {\rm as}\ k\to\infty.
%\end{equation}
%Since
%\begin{equation*}
%h(y):=\int_{\R^2}(x+y)^2Q^2(x){\rm d}x\ {\rm in}\ \R^2
%\end{equation*}
%possesses a unique minimal in $\R^2$ which is the origin, combining with \eqref{2.42}-\eqref{2.44}, we obtain that $y_1=0$. Consequently, based on (iii) of Lemma \ref{lem:2.4} and the above conclusions, we conclude that
%\begin{equation*}
%\varepsilon_ku_k(\varepsilon_kx+x_k)\to Q(x)\ {\rm in}\ X\cap L^\infty(\R^2)
%\end{equation*}
%and
%$x_k/\varepsilon_k\to0$ as $k\to\infty$. In addition, according to \eqref{A.7} and Lemma \ref{lem:A.2}, we know that \eqref{1.11} holds. This completes the proof of Theorem \ref{thm:1.2}.

\section{Local Uniqueness of Minimizers for $e(\rho)$}

Based on the limiting behavior and asymptotic properties of positive minimizers established in Theorem \ref{thm:1.2}, this section is devoted to the proof of  Theorem \ref{thm:1.3} on the local uniqueness of positive minimizers $u_\rho$ for $e(\rho)$ as $\rho\nearrow\rho^*$.
Throughout the whole section, we denote $u_{k}$ to be a positive minimizer of $e(\rho_{k})$ for $0<\rho_{k}<\rho^*$, and define
\begin{equation}\label{4.2}
    \varepsilon_{k}=\frac{2(\rho^*-\rho_{k})^{\frac12}}{\rho^*}>0,
\end{equation}
where $\rho_{k}\nearrow\rho^*$ as $k\rightarrow\infty$.
Note that the minimizer $u_k$ of $e(\rho_k)$ satisfies
\begin{equation}\label{4.3}
-\Delta u_k+{\rm ln}\big(1+|x|^2\big)u_k+\Big(\int_{\R^2}{\rm ln}|x-y|u_k^2(y){\rm d}y\Big)u_k=\mu_k u_k+u_k^3\ \ {\rm in}\ \ \R^2,
\end{equation}
where the associated Lagrange multiplier $\mu_k\in\R$ satisfies
\begin{equation}\label{4.4}
\mu_k\rho_k=2e(\rho_k)+\frac12\int_{\R^2}\int_{\R^2}{\rm ln}|x-y|u_k^2(x)u_k^2(y){\rm d}x{\rm d}y-\frac12\int_{\R^2}u_k^4(x){\rm d}x.
\end{equation}
Similar to \eqref{3.12}, it holds that
\begin{equation}\label{4.5}
\mu_k\varepsilon_k^2\to-1\ \ {\rm as}\ \ k\to\infty,
\end{equation}
where $\varepsilon_k>0$ is as in \eqref{4.2}.

Set
\begin{equation}\label{4.6}
    w_{k}(x):=\varepsilon_{k}u_{k}(\varepsilon_{k}x+x_{k})>0\ \ \ {\rm in}\ \ \R^2,
\end{equation}
where $x_{k}$ is the unique maximal point of $u_{k}$ as $k\rightarrow\infty$.
We then get from   \eqref{4.3} and   \eqref{4.6}  that $w_k>0$ satisfies
\begin{equation*}\begin{aligned}
&-\Delta w_k+\varepsilon_k^2{\rm ln}\big(1+\left|\varepsilon_k x+x_k\right|^2\big)w_k+\varepsilon_k^2\Big(\int_{\R^2}{\rm ln}|x-y|w_k^2(y){\rm d}y\Big)w_k+\rho_k\varepsilon_k^2{\rm ln}\varepsilon_kw_k\\
&=\mu_k\varepsilon_k^2w_k+w_k^3\ \ \ {\rm in}\ \ \R^2.
\end{aligned}\end{equation*}
Note from Lemma \ref{lem:3.2} (1) that there exist constants $C>0$ and $R>0$ large enough such that $w_k$ satisfies
\begin{equation}\label{4.8}
\big|w_k(x)\big|\leqslant Ce^{-\frac{2|x|}{3}},\quad  \quad \big|\nabla w_k(x)\big|\leqslant Ce^{-\frac{|x|}{2}} \ \  {\rm uniformly\ \ for}\ \ |x|\geqslant R
\end{equation}
%and
%\begin{equation}\label{4.8}
%\leqslant Ce^{-2|x|/3}\ \ {\rm for}\ \ |x|\geqslant R
%\end{equation}
as $k\to\infty$. We also define the linearized operator $\mathcal L$ of \eqref{1.5}:
\begin{equation*}
\mathcal L:=-\Delta+(1-3Q^2)\ \ {\rm in}\ \ \R^2,
\end{equation*}
where $Q(x)=Q(|x|)>0$ is the unique positive solution of \eqref{1.5} and satisfies the exponential decay \eqref{1.6}. Recall from \cite{Kwong,Ni} that
\begin{equation}\label{4.1}
{\rm ker}(\mathcal L)={\rm span}\Big\{\frac{\partial Q}{\partial x_1},\frac{\partial Q}{\partial x_2}\Big\}.
\end{equation}
We are now ready to address the local uniqueness of positive minimizers $u_\rho$ for $e(\rho)$ as $\rho\nearrow\rho^*$.\vskip 0.10truein

\noindent{\bf Proof of Theorem \ref{thm:1.3}.}
Argue by contradiction, suppose that there exist two different positive minimizers $u_{1,k}$ and $u_{2,k}$ for $e(\rho_k)$, where $\rho_k\nearrow\rho^*$ as $k\to\infty$. Let $x_{1,k}$ and $x_{2,k}$ be the unique maximal point of
$u_{1,k}$ and $u_{2,k}$ as $k\to\infty$, respectively.

Denote $$V(x):={\rm ln}(1+|x|^2).$$ Following \eqref{4.3}, we have
\begin{equation}\label{4.10}
-\Delta u_{i,k}+V(x)u_{i,k}+\Big(\int_{\R^2}{\rm ln}|x-y|u_{i,k}^2(y){\rm d}y\Big)u_{i,k}=\mu_{i,k} u_{i,k}+u_{i,k}^3\ \ {\rm in}\ \ \R^2,\ \ i=1,\ 2.
\end{equation}
Define
\begin{equation}\label{4:3}
\bar u_{i,k}(x):=\varepsilon_k u_{i,k}(x)>0\ \ \ {\rm in}\ \ \R^2,\ \ i=1,\ 2.
\end{equation}
One can check that $\bar u_{i,k}>0$ satisfies
\begin{equation}\label{4.11}
-\varepsilon_k^2\Delta\bar u_{i,k}+\varepsilon_k^2V(x)\bar u_{i,k}+\Big(\int_{\R^2}{\rm ln}|x-y|\bar u_{i,k}^2(y){\rm d}y\Big)\bar u_{i,k}=\mu_{i,k}\varepsilon_k^2\bar u_{i,k}+\bar u_{i,k}^3\ \ {\rm in}\ \ \R^2,\ \ i=1,\ 2.
\end{equation}
Because $u_{1,k}\not\equiv u_{2,k}$ in $\R^2$, we define
\begin{equation}\label{4.12}
\bar \eta_k(x):=\frac{u_{2,k}(x)-u_{1,k}(x)}{\|u_{2,k}-u_{1,k}\|_{L^\infty(\R^2)}}=\frac{\bar u_{2,k}(x)-\bar u_{1,k}(x)}{\|\bar u_{2,k}-\bar u_{1,k}\|_{L^\infty(\R^2)}}\ \ {\rm in}\  \ \R^2.
\end{equation}
It then follows from \eqref{4.11} and \eqref{4.12} that $\bar\eta_k$ satisfies the following equation:
\begin{equation}\label{4.13}
-\varepsilon_k^2\Delta\bar\eta_k+\varepsilon_k^2V(x)\bar\eta_k=\mu_{2,k}\varepsilon_k^2\bar\eta_k+\bar g_k(x)+\bar f_k(x)\ \ {\rm in}\ \ \R^2,
\end{equation}
where
\begin{equation*}
\bar g_k(x)=-\Big(\int_{\R^2}{\rm ln}|x-y|\bar u_{2,k}^2(y){\rm d}y\Big)\bar\eta_k+\Big(\bar u_{2,k}^2+\bar u_{2,k}\bar u_{1,k}+\bar u_{1,k}^2\Big)\bar\eta_k,
\end{equation*}
and
\begin{equation*}
\begin{aligned}
\bar f_k(x)=&\frac{\mu_{2,k}-\mu_{1,k}}{\|\bar u_{2,k}-\bar u_{1,k}\|_{L^\infty(\R^2)}}\varepsilon_k^2\bar u_{1,k}-\Big[\int_{\R^2}{\rm ln}|x-y|\big(\bar u_{2,k}(y)+\bar u_{1,k}(y)\big)\bar\eta_k(y){\rm d}y\Big]\bar u_{1,k}\\
=&\frac1{2\rho_k\varepsilon_k^2}\int_{\R^2}\int_{\R^2}{\rm ln}|x-y|\big(\bar u_{2,k}(x)+\bar u_{1,k}(x)\big)\bar\eta_k(x)\bar u_{2,k}^2(y){\rm d}x{\rm d}y\bar u_{1,k}\\
&+\frac1{2\rho_k\varepsilon_k^2}\int_{\R^2}\int_{\R^2}{\rm ln}|x-y|\bar u_{1,k}^2(x)\big(\bar u_{2,k}(y)+\bar u_{1,k}(y)\big)\bar\eta_k(y){\rm d}x{\rm d}y\bar u_{1,k}\\
&-\frac1{2\rho_k\varepsilon_k^2}\int_{\R^2}\Big(\bar u_{2,k}^2(x)+\bar u_{1,k}^2(x)\Big)\big(\bar u_{2,k}(x)+\bar u_{1,k}(x)\big)\bar\eta_k(x){\rm d}x\bar u_{1,k}\\
&-\Big[\int_{\R^2}{\rm ln}|x-y|\big(\bar u_{2,k}(y)+\bar u_{1,k}(y)\big)\bar\eta_k(y){\rm d}y\Big]\bar u_{1,k}.\end{aligned}
\end{equation*}

We first claim that for any $x_0\in\R^2$, there exist a small constant $\delta>0$ and a constant $C>0$ such that
\begin{equation}\label{4.14}
\int_{\partial B_\delta(x_0)}\Big(\varepsilon_k^2\left|\nabla\bar\eta_k\right|^2+\varepsilon_k^2V(x)\bar\eta_k^2+\bar\eta_k^2\Big){\rm d}S\leqslant C\varepsilon_k^2\ \ {\rm as}\ \ k\to\infty.
\end{equation}
To prove the above claim, multiplying \eqref{4.13} by $\bar\eta_k$ and integrating over $\R^2$, we obtain that
\begin{equation}
\begin{aligned}\label{4.15}
&\varepsilon_k^2\int_{\R^2}\left|\nabla\bar\eta_k\right|^2{\rm d}x+\varepsilon_k^2\int_{\R^2}V(x)\bar\eta_k^2{\rm d}x-\mu_{2,k}\varepsilon_k^2\int_{\R^2}\bar\eta_k^2{\rm d}x
+\frac12\int_{\R^2}\int_{\R^2}{\rm ln}(1+|x-y|^2)\bar\eta_k^2(x)\bar u_{2,k}^2(y){\rm d}x{\rm d}y\\
=&\frac12\int_{\R^2}\int_{\R^2}{\rm ln}\Big(1+\frac1{|x-y|^2}\Big)\bar\eta_k^2(x)\bar u_{2,k}^2(y){\rm d}x{\rm d}y+\int_{\R^2}\Big(\bar u_{2,k}^2(x)+\bar u_{2,k}(x)\bar u_{1,k}(x)+\bar u_{1,k}^2(x)\Big)\bar\eta_k^2(x){\rm d}x\\
&+\frac1{2\rho_k\varepsilon_k^2}\int_{\R^2}\int_{\R^2}{\rm ln}|x-y|\big(\bar u_{2,k}(x)+\bar u_{1,k}(x)\big)\bar\eta_k(x)\bar u_{2,k}^2(y){\rm d}x{\rm d}y\int_{\R^2}\bar u_{1,k}(x)\bar\eta_k(x){\rm d}x\\
&+\frac1{2\rho_k\varepsilon_k^2}\int_{\R^2}\int_{\R^2}{\rm ln}|x-y|\bar u_{1,k}^2(x)\big(\bar u_{2,k}(y)+\bar u_{1,k}(y)\big)\bar\eta_k(y){\rm d}x{\rm d}y\int_{\R^2}\bar u_{1,k}(x)\bar\eta_k(x){\rm d}x\\
&-\frac1{2\rho_k\varepsilon_k^2}\int_{\R^2}\Big(\bar u_{2,k}^2(x)+\bar u_{1,k}^2(x)\Big)\big(\bar u_{2,k}(x)+\bar u_{1,k}(x)\big)\bar\eta_k(x){\rm d}x\int_{\R^2}\bar u_{1,k}(x)\bar\eta_k(x){\rm d}x\\
&-\int_{\R^2}\int_{\R^2}{\rm ln}|x-y|\bar u_{1,k}(x)\bar\eta_k(x)\big(\bar u_{2,k}(y)+\bar u_{1,k}(y)\big)\bar\eta_k(y){\rm d}x{\rm d}y.
\end{aligned}
\end{equation}
Lemma \ref{lem:A.3} in Appendix shows that there exists a constant $C>0$ such that
%\begin{eqnarray*}
%\begin{aligned}
%&\varepsilon_k^2\int_{\R^2}\left|\nabla\bar\eta_k\right|^2{\rm d}x+\varepsilon_k^2\int_{\R^2}V(x)\bar\eta_k^2{\rm d}x-\mu_{2,k}\varepsilon_k^2\int_{\R^2}\bar\eta_k^2{\rm d}x\\
%&\leqslant C\Big(\varepsilon_k\int_{\R^2}\bar\eta_k^2{\rm d}x+\varepsilon_k^4\int_{\R^2}\left|\nabla\bar\eta_k\right|^2{\rm d}x+\varepsilon_k^2\Big)\ \ \hbox{as}\ \ k\to\infty.
%\end{aligned}
%\end{eqnarray*}
%Following \eqref{4.5}, we obtain from above that
\begin{equation}\label{4.16}
I:=\varepsilon_k^2\int_{\R^2}\left|\nabla\bar\eta_k\right|^2{\rm d}x+\varepsilon_k^2\int_{\R^2}V(x)\bar\eta_k^2{\rm d}x+\int_{\R^2}\bar\eta_k^2{\rm d}x\leqslant C\varepsilon_k^2\ \ \hbox{as}\ \ k\to\infty.
\end{equation}
We thus derive from \cite[Lemma 4.5]{Cao} that for any $x_{0}\in\R^2$, there exist  a small constant $\delta>0$ and a constant $M>0$ such that
\begin{equation*}
\int_{\partial B_\delta(x_0)}\Big(\varepsilon_k^2\left|\nabla\bar\eta_k\right|^2+\varepsilon_k^2V(x)\bar\eta_k^2+\bar\eta_k^2\Big){\rm d}S\leqslant MI\leqslant MC\varepsilon_k^2\ \ \hbox{as}\ \ k\to\infty.
\end{equation*}

We next define
\begin{equation*}
\eta_k(x):=\bar\eta_k\big(\varepsilon_kx+x_{2,k}\big),\ \ k=1,2,\cdots,
\end{equation*}
and
\begin{equation}\label{4:2}
\tilde u_{i,k}(x):=\varepsilon_ku_{i,k}\big(\varepsilon_kx+x_{2,k}\big)>0,\ \ i=1,2.
\end{equation}
According to Theorem \ref{thm:1.2}, we get that $\tilde u_{i,k}\to Q$ in $L^\infty(\R^2)$ as $k\to\infty$. Based on the previous conclusions, we shall carry out the proof of Theorem \ref{thm:1.3} by deriving a contradiction through the following three steps.

Step 1. There exist some constants $b_0,b_1$ and $b_2$ such that up to a subsequence if necessary, $\eta_k\to\eta_0$ in $C_{loc}(\R^2)$ as $k\to\infty$, where $\eta_0(x)$ satisfies
\begin{equation}\label{4.17}
\eta_0(x)=b_0\big(Q+x\cdot\nabla Q\big)+\Sigma_{i=1}^2b_i\frac{\partial Q}{\partial x_i} \ \ {\rm in}\ \ \R^2.
\end{equation}

Note first from \eqref{4.10} that $\eta_k$ satisfies
\begin{equation}\label{4.18}
-\Delta\eta_k+C_k(x)\eta_k=D_k(x)\ \ {\rm in}\ \ \R^2,
\end{equation}
where
\begin{eqnarray}\label{4:0}
C_k(x)=\varepsilon_k^2V(\varepsilon_kx+x_{2,k})+\frac12\varepsilon_k^2\int_{\R^2}{\ln}\big(1+|x-y|^2\big)\tilde u_{2,k}^2(y){\rm d}y+\rho_k\varepsilon_k^2{\rm ln}\varepsilon_k-\mu_{2,k}\varepsilon_k^2+A_k(x),
\end{eqnarray}
\begin{equation}\label{140}
A_k(x)=-\frac12\varepsilon_k^2\int_{\R^2}{\ln}\Big(1+\frac1{|x-y|^2}\Big)\tilde u_{2,k}^2(y){\rm d}y-\big(\tilde u_{2,k}^2+\tilde u_{2,k}\tilde u_{1,k}+\tilde u_{1,k}^2\big),
\end{equation}
and
\begin{equation}\label{4:1}
\begin{aligned}
D_k(x)=&\frac{\mu_{2,k}-\mu_{1,k}}{\|\tilde{u}_{2,k}-\tilde{u}_{1,k}\|_{L^\infty(\R^2)}}\varepsilon_{k}^2\tilde u_{1,k}-\varepsilon_k^2\Big[\int_{\R^2}{\rm ln}|x-y|\big(\tilde u_{2,k}(y)+\tilde u_{1,k}(y)\big)\eta_k(y){\rm d}y\Big]\tilde u_{1,k}\\
=&\frac1{2\rho_k}\varepsilon_k^2\int_{\R^2}\int_{\R^2}{\rm ln}|x-y|\big(\tilde u_{2,k}(x)+\tilde u_{1,k}(x)\big)\eta_k(x)\tilde u_{2,k}^2(y){\rm d}x{\rm d}y\tilde u_{1,k}\\
&+\frac1{2\rho_k}\varepsilon_k^2\int_{\R^2}\int_{\R^2}{\rm ln}|x-y|\tilde u_{1,k}^2(x)\big(\tilde u_{2,k}(y)+\tilde u_{1,k}(y)\big)\eta_k(y){\rm d}x{\rm d}y\tilde u_{1,k}\\
&-\frac1{2\rho_k}\left[\int_{\R^2}\big(\tilde u_{2,k}^2+\tilde u_{1,k}^2\big)\big(\tilde u_{2,k}+\tilde u_{1,k}\big)\eta_k{\rm d}x\right]\tilde u_{1,k}\\
&-\varepsilon_k^2\left[\int_{\R^2}{\rm ln}|x-y|\big(\tilde u_{2,k}(y)+\tilde u_{1,k}(y)\big)\eta_k(y){\rm d}y\right]\tilde u_{1,k}.
\end{aligned}
\end{equation}
Since $|\eta_k|\leqslant1$, the standard elliptic regularity then implies that $\|\eta_k\|_{C_{loc}^{1,\alpha}(\R^2)}\leqslant C$ for some $\alpha\in(0,1)$, where the constant $C>0$ is independent of $k$, see Lemma \ref{lem:A.4} in the Appendix for details. Therefore, there exists a subsequence of  $\{\rho_k\}$ (still denoted by $\{\rho_k\}$) and a function $\eta_0$ such that $\eta_k\to\eta_0$ in $C_{loc}(\R^2)$ as $k\to\infty$. We further deduce from Lemma \ref{lem:A.5} that $\eta_0$ satisfies
\begin{equation}\label{4.19}
\mathcal{L}\eta_0=-\Delta\eta_0+(1-3Q^2)\eta_0=-\frac2{\rho^*}\Big(\int_{\R^2}Q^3\eta_0{\rm d}x\Big)Q\quad {\rm in}\ \ \R^2.
\end{equation}
Since $\mathcal{L}(Q+x\cdot\nabla Q)=-2Q$, we conclude from \eqref{4.1} and \eqref{4.19} that \eqref{4.17} holds for some constants $b_0,b_1$ and $b_2$.

Step 2. The constants $b_0=b_1=b_2=0$ in \eqref{4.17}.

We first derive $b_j=0$ for $j=1,2$. Multiplying \eqref{4.11} by $\frac{\partial\bar u_{i,k}}{\partial x_j}$, where $i,j=1,2$, and integrating over $B_\delta(x_{2,k})$, where $\delta>0$ is small and given by \eqref{4.14}, direct computations yield that
\begin{equation}\label{4.20}
\begin{aligned}
&-\varepsilon_k^2\int_{B_\delta(x_{2,k})}\frac{\partial\bar u_{i,k}}{\partial x_j}\Delta\bar u_{i,k}{\rm d}x+\varepsilon_k^2\int_{B_\delta(x_{2,k})}V(x)\bar u_{i,k}\frac{\partial\bar u_{i,k}}{\partial x_j}{\rm d}x\\
&+\int_{B_\delta(x_{2,k})}\int_{\R^2}{\rm ln}|x-y|\bar u_{i,k}(x)\frac{\partial\bar u_{i,k}(x)}{\partial x_j}\bar u_{i,k}^2(y){\rm d}y{\rm d}x\\
&=\mu_{i,k}\varepsilon_k^2\int_{B_\delta(x_{2,k})}\bar u_{i,k}\frac{\partial\bar u_{i,k}}{\partial x_j}{\rm d}x+\int_{B_\delta(x_{2,k})}\bar u_{i,k}^3\frac{\partial\bar u_{i,k}}{\partial x_j}{\rm d}x\\
&=\frac12\mu_{i,k}\varepsilon_k^2\int_{\partial B_\delta(x_{2,k})}\bar u_{i,k}^2\nu_j{\rm d}S+\frac14\int_{\partial B_\delta(x_{2,k})}\bar u_{i,k}^4\nu_j{\rm d}S,
\end{aligned}\end{equation}
where $\nu:=(\nu_1,\nu_2)$ denotes the outward unit normal vector of $\partial B_\delta(x_{2,k})$. Note that
\begin{equation}\label{4.21}
\begin{aligned}
-\varepsilon_k^2\int_{B_\delta(x_{2,k})}\frac{\partial\bar u_{i,k}}{\partial x_j}\Delta\bar u_{i,k}{\rm d}x
&=-\varepsilon_k^2\int_{\partial B_\delta(x_{2,k})}\frac{\partial\bar u_{i,k}}{\partial x_j}\frac{\partial\bar u_{i,k}}{\partial\nu}{\rm d}S+\varepsilon_k^2\int_{B_\delta(x_{2,k})}\nabla\bar u_{i,k}\cdot\nabla\Big(\frac{\partial\bar u_{i,k}}{\partial x_j}\Big){\rm d}x\\
&=-\varepsilon_k^2\int_{\partial B_\delta(x_{2,k})}\frac{\partial\bar u_{i,k}}{\partial x_j}\frac{\partial\bar u_{i,k}}{\partial\nu}{\rm d}S+\frac12\varepsilon_k^2\int_{\partial B_\delta(x_{2,k})}\left|\nabla\bar u_{i,k}\right|^2\nu_j{\rm d}S,
\end{aligned}
\end{equation}
\begin{equation}\label{4.22}
\begin{aligned}
\varepsilon_k^2\int_{B_\delta(x_{2,k})}V(x)\bar u_{i,k}\frac{\partial\bar u_{i,k}}{\partial x_j}{\rm d}x=\frac12\varepsilon_k^2\int_{\partial B_\delta(x_{2,k})}V(x)\bar u_{i,k}^2\nu_j{\rm d}S-\frac12\varepsilon_k^2\int_{B_\delta(x_{2,k})}\frac{\partial V(x)}{\partial x_j}\bar u_{i,k}^2{\rm d}x,
\end{aligned}
\end{equation}
and
\begin{equation}\label{4.23}
\begin{aligned}
&\int_{B_\delta(x_{2,k})}\int_{\R^2}{\rm ln}|x-y|\bar u_{i,k}(x)\frac{\partial\bar u_{i,k}(x)}{\partial x_j}\bar u_{i,k}^2(y){\rm d}y{\rm d}x\\
=&\frac12\int_{\partial B_\delta(x_{2,k})}\int_{\R^2}{\rm ln}|x-y|\bar u_{i,k}^2(y)\bar u_{i,k}^2(x)\nu_j{\rm d}y{\rm d}S-\frac12\int_{B_\delta(x_{2,k})}\int_{\R^2}\frac{x_j-y_j}{|x-y|^2}\bar u_{i,k}^2(x)\bar u_{i,k}^2(y){\rm d}y{\rm d}x\\
=&\frac12\int_{\partial B_\delta(x_{2,k})}\int_{\R^2}{\rm ln}|x-y|\bar u_{i,k}^2(y)\bar u_{i,k}^2(x)\nu_j{\rm d}y{\rm d}S+\frac12\int_{B_\delta^c(x_{2,k})}\int_{\R^2}\frac{x_j-y_j}{|x-y|^2}\bar u_{i,k}^2(x)\bar u_{i,k}^2(y){\rm d}y{\rm d}x,
\end{aligned}
\end{equation}
where we have used the following fact
\begin{equation*}
\int_{\R^2}\int_{\R^2}\frac{x_j-y_j}{|x-y|^2}\bar u_{i,k}^2(x)\bar u_{i,k}^2(y){\rm d}x{\rm d}y=0.
\end{equation*}

Submitting \eqref{4.21}--\eqref{4.23} into \eqref{4.20}, we then deduce that
\begin{equation}\label{123}
\begin{aligned}
&\varepsilon_k^2\int_{B_\delta(x_{2,k})}\frac{\partial V(x)}{\partial x_j}\bar u_{i,k}^2(x){\rm d}x\\
=&-2\varepsilon_k^2\int_{\partial B_\delta(x_{2,k})}\frac{\partial\bar u_{i,k}}{\partial x_j}\frac{\partial\bar u_{i,k}}{\partial\nu}{\rm d}S+\varepsilon_k^2\int_{\partial B_\delta(x_{2,k})}|\nabla\bar u_{i,k}|^2\nu_j{\rm d}S+\varepsilon_k^2\int_{\partial B_\delta(x_{2,k})}V(x)\bar u_{i,k}^2\nu_j{\rm d}S\\
&+\int_{\partial B_\delta(x_{2,k})}\int_{\R^2}{\rm ln}|x-y|\bar u_{i,k}^2(y)\bar u_{i,k}^2(x)\nu_j{\rm d}y{\rm d}S+\int_{B_\delta^c(x_{2,k})}\int_{\R^2}\frac{x_j-y_j}{|x-y|^2}\bar u_{i,k}^2(x)\bar u_{i,k}^2(y){\rm d}y{\rm d}x\\
&-\mu_{i,k}\varepsilon_k^2\int_{\partial B_\delta(x_{2,k})}\bar u_{i,k}^2\nu_j{\rm d}S-\frac12\int_{\partial B_\delta(x_{2,k})}\bar u_{i,k}^4\nu_j{\rm d}S.
\end{aligned}
\end{equation}
Following \eqref{123}, it yields that
\begin{equation}
\begin{aligned}\label{4.24}
&\varepsilon_k^2\int_{B_\delta(x_{2,k})}\frac{\partial V(x)}{\partial x_j}\big(\bar u_{2,k}(x)+\bar u_{1,k}(x)\big)\bar\eta_k{\rm d}x\\
=&-2\varepsilon_k^2\int_{\partial B_\delta(x_{2,k})}\Big(\frac{\partial\bar\eta_k}{\partial x_j}\frac{\partial\bar u_{2,k}}{\partial\nu}+\frac{\partial\bar u_{1,k}}{\partial x_j}\frac{\partial\bar\eta_k}{\partial\nu}\Big){\rm d}S\\
&+\varepsilon_k^2\int_{\partial B_\delta(x_{2,k})}\big(\nabla\bar u_{2,k}+\nabla\bar u_{1,k}\big)\cdot\nabla\bar\eta_k\nu_j{\rm d}S\\
&+\varepsilon_k^2\int_{\partial B_\delta(x_{2,k})}V(x)\big(\bar u_{2,k}+\bar u_{1,k}\big)\bar\eta_k\nu_j{\rm d}S\\
&+\int_{\partial B_\delta(x_{2,k})}\int_{\R^2}{\rm ln}|x-y|\big(\bar u_{2,k}(y)+\bar u_{1,k}(y)\big)\bar\eta_k(y)\bar u_{2,k}^2(x)\nu_j{\rm d}y{\rm d}S\\
&+\int_{\partial B_\delta(x_{2,k})}\int_{\R^2}{\rm ln}|x-y|\bar u_{1,k}^2(y)\big(\bar u_{2,k}(x)+\bar u_{1,k}(x)\big)\bar\eta_k(x)\nu_j{\rm d}y{\rm d}S\\
&+\int_{B_\delta^c(x_{2,k})}\int_{\R^2}\frac{x_j-y_j}{|x-y|^2}\big(\bar u_{2,k}(x)+\bar u_{1,k}(x)\big)\bar\eta_k(x)\bar u_{2,k}^2(y){\rm d}y{\rm d}x\\
&+\int_{ B_\delta^c(x_{2,k})}\int_{\R^2}\frac{x_j-y_j}{|x-y|^2}\bar u_{1,k}^2(x)\big(\bar u_{2,k}(y)+\bar u_{1,k}(y)\big)\bar\eta_k(y){\rm d}y{\rm d}x\\
&-\varepsilon_k^2\mu_{2,k}\int_{\partial B_\delta(x_{2,k})}\big(\bar u_{2,k}(x)+\bar u_{1,k}(x)\big)\bar\eta_k(x)\nu_j{\rm d}S\\
&-\frac12\int_{\partial B_\delta(x_{2,k})}\big(\bar u_{2,k}^2+\bar u_{1,k}^2\Big)\big(\bar u_{2,k}+\bar u_{1,k}\big)\bar\eta_k\nu_j{\rm d}S\\
&-\frac{\mu_{2,k}-\mu_{1,k}}{\|\bar u_{2,k}-\bar u_{1,k}\|_{L^\infty(\R^2)}}\varepsilon_k^2\int_{\partial B_\delta(x_{2,k})}\bar u_{1,k}^2\nu_j{\rm d}S:=\alpha_{j,k}.
\end{aligned}
\end{equation}
Since $V(x)={\rm ln}(1+|x|^2)$, according to  Lemma \ref{lem:A.6}, it follows from \eqref{3.35}, \eqref{4.8} and \eqref{4.24} that there exists a constant $a>0$ such that
\begin{equation}\label{4.25}
\begin{aligned}
o\big(e^{-\frac{a\delta}{\varepsilon_k}}\big)=& \varepsilon_k^2\int_{B_\delta(x_{2,k})}\frac{\partial V(x)}{\partial x_j}\big(\bar u_{2,k}+\bar u_{1,k}\big)\bar\eta_k{\rm d}x\\
=&\varepsilon_k^4\int_{B_\frac{\delta}{\varepsilon_k}(0)}\frac{\partial V(\varepsilon_k x+x_{2,k})}{\partial x_j}\big(\tilde u_{2,k}(x)+\tilde u_{1,k}(x)\big)\eta_k(x){\rm d}x\\
=&2\varepsilon_k^4\int_{B_\frac{\delta}{\varepsilon_k}(0)}\frac{\varepsilon_k x_j+x_{2,k}^j}{1+|\varepsilon_k x+x_{2,k}|^2}\big(\tilde u_{2,k}(x)+\tilde u_{1,k}(x)\big)\eta_k(x){\rm d}x\\
=&O(\varepsilon_k^5)\int_{B_\frac{\delta}{\varepsilon_k}(0)}\big(x_j+\frac{x_{2,k}^j}{\varepsilon_k}\Big)\big(\tilde u_{2,k}(x)+\tilde u_{1,k}(x)\big)\eta_k(x){\rm d}x\\
=&O(\varepsilon_k^5)\int_{B_\frac{\delta}{\varepsilon_k}(0)}\Big[\Big(x_j+\frac{x_{2,k}^j}{\varepsilon_k}\Big)\big(\tilde u_{2,k}(x)+\bar u_{1,k}(\varepsilon_kx+x_{1,k})\big)\eta_k(x)\\
&\quad +\Big(x_j+\frac{x_{2,k}^j}{\varepsilon_k}\Big)\big(\bar u_{1,k}(\varepsilon_kx+x_{2,k})-\bar u_{1,k}(\varepsilon_kx+x_{1,k})\big)\eta_k(x)\Big]{\rm d}x\\
=&O(\varepsilon_k^5) \int_{B_\frac{\delta}{\varepsilon_k}(0)}\Big(x_j+\frac{x_{2,k}^j}{\varepsilon_k}\Big)\big(\tilde u_{2,k}(x)+\bar u_{1,k}(\varepsilon_kx+x_{1,k})\big)\eta_k(x){\rm d}x\ \ {\rm as}\ \ k\to\infty,
\end{aligned}
\end{equation}
where $x_{2,k}:=(x_{2,k}^1,x_{2,k}^2)\in\R^2$.

Since $\eta_k\to\eta_0$ in $C_{loc}(\R^2)$ as $k\to\infty$, in view of the exponential decays \eqref{1.6} and \eqref{A.16}, we derive from \eqref{3.35}, \eqref{4.8} and \eqref{A.10} that
\begin{equation*}
\begin{aligned}
&\Big|\int_{B_\frac{\delta}{\varepsilon_k}(0)}\big(x_j+\frac{x_{2,k}^j}{\varepsilon_k}\big)\big(\tilde u_{2,k}(x)+\bar u_{1,k}(\varepsilon_kx+x_{1,k})\big)\eta_k(x){\rm d}x-2\int_{\R^2}x_jQ(x)\eta_0(x){\rm d}x\Big|\\
%=&\Big|\frac{x_{2,k}^j}{\varepsilon_k}\int_{B_\frac{\delta}{\varepsilon_k}(0)}\Big(\tilde u_{2,k}(x)+\bar u_{1,k}(\varepsilon_kx+x_{1,k})\Big)\eta_k(x){\rm d}x\\
%&\quad+\int_{\R^2}x_j\Big[\big(\tilde u_{2,k}(x)+\bar u_{1,k}(\varepsilon_kx+x_{1,k})\big)\eta_k(x)-2Q(x)\eta_0(x)\Big]{\rm d}x\Big|\\
\leqslant&\left|\frac{x_{2,k}^j}{\varepsilon_k}\right|\int_{\R^2}\left|\big(\tilde u_{2,k}(x)+\bar u_{1,k}(\varepsilon_kx+x_{1,k})\big)\eta_k(x)\right|{\rm d}x+\int_{\R^2}|x_j|\left|\tilde u_{2,k}(x)+\bar u_{1,k}(\varepsilon_kx+x_{1,k})-2Q(x)\right||\eta_k(x)|{\rm d}x\\
&\quad+2\int_{\R^2}|x_j|Q|\eta_k-\eta_0|{\rm d}x+\int_{B_{\delta/\varepsilon_k}^c(0)}\left|\Big(x_j+\frac{x_{2,k}^j}{\varepsilon_k}\Big)\big(\tilde u_{2,k}(x)+\bar u_{1,k}(\varepsilon_kx+x_{1,k})\big)\eta_k(x)\right|{\rm d}x\to0\ \ {\rm as}\ \ k\to\infty,
\end{aligned}
\end{equation*}
which implies that
\begin{equation}\label{139}
\int_{B_\frac{\delta}{\varepsilon_k}(0)}\Big(x_j+\frac{x_{2,k}^j}{\varepsilon_k}\Big)\big(\tilde u_{2,k}(x)+\bar u_{1,k}(\varepsilon_kx+x_{1,k})\big)\eta_k(x){\rm d}x\to2\int_{\R^2}x_jQ(x)\eta_0(x){\rm d}x\ \ {\rm as}\ \ k\to\infty.
\end{equation}
We thus deduce from \eqref{4.17}, \eqref{4.25} and \eqref{139} that
\begin{equation*}
\begin{aligned}
0=2\int_{\R^2}x_jQ\eta_0{\rm d}x =&2\int_{\R^2}x_jQ\Big[b_0(Q+x\cdot\nabla Q)+\Sigma_{i=1}^2b_i\frac{\partial Q}{\partial x_i}\Big]{\rm d}x=-b_j\int_{\R^2}Q^2{\rm d}x=-b_j\rho^*,\ \ j=1,\ 2,
\end{aligned}
\end{equation*}
which means that $b_j=0$ for $j=1,2$.

We next prove $b_0=0$. Using integration by parts, we get that
\begin{equation}\label{4.26}
\begin{aligned}
&-\varepsilon_k^2\int_{B_\delta(x_{2,k})}\Delta\bar u_{i,k}\Big[\big(x-x_{2,k}\big)\cdot\nabla\bar u_{i,k}\Big]{\rm d}x\\
=&-\varepsilon_k^2\int_{\partial B_\delta(x_{2,k})}\frac{\partial \bar u_{i,k}}{\partial \nu}\Big[\big(x-x_{2,k}\big)\cdot\nabla\bar u_{i,k}\Big]{\rm d}S+\varepsilon_k^2\int_{B_\delta(x_{2,k})}\nabla\bar u_{i,k}\cdot\nabla\Big[\big(x-x_{2,k}\big)\cdot\nabla\bar u_{i,k}\Big]{\rm d}x\\
=&-\varepsilon_k^2\int_{\partial B_\delta(x_{2,k})}\frac{\partial \bar u_{i,k}}{\partial\nu}\Big[\big(x-x_{2,k}\big)\cdot\nabla\bar u_{i,k}\Big]{\rm d}S+\frac12\varepsilon_k^2\int_{\partial B_\delta(x_{2,k})}\big(x-x_{2,k}\big)\cdot\nu\left|\nabla\bar u_{i,k}\right|^2{\rm d}S.
\end{aligned}
\end{equation}
Multiplying \eqref{4.11} by $(x-x_{2,k})\cdot\nabla\bar u_{i,k}$, $i=1,2$, and integrating over $B_\delta(x_{2,k})$, where $\delta>0$ is as in \eqref{4.14}, we derive that
\begin{equation}\nonumber
\begin{aligned}
&-\varepsilon_k^2\int_{B_\delta(x_{2,k})}\Delta\bar u_{i,k}\Big[\big(x-x_{2,k}\big)\cdot\nabla\bar u_{i,k}\Big]{\rm d}x\\
=&\varepsilon_k^2\int_{B_\delta(x_{2,k})}\Big(\mu_{i,k}-V(x)\Big)\bar u_{i,k}(x)\Big[\big(x-x_{2,k}\big)\cdot\nabla\bar u_{i,k}(x)\Big]{\rm d}x\\
&-\int_{B_\delta(x_{2,k})}\int_{\R^2}{\ln}|x-y|\bar u_{i,k}(x)\Big[\big(x-x_{2,k}\big)\cdot\nabla\bar u_{i,k}(x)\Big]\bar u_{i,k}^2(y){\rm d}y{\rm d}x
\end{aligned}
\end{equation}
\begin{equation}\label{4.27}
\begin{aligned}
&+\int_{B_\delta(x_{2,k})}\bar u_{i,k}^3(x)\Big[\big(x-x_{2,k}\big)\cdot\nabla\bar u_{i,k}(x)\Big]{\rm d}x\\
=&-\frac12\varepsilon_k^2\int_{B_\delta(x_{2,k})}\Big[2\big(\mu_{i,k}-V(x)\big)-\big(x-x_{2,k}\big)\cdot\nabla V(x)\Big]\bar u_{i,k}^2(x){\rm d}x\\
&+\frac12\varepsilon_k^2\int_{\partial B_\delta(x_{2,k})}\big(\mu_{i,k}-V(x)\big)\big(x-x_{2,k}\big)\cdot\nu\bar u_{i,k}^2(x){\rm d}S\\
&+\frac12\int_{ B_\delta(x_{2,k})}\int_{\R^2}\frac{\big(x-x_{2,k}\big)\cdot(x-y)}{|x-y|^2}\bar  u_{i,k}^2(x)\bar u_{i,k}^2(y){\rm d}y{\rm d}x\\
&+\int_{B_\delta(x_{2,k})}\int_{\R^2}{\rm ln}|x-y|\bar  u_{i,k}^2(x)\bar u_{i,k}^2(y){\rm d}y{\rm d}x\\
&-\frac12\int_{\partial B_\delta(x_{2,k})}\int_{\R^2}{\rm ln}|x-y|\bar u_{i,k}^2(x)\big(x-x_{2,k}\big)\cdot\nu\bar u_{i,k}^2(y){\rm d}y{\rm d}S\\
&-\frac12\int_{B_\delta(x_{2,k})}\bar u_{i,k}^4(x){\rm d}x+\frac14\int_{\partial B_\delta(x_{2,k})}\big(x-x_{2,k}\big)\cdot\nu\bar u_{i,k}^4(x){\rm d}S,
\end{aligned}
\end{equation}
where $\nu$ denotes the outward unit normal vector of $\partial B_{\delta}(x_{2,k})$.
Combining \eqref{4.26} with \eqref{4.27} gives that
\begin{equation}\label{124}
\begin{aligned}
&-\varepsilon_k^2\int_{\partial B_\delta(x_{2,k})}\frac{\partial\bar u_{i,k}}{\partial\nu}\Big[\big(x-x_{2,k}\big)\cdot\nabla\bar u_{i,k}\Big]{\rm d}S+\frac12\varepsilon_k^2\int_{\partial B_\delta(x_{2,k})}\big(x-x_{2,k}\big)\cdot\nu\left|\nabla\bar u_{i,k}\right|^2{\rm d}S\\
=&-\varepsilon_k^2\mu_{i,k}\int_{B_\delta(x_{2,k})}\bar u_{i,k}^2{\rm d}x+\varepsilon_k^2\int_{B_\delta(x_{2,k})}V(x)\bar u_{i,k}^2{\rm d}x+\frac12\varepsilon_k^2\int_{B_\delta(x_{2,k})}(x-x_{2,k})\cdot\nabla V(x)\bar u_{i,k}^2{\rm d}x\\
&+\frac{\varepsilon_k^2}2\int_{\partial B_\delta(x_{2,k})}\big(\mu_{i,k}-V(x)\big)\big(x-x_{2,k}\big)\cdot\nu\bar u_{i,k}^2(x){\rm d}S\\
&+\frac12\int_{ B_\delta(x_{2,k})}\int_{\R^2}\frac{\big(x-x_{2,k}\big)\cdot(x-y)}{|x-y|^2}\bar  u_{i,k}^2(x)\bar u_{i,k}^2(y){\rm d}y{\rm d}x\\
&+\int_{B_\delta(x_{2,k})}\int_{\R^2}{\rm ln}|x-y|\bar  u_{i,k}^2(x)\bar u_{i,k}^2(y){\rm d}y{\rm d}x-\frac12\int_{\partial B_\delta(x_{2,k})}\int_{\R^2}{\rm ln}|x-y|\bar u_{i,k}^2(x)\big(x-x_{2,k}\big)\cdot\nu\bar u_{i,k}^2(y){\rm d}y{\rm d}S\\
&-\frac12\int_{B_\delta(x_{2,k})}\bar u_{i,k}^4{\rm d}x+\frac14\int_{\partial B_\delta(x_{2,k})}\big(x-x_{2,k}\big)\cdot\nu\bar u_{i,k}^4{\rm d}S\\
=&-\varepsilon_k^2\mu_{i,k}\int_{\R^2}\bar u_{i,k}^2{\rm d}x+\varepsilon_k^2\int_{B_\delta(x_{2,k})}V(x)\bar u_{i,k}^2{\rm d}x+\frac12\varepsilon_k^2\int_{B_\delta(x_{2,k})}x\cdot\nabla V(x)\bar u_{i,k}^2(x){\rm d}x\\
&-\frac12\varepsilon_k^2\int_{B_\delta(x_{2,k})}x_{2,k}\cdot\nabla V(x)\bar u_{i,k}^2(x){\rm d}x+\frac12\int_{\R^2}\int_{\R^2}\frac{(x-x_{2,k})\cdot(x-y)}{|x-y|^2}\bar u_{i,k}^2(x)\bar u_{i,k}^2(y){\rm d}x{\rm d}y\\
&+\int_{\R^2}\int_{\R^2}{\rm ln}|x-y|\bar u_{i,k}^2(x)\bar u_{i,k}^2(y){\rm d}x{\rm d}y-\frac12\int_{\R^2}\bar u_{i,k}^4{\rm d}x\\
&+\varepsilon_k^2\mu_{i,k}\int_{B^c_\delta(x_{2,k})}\bar u_{i,k}^2(x){\rm d}x+\frac12\varepsilon_k^2\int_{\partial B_\delta(x_{2,k})}\big(\mu_{i,k}-V(x)\big)\bar u_{i,k}^2(x)(x-x_{2,k})\cdot\nu{\rm d}S\\
\end{aligned}
\end{equation}
\begin{equation*}
\begin{aligned}
&-\frac12\int_{B_\delta^c(x_{2,k})}\int_{\R^2}\frac{\big(x-x_{2,k}\big)\cdot(x-y)}{|x-y|^2}\bar u_{i,k}^2(x)\bar u_{i,k}^2(y){\rm d}y{\rm d}x-\int_{B_\delta^c(x_{2,k})}\int_{\R^2}{\rm ln}|x-y|\bar u_{i,k}^2(x)\bar u_{i,k}^2(y){\rm d}y{\rm d}x\\
&-\frac12\int_{\partial B_\delta(x_{2,k})}\int_{\R^2}{\rm ln}|x-y|\bar u_{i,k}^2(y)\bar u_{i,k}^2(x)\big(x-x_{2,k}\big)\cdot\nu{\rm d}y{\rm d}S\\
&+\frac12\int_{B_\delta^c(x_{2,k})}\bar u_{i,k}^4{\rm d}x+\frac14\int_{\partial B_\delta(x_{2,k})}\bar u_{i,k}^4(x)\big(x-x_{2,k}\big)\cdot\nu{\rm d}S.
\end{aligned}
\end{equation*}
Note that
\begin{equation}\label{4.28}\nonumber
\begin{aligned}
&\quad\int_{\R^2}\int_{\R^2}\frac{(x-x_{2,k})\cdot(x-y)}{|x-y|^2}\bar u_{i,k}^2(x)\bar u_{i,k}^2(y){\rm d}x{\rm d}y\\
&=\varepsilon_k^4\int_{\R^2}\int_{\R^2}\frac{(x-y)\cdot x}{|x-y|^2}\tilde u_{i,k}^2(x)\tilde u_{i,k}^2(y){\rm d}x{\rm d}y\\
&=\frac12\varepsilon_k^4\int_{\R^2}\int_{\R^2}\frac{(x-y)\cdot x+(y-x)\cdot y}{|x-y|^2}\tilde u_{i,k}^2(x)\tilde u_{i,k}^2(y){\rm d}x{\rm d}y\\
&=\frac12\varepsilon_k^4\int_{\R^2}\int_{\R^2}\tilde u_{i,k}^2(x)\tilde u_{i,k}^2(y){\rm d}x{\rm d}y=\frac12\rho_k^2\varepsilon_k^4.
\end{aligned}
\end{equation}
We then infer from \eqref{124} that
\begin{eqnarray}\label{4.29}
\begin{aligned}
&-\varepsilon_k^2\int_{\partial B_\delta(x_{2,k})}\frac{\partial \bar u_{i,k}}{\partial \nu}\Big[\big(x-x_{2,k}\big)\cdot\nabla\bar u_{i,k}\Big]{\rm d}S+\frac12\varepsilon_k^2\int_{\partial B_\delta(x_{2,k})}\left|\nabla\bar u_{i,k}\right|^2\big(x-x_{2,k}\big)\cdot\nu{\rm d}S\\
&=-\varepsilon_k^2\mu_{i,k}\int_{\R^2}\bar u_{i,k}^2{\rm d}x+\varepsilon_k^2\int_{B_\delta(x_{2,k})}V(x)\bar u_{i,k}^2{\rm d}x+\frac12\varepsilon_k^2\int_{B_\delta(x_{2,k})}x\cdot\nabla V(x)\bar u_{i,k}^2{\rm d}x\\
&\quad-\frac12\varepsilon_k^2\int_{B_\delta(x_{2,k})}x_{2,k}\cdot\nabla V(x)\bar u_{i,k}^2(x){\rm d}x+\int_{\R^2}\int_{\R^2}{\rm ln}|x-y|\bar u_{i,k}^2(x)\bar u_{i,k}^2(y){\rm d}x{\rm d}y\\
&\quad-\frac12\int_{\R^2}\bar u_{i,k}^4{\rm d}x+\frac14\rho_k^2\varepsilon_k^4+\beta_{i,k},
\end{aligned}
\end{eqnarray}
where $\beta_{i,k}$ satisfies
\begin{equation}\label{4.30}
\begin{aligned}
\beta_{i,k}=&\varepsilon_k^2\mu_{i,k}\int_{B_\delta^c(x_{2,k})}\bar u_{i,k}^2(x){\rm d}x+\frac12\varepsilon_k^2\int_{\partial B_\delta(x_{2,k})}\big(\mu_{i,k}-V(x)\big)\bar u_{i,k}^2(x)\big(x-x_{2,k}\big)\cdot\nu{\rm d}S\\
&-\frac12\int_{B_\delta^c(x_{2,k})}\int_{\R^2}\frac{\big(x-x_{2,k}\big)\cdot(x-y)}{|x-y|^2}\bar u_{i,k}^2(x)\bar u_{i,k}^2(y){\rm d}y{\rm d}x\\
&-\int_{B_\delta^c(x_{2,k})}\int_{\R^2}{\rm ln}|x-y|\bar u_{i,k}^2(x)\bar u_{i,k}^2(y){\rm d}y{\rm d}x\\
&-\frac12\int_{\partial B_\delta(x_{2,k})}\int_{\R^2}{\rm ln}|x-y|\bar u_{i,k}^2(x)\bar u_{i,k}^2(y)\big(x-x_{2,k}\big)\cdot\nu{\rm d}y{\rm d}S\\
&+\frac12\int_{B_\delta^c(x_{2,k})}\bar u_{i,k}^4{\rm d}x+\frac14\int_{\partial B_\delta(x_{2,k})}\bar u_{i,k}^4(x)\big(x-x_{2,k}\big)\cdot\nu{\rm d}S.
\end{aligned}
\end{equation}

Noting from \eqref{4.4} that
\begin{equation*}
\begin{aligned}
&\frac12\int_{\R^2}\int_{\R^2}{\rm ln}|x-y|\bar u_{i,k}^2(x)\bar u_{i,k}^2(y){\rm d}x{\rm d}y-\frac12\int_{\R^2}\bar u_{i,k}^4{\rm d}x-\varepsilon_k^4\mu_{i,k}\rho_k=-2\varepsilon_k^4e(\rho_k),
\end{aligned}
\end{equation*}
we obtain from \eqref{4.29} that
\begin{equation}\label{101}
\begin{aligned}
&2\varepsilon_k^4e(\rho_k)-\varepsilon_k^2\int_{B_\delta(x_{2,k})}V(x)\bar u_{i,k}^2{\rm d}x-\frac12\varepsilon_k^2\int_{B_\delta(x_{2,k})}x\cdot\nabla V(x)\bar u_{i,k}^2{\rm d}x-\frac14\rho_k^2\varepsilon_k^4\\
&-\frac12\int_{\R^2}\int_{\R^2}{\rm ln}|x-y|\bar u_{i,k}^2(x)\bar u_{i,k}^2(y){\rm d}x{\rm d}y\\
\end{aligned}
\end{equation}
\begin{equation*}
\begin{aligned}
=&\varepsilon_k^2\int_{\partial B_\delta(x_{2,k})}\frac{\partial \bar u_{i,k}}{\partial\nu}\Big[\big(x-x_{2,k}\big)\cdot\nabla\bar u_{i,k}\Big]{\rm d}S-\frac12\varepsilon_k^2\int_{\partial B_\delta(x_{2,k})}\left|\nabla\bar u_{i,k}\right|^2\big(x-x_{2,k}\big)\cdot\nu{\rm d}S\\
&-\frac12\varepsilon_k^2\int_{B_\delta(x_{2,k})}x_{2,k}\cdot\nabla V(x)\bar u_{i,k}^2{\rm d}x+\beta_{i,k}.
\end{aligned}
\end{equation*}
Since
\begin{eqnarray*}
\int_{\R^2}\int_{\R^2}{\rm ln}|x-y|\bar u_{i,k}^2(x)\bar u_{i,k}^2(y){\rm d}x{\rm d}y=\varepsilon_{k}^4{\rm ln}\varepsilon_k\rho_k^2+\varepsilon_{k}^4\int_{\R^2}\int_{\R^2}{\rm ln}|x-y|\tilde u_{i,k}^2(x)\tilde u_{i,k}^2(y){\rm d}x{\rm d}y,
\end{eqnarray*}
we thus deduce from \eqref{4.14} and \eqref{101} that
\begin{equation}\label{4.31}
\begin{aligned}
\gamma_k:=&-\varepsilon_k^2\int_{B_\delta(x_{2,k})}V(x)\big(\bar u_{2,k}+\bar u_{1,k}\big)\bar\eta_k{\rm d}x-\frac12\varepsilon_k^2\int_{B_\delta(x_{2,k})}x\cdot\nabla V(x)\big(\bar u_{2,k}+\bar u_{1,k}\big)\bar\eta_k{\rm d}x\\
&-\frac12\varepsilon_k^4\int_{\R^2}\int_{\R^2}{\rm ln}|x-y|\big(\tilde u_{2,k}(x)+\tilde u_{1,k}(x)\big)\eta_k(x)\tilde u_{2,k}^2(y){\rm d}x{\rm d}y\\
&-\frac12\varepsilon_k^4\int_{\R^2}\int_{\R^2}{\rm ln}|x-y|\tilde u_{1,k}^2(x)\big(\tilde u_{2,k}(y)+\tilde u_{1,k}(y)\big)\eta_k(y){\rm d}x{\rm d}y\\
=&\frac{\beta_{2,k}-\beta_{1,k}}{\|\bar u_{2,k}-\bar u_{1,k}\|_{L^\infty(\R^2)}}-\frac12\varepsilon_k^2\int_{B_\delta(x_{2,k})}x_{2,k}\cdot\nabla V(x)\big(\bar u_{2,k}+\bar u_{1,k}\big)\bar\eta_k{\rm d}x\\
&+\varepsilon_k^2\int_{\partial B_\delta(x_{2,k})}\frac{\partial\bar\eta_k}{\partial \nu}\Big[\big(x-x_{2,k}\big)\cdot\nabla\bar u_{2,k}\Big]{\rm d}S+\varepsilon_k^2\int_{\partial B_\delta(x_{2,k})}\frac{\partial\bar u_{1,k}}{\partial \nu}\Big[\big(x-x_{2,k}\big)\cdot\nabla\bar\eta_k\Big]{\rm d}S\\
&-\frac12\varepsilon_k^2\int_{\partial B_\delta(x_{2,k})}\big(\nabla\bar u_{2,k}+\nabla\bar u_{1,k}\big)\cdot\nabla\bar\eta_k\big(x-x_{2,k}\big)\cdot\nu{\rm d}S\\
=&\frac{\beta_{2,k}-\beta_{1,k}}{\|\bar u_{2,k}-\bar u_{1,k}\|_{L^\infty(\R^2)}}-\frac12\varepsilon_k^2\int_{B_\delta(x_{2,k})}x_{2,k}\cdot\nabla V(x)\big(\bar u_{2,k}+\bar u_{1,k}\big)\bar\eta_k{\rm d}x+o(e^{-\frac{a\delta}{\varepsilon_k}}),
\end{aligned}
\end{equation}
where the exponential decay of $|\nabla\tilde u_{i,k}|$ as $|x|\to\infty$ is used.

Furthermore, applying  Lemma \ref{lem:A.7}, we conclude from \eqref{4.31} that there exists a constant $b\in(0,a)$ such that
\begin{equation}\label{102}
\begin{aligned}
&\frac12\varepsilon_k^4\int_{\R^2}\int_{\R^2}{\rm ln}|x-y|\big(\tilde u_{2,k}(x)+\tilde u_{1,k}(x)\big)\eta_k(x)\tilde u_{2,k}^2(y){\rm d}x{\rm d}y\\
&+\frac12\varepsilon_k^4\int_{\R^2}\int_{\R^2}{\rm ln}|x-y|\tilde u_{1,k}^2(x)\big(\tilde u_{2,k}(y)+\tilde u_{1,k}(y)\big)\eta_k(y){\rm d}x{\rm d}y+o(e^{-\frac{b\delta}{\varepsilon_k}})\\
=&-\varepsilon_k^2\int_{B_\delta(x_{2,k})}V(x)\big(\bar u_{2,k}+\bar u_{1,k}\big)\bar\eta_k{\rm d}x-\frac12\varepsilon_k^2\int_{B_\delta(x_{2,k})}x\cdot\nabla V(x)\big(\bar u_{2,k}+\bar u_{1,k}\big)\bar\eta_k{\rm d}x\\
=&-\varepsilon_k^4\int_{B_{\frac{\delta}{\varepsilon_k}}(0)}V(\varepsilon_k x+x_{2,k})\big(\tilde u_{2,k}(x)+\tilde u_{1,k}(x)\big)\eta_k(x){\rm d}x\\
&-\frac12\varepsilon_k^4\int_{B_{\frac{\delta}{\varepsilon_k}}(0)}\big(\varepsilon_k x+x_{2,k}\big)\cdot\nabla V(\varepsilon_k x+x_{2,k})\big(\tilde u_{2,k}(x)+\tilde u_{1,k}(x)\big)\eta_k(x){\rm d}x\\
=&-\varepsilon_k^4\int_{B_{\frac{\delta}{\varepsilon_k}}(0)}{\rm ln}\Big(1+|\varepsilon_k x+x_{2,k}|^2\Big)\big(\tilde u_{2,k}(x)+\tilde u_{1,k}(x)\big)\eta_k(x){\rm d}x\\
&-\varepsilon_k^4\int_{B_{\frac{\delta}{\varepsilon_k}}(0)}\frac{|\varepsilon_k x+x_{2,k}|^2}{1+|\varepsilon_k x+x_{2,k}|^2}\big(\tilde u_{2,k}(x)+\tilde u_{1,k}(x)\big)\eta_k(x){\rm d}x\\
=&O(\varepsilon_k^6)\int_{B_{\frac{\delta}{\varepsilon_k}}(0)}\Big| x+\frac{x_{2,k}}{\varepsilon_k}\Big|^2\Big(\tilde u_{2,k}(x)
+\tilde u_{1,k}(x)\Big)\eta_k(x){\rm d}x\ \ {\rm as}\ \ k\to\infty.
%&=&-\big(1+O(\delta)\big)\varepsilon_k^6\int_{B_{\delta/\varepsilon_k}(0)}\Big( x+\frac{x_{2,k}}{\varepsilon_k}\Big)^2\Big(\tilde u_{2,k}(x)+\tilde u_{1,k}\Big)\eta_k(x){\rm d}x,
\end{aligned}
\end{equation}

Since $\eta_k\to\eta_0$ in $C_{loc}(\R^2)$ as $k\to\infty$, in view of the exponential decay \eqref{1.6} and \eqref{A.16}, we derive from \eqref{3.35} and \eqref{A.10} that
\begin{equation*}
\int_{B_{\frac{\delta}{\varepsilon_k}}(0)}\Big| x+\frac{x_{2,k}}{\varepsilon_k}\Big|^2\big(\tilde u_{2,k}(x)
+\tilde u_{1,k}(x)\big)\eta_k(x){\rm d}x\to2\int_{\R^2}|x|^2Q(x)\eta_0(x){\rm d}x\ \ {\rm as}\ \ k\to\infty.
\end{equation*}
Together with \eqref{102}, this yields that
\begin{eqnarray}\label{142}
&&\int_{\R^2}\int_{\R^2}{\rm ln}|x-y|\eta_k(x)\big(\tilde u_{2,k}(x)+\tilde u_{1,k}(x)\big)\tilde u_{2,k}^2(y){\rm d}x{\rm d}y\\\nonumber
&&+\int_{\R^2}\int_{\R^2}{\rm ln}|x-y|\tilde u_{1,k}^2(x)\eta_k(y)\big(\tilde u_{2,k}(y)+\tilde u_{1,k}(y)\big){\rm d}x{\rm d}y=O(\varepsilon_k^2)\ \ {\rm as}\ \ k\to\infty.
\end{eqnarray}
Since \begin{equation*}
\int_{\R^2}\int_{\R^2}\frac{(x-y)\cdot x}{|x-y|^2}Q^2(x)Q^2(y){\rm d}x{\rm d}y=\frac12(\rho^*)^2,
\end{equation*}
applying Lemma \ref{lem:A.8},  we derive from \eqref{4.17} and \eqref{142} that
\begin{eqnarray*}
\begin{aligned}
0=&2\int_{\R^2}\int_{\R^2}{\rm ln}|x-y|\eta_0(x)Q(x)Q^2(y){\rm d}x{\rm d}y\\
=&2b_0\int_{\R^2}\int_{\R^2}{\rm ln}|x-y|Q(x)\big[Q(x)+x\cdot\nabla Q(x)\big]Q^2(y){\rm d}x{\rm d}y\\
=&-b_0\int_{\R^2}\int_{\R^2}\frac{(x-y)\cdot x}{|x-y|^2}Q^2(x)Q^2(y){\rm d}x{\rm d}y=-\frac12b_0(\rho^*)^2,
\end{aligned}
\end{eqnarray*}
which implies that $b_0=0$. Step 2 is thus proved.

Step 3. $\eta_0\equiv0$ cannot occur.

By the exponential decay of $\tilde u_{i,k}$ for $i=1,2$, there exists a  point $z_k\in\R^2$ such that $|\eta_k(z_k)|=\|\eta_k\|_{\infty}=1$. Since both $\tilde{u}_{i,k}$ and $Q$ decay exponentially as $|x|\rightarrow\infty$,
applying the maximal principle to \eqref{4.18} yields that $|z_{k}|\leqslant C$ holds uniformly in $k$. Therefore,
we have $\eta_{k}\rightarrow\eta_{0}\not\equiv 0$ in $C_{loc}(\R^2)$ as $k\to\infty$, which however contradicts with the fact that $\eta_{0}\equiv 0$ in $\R^2$.
%Taking $z_k$ in \eqref{3.13}, we know that
%$-A(z_k)+D(z_k)\geqslant2^{-1}$ as $k\to\infty$. Arguing as the proof \eqref{3.14}, it holds that
%\begin{equation*}
%A_k(x)\to-3Q^2(x)\ {\rm uniformly\ in}\ \R^2\ {\rm as}\ k\to\infty
%\end{equation*}
%and
%\begin{equation*}
%D_k(x)\to-\frac2{\rho^*}Q(x)\int_{\R^2}Q^3\eta_0\ {\rm uniformly\ in}\ \R^2\ {\rm as}\ k\to\infty.
%\end{equation*}
%Moreover, we have that $Q(z_k)\geqslant C$ for some $C>0$ as $k\to\infty$, which implies that $\{z_k\}$ is bounded, due to the fact that $Q$ decays exponentially at infinity. Therefore, we deduce that $\bar\eta_k\to\eta_0\neq0$ uniformly in $\R^2$, which however contradicts to the fact that $\eta_0=0$ on $\R^2$ in Step 2. This completes the proof of Theorem \ref{thm:1.3}.
\qed
\medskip

\appendix
\section{Appendix}
For the reader's convenience, in this appendix we shall give the detailed proofs of some results used in Theorems  \ref{thm:1.3} and \ref{thm:1.2}.

\begin{lemma}\label{lem:A.2}
Let $u_\rho>0$ be a positive minimizer of $e(\rho)$ for $\rho\in(0,\rho^*)$, and define
\begin{equation*}
w_\rho(x):=\varepsilon_\rho u_\rho(\varepsilon_\rho x+x_\rho)>0\ \ {\rm in}\ \  \R^2,
\end{equation*}
where $\varepsilon_\rho>0$ is as in \eqref{125} and $x_\rho$ is a  maximal point of $u_\rho$. Then there exists a large constant $R>0$ such that
\begin{equation*}
|\nabla w_\rho(x)|\leqslant Ce^{-\frac{|x|}{2\sqrt{\rho^*}}}\quad {\rm uniformly\  for}\ \ |x|\geqslant R\ \ {\rm as}\ \ \rho\nearrow\rho^*.
\end{equation*}
\end{lemma}
\noindent{\bf Proof.}
Denoting $\partial_j w_\rho=\partial w_\rho/\partial x_j$ for $j=1,2$, it follows from \eqref{1.12} that
\begin{eqnarray}\label{A.4}\nonumber
&&-\Delta\partial_jw_\rho+\varepsilon_\rho^2{\rm ln}\big[1+\left|\varepsilon_\rho x+x_\rho\right|^2\big]\partial_jw_\rho+\frac{2\varepsilon_\rho^3\big(\varepsilon_\rho x_j+x_{\rho}^j\big)}{1+\left|\varepsilon_\rho x+x_\rho\right|^2}w_\rho\\
&&+\varepsilon_\rho^2\Big(\int_{\R^2}{\rm ln}|x-y|w_\rho^2(y){\rm d}y\Big)\partial_jw_\rho+\varepsilon_\rho^2\Big(\int_{\R^2}\frac{x_j-y_j}{|x-y|^2}w_\rho^2(y){\rm d}y\Big)w_\rho\\\nonumber
&&+\rho\varepsilon_\rho^2{\rm ln}\varepsilon_\rho\partial_jw_\rho=\mu_\rho\varepsilon_\rho^2\partial_jw_\rho+3w_\rho^2\partial_jw_\rho\ \ {\rm in}\ \ \R^2,
\end{eqnarray}
where $x_\rho:=(x_{\rho}^1,x_{\rho}^2)\in\R^2$.
Multiplying \eqref{A.4} by $\partial_jw_\rho$ yields that
\begin{equation}\label{A.5}
\begin{aligned}
&-\frac12\Delta\left|\partial_jw_\rho\right|^2+\left|\nabla\partial_jw_\rho\right|^2+\varepsilon_\rho^2{\rm ln}\big[1+\left|\varepsilon_\rho x+x_\rho\right|^2\big]\left|\partial_jw_\rho\right|^2\\
&+\frac{2\varepsilon_\rho^3\big(\varepsilon_\rho x_j+x_{\rho}^j\big)}{1+|\varepsilon_\rho x+x_\rho|^2}w_\rho\partial_jw_\rho+\varepsilon_\rho^2\Big(\int_{\R^2}{\rm ln}|x-y|w_\rho^2(y){\rm d}y\Big)\left|\partial_jw_\rho\right|^2\\
&+\varepsilon_\rho^2\Big(\int_{\R^2}\frac{x_j-y_j}{|x-y|^2}w_\rho^2(y){\rm d}y\Big)w_\rho\partial_jw_\rho+\rho\varepsilon_\rho^2{\rm ln}\varepsilon_\rho\left|\partial_jw_\rho\right|^2\\
&=\mu_\rho\varepsilon_\rho^2\left|\partial_jw_\rho\right|^2+3w_\rho^2\left|\partial_jw_\rho\right|^2\ \ {\rm in}\ \ \R^2.
\end{aligned}
\end{equation}

By Cauchy's inequality, we obtain from \eqref{3.14} and \eqref{3:4} that there exists a large constant $R>0$ such that
\begin{equation*}
\frac{2\varepsilon_\rho^3\big|\varepsilon_\rho x_j+x_{\rho}^j\big|}{1+|\varepsilon_\rho x+x_\rho|^2}\left|w_\rho\partial_jw_\rho\right|\leqslant |w_\rho|^2+\varepsilon_\rho^6\left|\partial_jw_\rho\right|^2\leqslant Ce^{-\frac{4|x|}{3\sqrt{\rho^*}}}+\varepsilon_\rho^6\left|\partial_jw_\rho\right|^2
\end{equation*}
and
\begin{equation*}
\begin{aligned}
\varepsilon_\rho^2\Big(\int_{\R^2}\frac{\left|x_j-y_j\right|}{|x-y|^2}w_\rho^2(y){\rm d}y\Big)\left|w_\rho\partial_jw_\rho\right|\leqslant\varepsilon_\rho^2\Big(\int_{\R^2}\frac1{|x-y|}w_\rho^2(y){\rm d}y\Big)\left|w_\rho\partial_jw_\rho\right|\leqslant Ce^{-\frac{4|x|}{3\sqrt{\rho^*}}}+\varepsilon_\rho^4\left|\partial_jw_\rho\right|^2\
\end{aligned}
\end{equation*}
uniformly for $|x|\geqslant R$ as $\rho\nearrow\rho^*$.
In addition, we infer from \eqref{3.14} that
\begin{equation*}
\varepsilon_\rho^2\left[\int_{\R^2}{\rm ln}\Big(1+\frac1{|x-y|^2}\Big)w_\rho^2(y){\rm d}y\right]\left|\partial_jw_\rho\right|^2\leqslant C\varepsilon_\rho^2\|w_\rho\|^2\left|\partial_jw_\rho\right|^2\ \ {\rm in}\ \ \R^2,
\end{equation*}
where and below $\|\cdot\|$ is as in \eqref{1:1}.
Since $\mu_\rho\varepsilon_\rho^2\to-\frac{1}{\rho^*}$ as $\rho\nearrow\rho^*$, and
\begin{eqnarray*}
&&\left|\nabla\partial_jw_\rho\right|^2+\varepsilon_\rho^2{\rm ln}\big[1+|\varepsilon_\rho x+x_\rho|^2\big]\left|\partial_jw_\rho\right|^2\\
&&+\frac12\varepsilon_\rho^2\left[\int_{\R^2}{\rm ln}(1+|x-y|^2)w_\rho^2(y){\rm d}y\right]\left|\partial_jw_\rho\right|^2\geqslant0\ \ {\rm in}\ \ \R^2,
\end{eqnarray*}
we conclude from \eqref{A.5} that
\begin{eqnarray}\label{A.6}
\Big(-\frac12\Delta+\frac3{4\rho^*}-3w_\rho^2\Big)\left|\nabla w_\rho\right|^2\leqslant Ce^{-\frac{4|x|}{3\sqrt{\rho^*}}}\ \ {\rm uniformly\  for}\ \ |x|\geqslant R\ \ {\rm as}\ \ \rho\nearrow\rho^*.
\end{eqnarray}

Following De Giorgi-Nash-Moser theory \cite[Theorem 4.1]{Han}, we deduce from \eqref{A.6} that for any $\xi\in\R^2$, there exists a constant $C>0$ such that
\begin{equation}\label{A.7}
\max_{B_1(\xi)}|\nabla w_\rho(x)|^2\leqslant C\Big(\int_{B_2(\xi)}|\nabla w_\rho|^2{\rm d}x+\Big\|e^{-\frac{4|x|}{3\sqrt{\rho^*}}}\Big\|_{L^2(B_2(\xi))}\Big).
\end{equation}
Moreover, because $w_\rho\to \frac{1}{\sqrt{\rho^*}} Q\big(\frac{1}{\sqrt{\rho^*}}|x|\big)$ in $H^1(\R^2)$ as $\rho\nearrow\rho^*$, it follows from \eqref{A.7} that
\begin{equation}\label{A.8}
\lim_{|x|\to\infty}|\nabla w_\rho(x)|=0\ \ {\rm as}\ \ \rho\nearrow\rho^*.
\end{equation}
Substituting  \eqref{A.8} into \eqref{A.6}, we  infer that for
sufficiently large $R>0$,
\begin{equation}\label{A:22}
\Big(-\Delta+\frac4{3\rho^*}\Big)\left|\nabla w_\rho\right|^2\leqslant Ce^{-\frac{4|x|}{3\sqrt{\rho^*}}}\leqslant Ce^{-\frac{|x|}{\sqrt{\rho^*}}}\ \ {\rm uniformly\  for}\ \ |x|\geqslant R\ \ {\rm as}\ \ \rho\nearrow\rho^*.
\end{equation}
Applying the comparison principle to \eqref{A:22} yields that
\begin{equation*}
|\nabla w_\rho(x)|\leqslant Ce^{-\frac{|x|}{2\sqrt{\rho^*}}}\ \ {\rm  uniformly\  for}\ \ |x|\geqslant R\ \ {\rm as}\ \ \rho\nearrow\rho^*.
\end{equation*}
This completes the proof of Lemma \ref{lem:A.2}.
\qed
\medskip

\begin{lemma}\label{lem:A.3}
Let $\varepsilon_k>0$ and $\bar\eta_k$ be as in \eqref{4.2} and \eqref{4.12}, respectively. Then there exists a constant $C>0$, independent of $k$, such that
\begin{eqnarray*}
\begin{aligned}
\varepsilon_k^2\int_{\R^2}\left|\nabla\bar\eta_k\right|^2{\rm d}x+\varepsilon_k^2\int_{\R^2}V(x)\bar\eta_k^2{\rm d}x+\int_{\R^2}\bar\eta_k^2{\rm d}x\leqslant C\varepsilon_k^2\ \ \hbox{as}\ \ k\to\infty.
 \end{aligned}
\end{eqnarray*}
\end{lemma}
\noindent{\bf Proof.}
Define
\begin{equation*}
\eta_k(x):=\bar\eta_k\big(\varepsilon_kx+x_{2,k}\big)\ \ \hbox{in}\ \ \R^2,
\end{equation*}
and
\begin{equation*}
\tilde u_{i,k}(x):=\varepsilon_ku_{i,k}\big(\varepsilon_kx+x_{2,k}\big)=\bar{u}_{i,k}\big(\varepsilon_{k}x+x_{2,k}\big)>0\ \ \hbox{in}\ \ \R^2,\ \ i=1,2,
\end{equation*}
where $\bar{u}_{i,k}$ is as in \eqref{4:3} and $x_{2,k}$ is the unique maximal point of $u_{2,k}$ as $k\rightarrow\infty$.
%According to \eqref{4.8} and \eqref{3.35}, we know that $\tilde u_{1,k}$ also satisfies the exponential decay at infinity as $k\to\infty$.  We calculate from \eqref{3.14} that
We now estimate the right hand side of \eqref{4.15} as follows.

The exponential decay of $\tilde u_{i,k}$ in \eqref{4.8} yields that
\begin{eqnarray*}
\begin{aligned}
&\quad\int_{\R^2}\big(\bar u_{2,k}^2+\bar u_{2,k}\bar u_{1,k}+\bar u_{1,k}^2\big)\bar\eta_k^2{\rm d}x-\frac{1}{2\rho_{k}\varepsilon_{k}^2}\int_{\R^2}\big(\bar u_{2,k}^2+\bar u_{1,k}^2\big)\big(\bar u_{2,k}+\bar u_{1,k}\big)\bar\eta_k{\rm d}x\int_{\R^2}\bar u_{1,k}\bar\eta_k{\rm d}x\\
&\leqslant C\varepsilon_k^2   \ \  \mbox{as}\ \  k\to\infty.
\end{aligned}
\end{eqnarray*}
It follows from \eqref{3.14} that there exists a constant $C>0$, independent of $x\in\R^2$ and $k$, such that
\begin{equation*}
    \int_{\R^2}\frac1{|x-y|}\tilde u_{2,k}^2(y){\rm d}y\leqslant C\|\tilde u_{2,k}\|^2\leqslant C,
\end{equation*}
from which we deduce that
\begin{equation*}
\begin{aligned}
&\frac12\int_{\R^2}\int_{\R^2}{\rm ln}\Big(1+\frac1{|x-y|^2}\Big)\bar\eta_k^2(x)\bar u_{2,k}^2(y){\rm d}x{\rm d}y\\
\leqslant&\int_{\R^2}\int_{\R^2}\frac1{|x-y|}\bar\eta_k^2(x)\bar u_{2,k}^2(y){\rm d}x{\rm d}y\\
\leqslant&\varepsilon_k^3\int_{\R^2}\int_{\R^2}\frac1{|x-y|}\eta_k^2(x)\tilde u_{2,k}^2(y){\rm d}x{\rm d}y\\
=&\varepsilon_k^3\int_{\R^2}\eta_k^2(x)\int_{\R^2}\frac1{|x-y|}\tilde u_{2,k}^2(y){\rm d}y{\rm d}x\\
\leqslant&C\varepsilon_k\|\tilde u_{2,k}\|^2\int_{\R^2}\bar\eta_k^2(x){\rm d}x
\leqslant C\varepsilon_k\int_{\R^2}\bar\eta_k^2(x){\rm d}x\ \ \hbox{as}\ \  k\to\infty.
\end{aligned}
\end{equation*}

In addition, since $|\eta_k|\leqslant1$, we get from \eqref{3.14} and \eqref{4.8} that
\begin{equation*}\label{A:0}
\begin{aligned}
&\frac{1}{2\rho_{k}\varepsilon_{k}^2}\int_{\R^2}\int_{\R^2}{\rm ln}|x-y|\big(\bar u_{2,k}(x)+\bar u_{1,k}(x)\big)\bar\eta_k(x)\bar u_{2,k}^2(y){\rm d}x{\rm d}y\int_{\R^2}\bar u_{1,k}(x)\bar\eta_k(x){\rm d}x\\
=&\frac{\varepsilon_k^4{\rm ln}\varepsilon_k}{2\rho_{k}}\int_{\R^2}\int_{\R^2}\big(\tilde u_{2,k}(x)+\tilde u_{1,k}(x)\big)\eta_k(x)\tilde u_{2,k}^2(y){\rm d}x{\rm d}y\int_{\R^2}\tilde u_{1,k}(x)\eta_k(x){\rm d}x\\
&+\frac{\varepsilon_k^4}{4\rho_{k}}\int_{\R^2}\int_{\R^2}{\rm ln}(1+|x-y|^2)\big(\tilde u_{2,k}(x)+\tilde u_{1,k}(x)\big)\eta_k(x)\tilde u_{2,k}^2(y){\rm d}x{\rm d}y\int_{\R^2}\tilde u_{1,k}(x)\eta_k(x){\rm d}x\\
&-\frac{\varepsilon_k^4}{4\rho_{k}}\int_{\R^2}\int_{\R^2}{\rm ln}\Big(1+\frac1{|x-y|^2}\Big)\big(\tilde u_{2,k}(x)+\tilde u_{1,k}(x)\big)\eta_k(x)\tilde u_{2,k}^2(y){\rm d}x{\rm d}y\int_{\R^2}\tilde u_{1,k}(x)\eta_k(x){\rm d}x\\
\leqslant&\left|\frac{1}{2}\varepsilon_k^4{\rm ln}\varepsilon_k\int_{\R^2}\big(\tilde u_{2,k}(x)+\tilde u_{1,k}(x)\big)\frac{\tilde u_{2,k}(x)-\tilde u_{1,k}(x)}{\|\tilde u_{2,k}-\tilde u_{1,k}\|_{L^\infty(\R^2)}}{\rm d}x\int_{\R^2}\tilde u_{1,k}(x)\eta_k(x){\rm d}x\right|\\
&+\frac{\varepsilon_k^4}{4\rho_{k}}\int_{\R^2}\int_{\R^2}\Big[{\rm ln}\big(1+2|x|^2+2|y|^2\big)
+{\rm ln}\Big(1+\frac1{|x-y|^2}\Big)\Big]|\tilde u_{2,k}(x)+\tilde u_{1,k}(x)|\tilde u_{2,k}^2(y){\rm d}x{\rm d}y\\
&\quad\cdot\left|\int_{\R^2}\tilde u_{1,k}(x)\eta_k(x){\rm d}x\right|\\
\leqslant&\left|\frac{(\rho_{k}-\rho_{k})\varepsilon_k^4{\rm ln}\varepsilon_k}{2\|\tilde u_{2,k}-\tilde u_{1,k}\|_{L^\infty(\R^2)}}\int_{\R^2}\tilde u_{1,k}(x)\eta_k(x){\rm d}x\right|\\
\end{aligned}
\end{equation*}
\begin{equation}
\begin{aligned}
&+\frac{\varepsilon_k^4}{4\rho_{k}}\int_{\R^2}\int_{\R^2}\Big[{\rm ln}(1+|x|^2)^2+\ln (1+|y|^2)^2
+{\rm ln}\Big(1+\frac1{|x-y|}\Big)^2\Big]|\tilde u_{2,k}(x)
+\tilde u_{1,k}(x)|\tilde u_{2,k}^2(y){\rm d}x{\rm d}y\\
&\quad\cdot\left|\int_{\R^2}\tilde u_{1,k}(x)\eta_k(x){\rm d}x\right|\\
\end{aligned}
\end{equation}
\begin{equation*}
\begin{aligned}
\leqslant&\frac{\varepsilon_k^4}{2\rho_{k}}\int_{\R^2}\int_{\R^2}\Big[{\rm ln}(1+|x|^2)+\ln (1+|y|^2)
+\frac1{|x-y|}\Big]|\tilde u_{2,k}(x)
+\tilde u_{1,k}(x)|\tilde u_{2,k}^2(y){\rm d}x{\rm d}y\left|\int_{\R^2}\tilde u_{1,k}(x)\eta_k(x){\rm d}x\right|\\
\leqslant&\frac{\varepsilon_k^4}{2\rho_{k}}\Big[\rho_{k}\int_{\R^2}{\rm ln}(1+|x|^2)\left|\tilde u_{2,k}(x)+\tilde u_{1,k}(x)\right|{\rm d}x+\int_{\R^2}\left|\tilde u_{2,k}(x)+\tilde u_{1,k}(x)\right|{\rm d}x \int_{\R^2}{\rm ln}(1+|y|^2)\tilde u_{2,k}^2(y){\rm d}y\\
&+\int_{\R^2}\left|\tilde u_{2,k}(x)
+\tilde u_{1,k}(x)\right|\int_{\R^2}\frac1{|x-y|}\tilde u_{2,k}^2(y){\rm d}y{\rm d}x\Big]\left|\int_{\R^2}\tilde u_{1,k}(x)\eta_k(x){\rm d}x\right|\leqslant C\varepsilon_k^4\ \  \mbox{as}\ \  k\to\infty,
\end{aligned}
\end{equation*}
and
\begin{eqnarray*}
\frac{1}{2\rho_{k}\varepsilon_{k}^2}\int_{\R^2}\int_{\R^2}{\rm ln}|x-y|\bar u_{1,k}^2(x)\big(\bar u_{2,k}(y)+\bar u_{1,k}(y)\big)\bar\eta_k(y){\rm d}x{\rm d}y\int_{\R^2}\bar u_{1,k}(x)\bar\eta_k(x){\rm d}x\leqslant C\varepsilon_k^4\ \  \mbox{as}\ \  k\to\infty.
\end{eqnarray*}

Similarly, we also have
\begin{equation}\label{A:17}
\begin{aligned}
&\int_{\R^2}\int_{\R^2}{\rm ln}|x-y|\bar u_{1,k}(x)\bar\eta_k(x)\big(\bar u_{2,k}(y)+\bar u_{1,k}(y)\big)\bar\eta_k(y){\rm d}x{\rm d}y\\
\leqslant&\varepsilon_k^4\int_{\R^2}{\rm ln}(1+|x|^2)\left|\tilde u_{1,k}(x)\right|{\rm d}x\int_{\R^2}\left|\tilde u_{2,k}(y)+\tilde u_{1,k}(y)\right|{\rm d}y\\
&+\varepsilon_k^4\int_{\R^2}\left|\tilde u_{1,k}(x)\right|{\rm d}x\int_{\R^2}{\rm ln}(1+|y|^2)\left|\tilde u_{2,k}(y)+\tilde u_{1,k}(y)\right|{\rm d}y\\
&+\varepsilon_k^4\int_{\R^2}|\tilde{u}_{1,k}(x)\eta_k(x)|\Big(\int_{\R^2}\frac1{|x-y|}\left|\tilde u_{2,k}(y)+\tilde u_{1,k}(y)\right|^2{\rm d}y\Big)^{\frac{1}{2}}\Big(\int_{\R^2}\frac1{|x-y|}\eta_k^2(y){\rm d}y\Big)^{\frac{1}{2}}{\rm d}x.\\
\end{aligned}
\end{equation}
The same argument of \eqref{3.14} gives that
\begin{equation}\label{A:4}
    \Big(\int_{\R^2}\frac1{|x-y|}\left|\tilde u_{2,k}(y)+\tilde u_{1,k}(y)\right|^2{\rm d}y\Big)^{\frac{1}{2}}\Big(\int_{\R^2}\frac1{|x-y|}\eta_k^2(y){\rm d}y\Big)^{\frac{1}{2}}
    \leqslant C\|\tilde{u}_{1,k}+\tilde{u}_{2,k}\|\|\eta_{k}\|.
\end{equation}
We then deduce from \eqref{A:17} and \eqref{A:4} that
\begin{equation*}
    \begin{aligned}
&\int_{\R^2}\int_{\R^2}{\rm ln}|x-y|\bar u_{1,k}(x)\bar\eta_k(x)\big(\bar u_{2,k}(y)+\bar u_{1,k}(y)\big)\bar\eta_k(y){\rm d}x{\rm d}y\\
\leqslant&C\varepsilon_k^4+C\varepsilon_k^4\rho_k^{\frac{1}{2}}\|\tilde u_{2,k}+\tilde u_{1,k}\|\|\eta_k\|^2\\
\leqslant&C\varepsilon_k^4+C\varepsilon_k^2\int_{\R^2}\bar\eta_k^2{\rm d}x+C\varepsilon_k^4\int_{\R^2}|\nabla\bar\eta_k|^2{\rm d}x  \ \  \mbox{as}\ \  k\to\infty.
\end{aligned}
\end{equation*}

Applying \eqref{4.15}, we infer from above that there exists a constant $C>0$, independent of $k$, such that
\begin{equation*}
\begin{aligned}
&\varepsilon_k^2\int_{\R^2}\left|\nabla\bar\eta_k\right|^2{\rm d}x+\varepsilon_k^2\int_{\R^2}V(x)\bar\eta_k^2{\rm d}x-\mu_{2,k}\varepsilon_k^2\int_{\R^2}\bar\eta_k^2{\rm d}x\\
\leqslant&C\Big(\varepsilon_k\int_{\R^2}\bar\eta_k^2{\rm d}x+\varepsilon_k^4\int_{\R^2}\left|\nabla\bar\eta_k\right|^2{\rm d}x+\varepsilon_k^2\Big)  \ \  \mbox{as}\ \  k\to\infty.
\end{aligned}
\end{equation*}
Following \eqref{4.5}, we obtain from above that
\begin{equation*}
\varepsilon_k^2\int_{\R^2}\left|\nabla\bar\eta_k\right|^2{\rm d}x+\varepsilon_k^2\int_{\R^2}V(x)\bar\eta_k^2{\rm d}x+\int_{\R^2}\bar\eta_k^2{\rm d}x\leqslant C\varepsilon_k^2\ \ \hbox{as}\ \ k\to\infty.
\end{equation*}
This completes the proof of Lemma \ref{lem:A.3}.
\qed
\medskip

\begin{lemma}\label{lem:A.4}
Let $\eta_k$ be a solution of \eqref{4.18}, i.e., $\eta_k$ satisfies
\begin{equation*}
-\Delta\eta_k+C_k(x)\eta_k=D_k(x)\ \ \hbox{in} \ \ \R^2,
\end{equation*}
where $C_k(x)$ and $D_k(x)$ are as in \eqref{4:0} and \eqref{4:1}, respectively. Then
the sequence $\{\eta_k\}$ is bounded uniformly in $C^{1,\alpha}_{loc}(\R^2)$ as $k\to\infty$ for some $\alpha\in(0,1)$.
\end{lemma}
\noindent{\bf Proof.}
Multiplying \eqref{4.18} by $\eta_k$, we deduce that
\begin{eqnarray}\label{A.9}\nonumber
&&-\frac12\Delta\eta_k^2+|\nabla\eta_k|^2+\varepsilon_k^2V(\varepsilon_kx+x_{2,k})\eta_k^2+\rho_k\varepsilon_k^2{\rm ln}\varepsilon_k\eta_k^2-\mu_{2,k}\varepsilon_k^2\eta_k^2\\
&&+\frac12\varepsilon_k^2\Big(\int_{\R^2}{\ln}(1+|x-y|^2)\tilde u_{2,k}^2(y){\rm d}y\Big)\eta_k^2+A_k(x)\eta_k^2=D_k(x)\eta_k\ \ {\rm in}\ \ \R^2,
\end{eqnarray}
where $A_k(x)$ and $D_k(x)$ are as in \eqref{140} and \eqref{4:1}, respectively.

We first claim that there exists a large constant $R>0$ such that
\begin{equation}\label{A.16}
\left|\eta_k(x)\right|\leqslant Ce^{-\frac{|x|}{3}}\ \ {\rm  uniformly}\ \mbox{for}\ \ |x|\geqslant R\ \  \mbox{as}\ \  k\to\infty.
\end{equation}
Indeed, one can obtain from \eqref{4.16} that there exists a constant $C>0$, independent of $k$, such that
\begin{equation}\label{A.10}
\|\eta_k\|^2=\varepsilon_k^{-2}\int_{\R^2}\bar\eta_k^2{\rm d}x+\int_{\R^2}|\nabla\bar\eta_k|^2dx\leqslant C\ \ \hbox{as}\ \ k\rightarrow\infty.
\end{equation}
 Similar to \eqref{A:0}, we derive that
\begin{equation}\label{A:5}
\begin{aligned}
    &\varepsilon_k^2\tilde u_{1,k}\eta_k\left[\int_{\R^2}\int_{\R^2}{\rm ln}|x-y|\big(\tilde u_{2,k}(x)+\tilde u_{1,k}(x)\big)\eta_k(x)\tilde u_{2,k}^2(y){\rm d}x{\rm d}y\right]\\
    \leqslant& C\varepsilon_k^2\left|\tilde u_{1,k}\eta_k\right|
\leqslant\varepsilon_k^4\eta_k^2+Ce^{-\frac{4|x|}{3}}\ \ {\rm uniformly}\ \ \mbox{for}\ \ |x|\geqslant R\ \  \mbox{as}\ \  k\to\infty,
\end{aligned}
\end{equation}
and
\begin{eqnarray}\label{A.13}
\begin{aligned}
&\varepsilon_k^2\tilde u_{1,k}\eta_k\left[\int_{\R^2}\int_{\R^2}{\rm ln}|x-y|\tilde u_{1,k}^2(x)\big(\tilde u_{2,k}(y)+\tilde u_{1,k}(y)\big)\eta_k(y){\rm d}x{\rm d}y\right]\\
\leqslant&\varepsilon_k^4\eta_k^2+Ce^{-\frac{4|x|}{3}}\ \ {\rm uniformly}\ \mbox{for}\ \ |x|\geqslant R\ \  \mbox{as}\ \  k\to\infty.
\end{aligned}
\end{eqnarray}
The same argument of \eqref{A:17} yields that
\begin{equation}\label{A.24}
\begin{aligned}
    &\varepsilon_k^2\tilde u_{1,k}\eta_k\left[\int_{\R^2}{\rm ln}|x-y|\big(\tilde u_{2,k}(y)+\tilde u_{1,k}(y)\big)\eta_k(y){\rm d}y\right]\\
    \leqslant&C\varepsilon_k^2\big[1+{\rm ln}(1+|x|^2)\big]\left|\tilde u_{1,k}\eta_k\right|
\leqslant \varepsilon_k^4\eta_k^2+Ce^{-\frac{2|x|}{3}}\ \ {\rm  uniformly}\ \mbox{for}\ \ |x|\geqslant R\ \  \mbox{as}\ \  k\to\infty.
\end{aligned}
\end{equation}
Moreover, it follows from \eqref{4.8} that
\begin{equation}\label{A.15}
\tilde u_{1,k}\eta_k\int_{\R^2}\big(\tilde u_{2,k}^2+\tilde u_{1,k}^2\big)\big(\tilde u_{2,k}+\tilde u_{1,k}\big)\eta_k{\rm d}x\leqslant \frac14\eta_k^2+Ce^{-\frac{4|x|}{3}}\ \ {\rm  uniformly}\ \ \mbox{for}\ \ |x|\geqslant R\ \ \mbox{as}\  \ k\to\infty,
\end{equation}
and
\begin{equation}\label{A:23}
    \big(\tilde{u}_{2,k}^2+\tilde{u}_{2,k}\tilde{u}_{1,k}+\tilde{u}_{1,k}^2\big)\eta_{k}^2\leqslant Ce^{-\frac{4|x|}{3}}\ \ {\rm  uniformly}\ \ \mbox{for}\ \ |x|\geqslant R\ \ \mbox{as}\  \ k\to\infty.
\end{equation}
Using \eqref{4.5}, we thus derive from \eqref{A.9} and \eqref{A:5}--\eqref{A:23} that there exists a large constant $R>0$ such that
\begin{equation}\label{A.01}
-\frac12\Delta\eta_k^2+\frac23\eta_k^2\leqslant Ce^{-\frac{2|x|}{3}}\ \ {\rm uniformly}\ \ \mbox{for}\ \ |x|\geqslant R\ \ \mbox{as}\ \  k\to\infty.
\end{equation}
Applying the comparison principle to \eqref{A.01}, the claim \eqref{A.16} is therefore proved.

We next prove that $\{\eta_k\}$ is bounded uniformly in $C_{loc}^{1,\alpha}(\R^2)$ as $k\rightarrow\infty$ for some $\alpha\in(0,1)$.
Denote
\begin{equation*}
F_k(x)=-C_k(x)\eta_k+D_k(x)\ \ {\rm in}\ \ \R^2,
\end{equation*}
so that $\eta_k$ satisfies
\begin{equation*}
-\Delta\eta_k=F_k\ \ {\rm in}\ \ \R^2.
\end{equation*}
%It follows from \eqref{3.14}, \eqref{A.16}, \eqref{A.10} and \eqref{4.8} that for $R>0$ large enough,
We now claim that for any $p\geqslant 2$,
$\{F_{k}\}$ is bounded uniformly in $L^p_{loc}(\R^2)$ as $k\to\infty$.
By the exponential decay of $\tilde{u}_{i,k}$, it is sufficient to prove that
$$\int_{\R^2}\ln |x-y|\tilde{u}_{2,k}^2(y){\rm d}y\eta_{k}(x)$$ is bounded uniformly in $L^p_{loc}(\R^2)$ as $k\to\infty$.

Direct calculations yield that for any $p\geqslant 2$,
\begin{equation}\label{A.17}
\begin{aligned}
&\int_{\R^2}\Big|\int_{\R^2}{\ln}|x-y|\tilde u_{2,k}^2(y){\rm d}y\Big|^p|\eta_k(x)|^p{\rm d}x\\
=&\int_{\R^2}\Big|\frac{1}{2}\int_{\R^2}\Big({\ln}(1+|x-y|^2)-\ln \Big(1+\frac{1}{|x-y|^2}\Big)\Big)\tilde u_{2,k}^2(y){\rm d}y\Big|^p|\eta_k(x)|^p{\rm d}x\\
\leqslant&\int_{\R^2}\Big[\frac{1}{2}\int_{\R^2}\Big({\ln}(1+2|x|^2+2|y|^2)+\ln \Big(1+\frac{1}{|x-y|^2}\Big)\Big)\tilde u_{2,k}^2(y){\rm d}y\Big]^p|\eta_k(x)|^p{\rm d}x\\
\leqslant&\int_{\R^2}\Big[\frac{1}{2}\int_{\R^2}\Big({\ln}(1+|x|^2)^2+\ln (1+|y|^2)^2+\ln\Big(1+\frac{1}{|x-y|}\Big)^2\Big)\tilde u_{2,k}^2(y){\rm d}y\Big]^p|\eta_k(x)|^p{\rm d}x\\
\leqslant&\int_{\R^2}\Big[\int_{\R^2}\Big({\ln}(1+|x|^2)+\ln (1+|y|^2)+\frac{1}{|x-y|}\Big)\tilde u_{2,k}^2(y){\rm d}y\Big]^p|\eta_k(x)|^p{\rm d}x\\
=&\int_{\R^2}\Big[\ln (1+|x|^2)\rho_{k}+\int_{\R^2}\ln(1+|y|^2)\tilde{u}_{2,k}^2(y){\rm d}y
+\int_{\R^2}\frac{1}{|x-y|}\tilde{u}_{2,k}^2(y){\rm d}y\Big]^p|\eta_{k}(x)|^p{\rm d}x.\\
%\leqslant&3\rho_k^2\int_{\R^2}\left[{\rm ln}(1+|x|^2)\right]^2\eta_k^2(x){\rm d}x\\\nonumber
%&+3\int_{\R^2}\eta_k^2\left[\Big(\int_{\R^2}{\rm ln}(1+|y|^2)\tilde u_{2,k}^2(y){\rm d}y\Big)^2+\Big(\int_{\R^2}\frac1{|x-y|}\tilde u_{2,k}^2(y){\rm d}y\Big)^2\right]{\rm d}x\leqslant C\ \ \mbox{as}\ \ k\rightarrow\infty.
\end{aligned}
\end{equation}
We obtain from \eqref{3.14} and \eqref{4.8} that
\begin{equation}\label{A:6}
    \int_{\R^2}\ln(1+|y|^2)\tilde{u}_{2,k}^2(y){\rm d}y
+\int_{\R^2}\frac{1}{|x-y|}\tilde{u}_{2,k}^2(y){\rm d}y\leqslant C+C\|\tilde{u}_{2,k}\|^2\leqslant C.
\end{equation}
Applying \eqref{A.16}, we then deduce from \eqref{A.17} and \eqref{A:6} that
\begin{equation*}
    \int_{\R^2}\Big|\int_{\R^2}{\ln}|x-y|\tilde u_{2,k}^2(y){\rm d}y\Big|^p|\eta_k|^p{\rm d}x
    \leqslant C\int_{\R^2}\Big[1+\ln (1+|x|^2)\Big]^p|\eta_{k}(x)|^p{\rm d}x\leqslant C.
\end{equation*}
 This implies that $\{F_{k}\}$ is bounded uniformly in $L^p_{loc}(\R^2)$ as $k\to\infty$ for $p\in[2,\infty)$.
The same argument of \eqref{3:5} yields that $\{\eta_k\}$ is bounded uniformly in $C_{loc}^{1,\alpha}(\R^2)$ as $k\to\infty$ for some $\alpha\in(0,1)$. This completes the proof of Lemma \ref{lem:A.4}.

%Similar to the arguments of \eqref{A.12}-\eqref{A.15}, we obtain from \eqref{4.8}, \eqref{A.10} and \eqref{A.16} that $\{F_k\}$ is bounded uniformly in $L^2_{loc}(\R^2)$ as $k\rightarrow\infty$. The $L^p$ estimate applied to \eqref{A:1} implies that $\{\eta_k\}$ is bounded uniformly in $H^2_{loc}(\R^2)$ as $k\to\infty$. The standard Sobolev
%embedding thus gives that $\{\eta_k\}$ is bounded bounded uniformly in $C_{loc}^{1,\alpha}(\R^2)$ as $k\to\infty$ for some $\alpha\in(0,1)$. This completes the proof of Lemma \ref{lem:A.4}.
\qed
\medskip

\begin{lemma}\label{lem:A.5}
Assume $\eta_k$ is a solution of \eqref{4.18} and $\eta_k\rightharpoonup\eta_0$ in $H^1(\R^2)$ as $k\to\infty$. Then $\eta_0$ is a weak solution of the following equation:
\begin{equation*}
-\Delta\eta_0+(1-3Q^2)\eta_0=-\frac2{\rho^*}\Big(\int_{\R^2}Q^3\eta_0{\rm d}x\Big)Q\ \ \ {\rm in}\ \ \R^2.
\end{equation*}
\end{lemma}
\noindent{\bf Proof.}
Since $x_{2,k}\to0$ and $\tilde u_{i,k}\to Q$ in $L^\infty(\R^2)$ as $k\to\infty$, we first derive from \eqref{1.6}, \eqref{A.16} and \eqref{A.10} that
%We first verify that
for any $\varphi\in H^1(\R^2)$,
\begin{equation}\label{A:3}
\begin{aligned}
&\int_{\R^2}\nabla\eta_k\cdot\nabla\varphi{\rm d}x+\varepsilon_k^2\int_{\R^2}V(\varepsilon_kx+x_{2,k})\eta_k\varphi{\rm d}x+\rho_k\varepsilon_k^2{\rm ln}\varepsilon_k\int_{\R^2}\eta_k\varphi{\rm d}x-\mu_{2,k}\varepsilon_k^2\int_{\R^2}\eta_k\varphi{\rm d}x\\
&-\int_{\R^2}\big(\tilde u_{2,k}^2+\tilde u_{2,k}\tilde u_{1,k}+\tilde u_{1,k}^2\big)\eta_k\varphi{\rm d}x+\frac1{2\rho_k}\int_{\R^2}\big(\tilde u_{2,k}^2+\tilde u_{1,k}^2\big)\big(\tilde u_{2,k}+\tilde u_{1,k}\big)\eta_k{\rm d}x\int_{\R^2}\tilde u_{1,k}\varphi{\rm d}x\\
\to&\int_{\R^2}\nabla\eta_0\cdot\nabla\varphi{\rm d}x+\int_{\R^2}\eta_0\varphi{\rm d}x-3\int_{\R^2}Q^2\eta_0\varphi{\rm d}x+\frac2{\rho^*}\Big(\int_{\R^2}Q^3\eta_0{\rm d}x\Big)\int_{\R^2}Q\varphi{\rm d}x\ \ \mbox{as}\ \ k\to\infty.
\end{aligned}
\end{equation}
Similar to \eqref{A:0}, we obtain from \eqref{3.14}, \eqref{4.8}, \eqref{A.16} and \eqref{A.10} that
\begin{equation}\label{A:10}
\begin{aligned}
&\varepsilon_k^2\int_{\R^2}\int_{\R^2}\big|{\ln}|x-y|\eta_k(x)\varphi(x)\big|\tilde u_{2,k}^2(y){\rm d}y{\rm d}x\\
%=&\frac{1}{2}\int_{\R^2}\int_{\R^2}\Big|\ln(1+|x-y|^2)-\ln \Big(1+\frac{1}{|x-y|^2}\Big)\Big|\big|\eta_k(x)\varphi(x)\big|\tilde u_{2,k}^2(y){\rm d}y{\rm d}x\\
%\leqslant&\frac{1}{2}\int_{\R^2}\int_{\R^2}\Big(\ln (1+2|x|^2+2|y|^2)+\ln(1+\frac{1}{|x-y|^2})\Big)\big|\eta_k(x)\varphi(x)\big|\tilde u_{2,k}^2(y){\rm d}y{\rm d}x\\
%\leqslant&\frac{1}{2}\int_{\R^2}\int_{\R^2}\Big(\ln (1+|x|^2)^2+\ln (1+|y|^2)^2+\ln(1+\frac{1}{|x-y|})^2\Big)\big|\eta_k(x)\varphi(x)\big|\tilde u_{2,k}^2(y){\rm d}y{\rm d}x\\
%\leqslant&\int_{\R^2}\int_{\R^2}\Big(\ln (1+|x|^2)+\ln (1+|y|^2)+\frac{1}{|x-y|}\Big)\big|\eta_k(x)\varphi(x)\big|\tilde u_{2,k}^2(y){\rm d}y{\rm d}x\\
\leqslant&\rho_k\int_{\R^2}{\rm ln}(1+|x|^2)\left|\eta_k(x)\varphi(x)\right|{\rm d}x
+\Big(\int_{\R^2}\eta_k^2{\rm d}x\Big)^{\frac{1}{2}}\Big(\int_{\R^2}\varphi^2{\rm d}x\Big)^{\frac{1}{2}}\int_{\R^2}{\rm ln}(1+|y|^2)\tilde u_{2,k}^2(y){\rm d}y\\
&+\int_{\R^2}|\eta_k(x)\varphi(x)|\int_{\R^2}\frac1{|x-y|}\tilde u_{2,k}^2(y){\rm d}y{\rm d}x\leqslant C\varepsilon_k^2\to 0\ \ \mbox{as}\ \ k\to\infty.
\end{aligned}
\end{equation}
Finally, the same arguments of \eqref{A:5}, \eqref{A.13} and \eqref{A.24} give that
\begin{equation}\label{A:2}
\begin{aligned}
&\frac1{2\rho_k}\varepsilon_k^2\left[\int_{\R^2}\int_{\R^2}{\rm ln}|x-y|\big(\tilde u_{2,k}(x)+\tilde u_{1,k}(x)\big)\eta_k(x)\tilde u_{2,k}^2(y){\rm d}x{\rm d}y\right]\int_{\R^2}\tilde u_{1,k}(x)\varphi(x){\rm d}x\\
&+\frac1{2\rho_k}\varepsilon_k^2\left[\int_{\R^2}\int_{\R^2}{\rm ln}|x-y|\tilde u_{1,k}^2(x)\big(\tilde u_{2,k}(y)+\tilde u_{1,k}(y)\big)\eta_k(y){\rm d}x{\rm d}y\right]\int_{\R^2}\tilde u_{1,k}(x)\varphi(x){\rm d}x\\
&-\varepsilon_k^2\int_{\R^2}\left[\int_{\R^2}{\rm ln}|x-y|\big(\tilde u_{2,k}(y)+\tilde u_{1,k}(y)\big)\eta_k(y){\rm d}y\right]\tilde u_{1,k}(x)\varphi(x){\rm d}x=o(1)\ \ \mbox{as}\ \ k\to\infty.
\end{aligned}
\end{equation}
The proof of Lemma \ref{lem:A.5} is therefore complete in view of \eqref{A:3}--\eqref{A:2}.
\qed
\medskip

\begin{lemma}\label{lem:A.6}
Let $\alpha_{j,k}$ be as in \eqref{4.24} for $j=1,2$. Then there exists a constant $a>0$ such that for $j=1,2$,
\begin{equation*}
\alpha_{j,k}=o\Big(e^{-\frac{a\delta}{\varepsilon_k}}\Big)\ \ as\ \ k\to\infty,
\end{equation*}
where $\varepsilon_k$ and $\delta>0$ are as in \eqref{4.2} and \eqref{4.14}, respectively.
\end{lemma}
\noindent{\bf Proof.}	
We now estimate each term of $\alpha_{j,k}$ as $k\to\infty$ as follows. Since $|\nabla \tilde{u}_{i,k}|$ decays exponentially as $k\to\infty$, it follows from \eqref{4.14} that there exists a constant $a>0$ such that
\begin{equation}
\begin{aligned}\label{143}
&\varepsilon_k^2\int_{\partial B_\delta(x_{2,k})}\left|\frac{\partial\bar\eta_k}{\partial x_j}\frac{\partial\bar u_{2,k}}{\partial\nu}\right|{\rm d}S\\
\leqslant&\varepsilon_k\Big(\int_{\partial B_\delta(x_{2,k})}\left|\nabla\bar u_{2,k}\right|^2{\rm d}S\Big)^{\frac{1}{2}}\Big(\varepsilon_k^2\int_{\partial B_\delta(x_{2,k})}\left|\nabla\bar\eta_k\right|^2{\rm d}S\Big)^{\frac{1}{2}}\\
\leqslant&C\varepsilon_k\Big(\varepsilon_k^{-1}\int_{\partial B_{\frac{\delta}{\varepsilon_k}}(0)}\left|\nabla\tilde u_{2,k}\right|^2{\rm d}S\Big)^{\frac{1}{2}}\Big(\varepsilon_k^2\int_{\partial B_\delta(x_{2,k})}\left|\nabla\bar\eta_k\right|^2{\rm d}S\Big)^{\frac{1}{2}}\\
\leqslant& C\varepsilon_k^2e^{-\frac{a\delta}{\varepsilon_k}}\ \ \mbox{as}\ \ k\to\infty,
\end{aligned}
\end{equation}
\begin{equation}\label{144}
\varepsilon_k^2\int_{\partial B_\delta(x_{2,k})}\left|\frac{\partial\bar u_{1,k}}{\partial x_j}\frac{\partial\bar\eta_k}{\partial\nu}\right|{\rm d}S\leqslant C\varepsilon_k^2e^{-\frac{a\delta}{\varepsilon_k}}\ \ \mbox{as}\ \ k\rightarrow\infty,
\end{equation}
and
\begin{equation}\label{145}
\varepsilon_k^2\int_{\partial B_\delta(x_{2,k})}\left|\Big(\nabla\bar u_{2,k}+\nabla\bar u_{1,k}\Big)\cdot\nabla\bar\eta_k\nu_j\right|{\rm d}S\leqslant C\varepsilon_k^2e^{-\frac{a\delta}{\varepsilon_k}}\ \ \mbox{as}\ \ k\rightarrow\infty.
\end{equation}

In addition, applying the exponential decay of $\tilde{u}_{i,k}$ in \eqref{4.8}, we infer from \eqref{4.5} and \eqref{4.14}  that
\begin{equation}\label{A:13}
\begin{aligned}
&\Big|\varepsilon_k^2\int_{\partial B_\delta(x_{2,k})}{\rm ln}(1+|x|^2)\big(\bar u_{2,k}+\bar u_{1,k}\big)\bar\eta_k\nu_j{\rm d}S-\varepsilon_k^2\mu_{2,k}\int_{\partial B_\delta(x_{2,k})}\big(\bar u_{2,k}+\bar u_{1,k}\big)\bar\eta_k(x)\nu_j{\rm d}S\\
&-\frac{1}{2}\int_{\partial B_\delta(x_{2,k})}\big(\bar u_{2,k}^2+\bar u_{1,k}^2\big)\big(\bar u_{2,k}+\bar u_{1,k}\big)\bar\eta_k\nu_j{\rm d}S\Big|=o\Big(e^{-\frac{a\delta}{\varepsilon_k}}\Big)\ \ \mbox{as}\ \ k\to\infty\color{red}{.}
\end{aligned}
\end{equation}
By the same argument of \eqref{A:0}, it yields from \eqref{4.8}, \eqref{4.14}, \eqref{A:4}, \eqref{A.16} and \eqref{A.10} that
\begin{equation}\label{A:12}
\begin{aligned}
&\left|\int_{\partial B_\delta(x_{2,k})}\int_{\R^2}{\rm ln}|x-y|\big(\bar u_{2,k}(y)+\bar u_{1,k}(y)\big)\bar\eta_k(y)\bar u_{2,k}^2(x)\nu_j{\rm d}y{\rm d}S\right|\\
%\leqslant&\left|\varepsilon_k^3{\rm ln}\varepsilon_k\int_{\partial B_{\frac{\delta}{\varepsilon_k}}(0)}\tilde u_{2,k}^2(x)\nu_j{\rm d}S\int_{\R^2}\big(\tilde u_{2,k}(y)+\tilde u_{1,k}(y)\big)\eta_k(y){\rm d}y\right|\\
%&+\left|\varepsilon_k^3\int_{\partial B_{\frac{\delta}{\varepsilon_k}}(0)}\int_{\R^2}\ln |x-y|\big(\tilde u_{2,k}(y)+\tilde u_{1,k}(y)\big)\eta_k(y)\tilde u_{2,k}^2(x)\nu_j{\rm d}y{\rm d}S\right|\\
%=&\left|\varepsilon_k^3{\rm ln}\varepsilon_k\int_{\partial B_{\frac{\delta}{\varepsilon_k}}(0)}\tilde u_{2,k}^2(x)\nu_j{\rm d}S\int_{\R^2}\big(\tilde u_{2,k}(y)+\tilde u_{1,k}(y)\big)\frac{\tilde u_{2,k}(y)-\tilde u_{1,k}(y)}{\|\tilde u_{2,k}-\tilde u_{1,k}\|_{L^\infty(\R^2)}}{\rm d}y\right|\\
%&+\frac{\varepsilon_k^3}{2}\left|\int_{\partial B_{\frac{\delta}{\varepsilon_k}}(0)}\int_{\R^2}\Big(\ln(1+|x-y|^2)-\ln\Big(1+\frac{1}{|x-y|^2}\Big) \Big)\big(\tilde u_{2,k}(y)+\tilde u_{1,k}(y)\big)\eta_k(y)\tilde u_{2,k}^2(x)\nu_j{\rm d}y{\rm d}S\right|\\
%\leqslant&\left|\frac{(\rho_{k}-\rho_{k})\varepsilon_k^3{\rm ln}\varepsilon_k}{\|\tilde u_{2,k}-\tilde u_{1,k}\|_{L^\infty(\R^2)}}\int_{\partial B_{\frac{\delta}{\varepsilon_k}}(0)}\tilde u_{2,k}^2(x)\nu_j{\rm d}S\right|\\
%&+\varepsilon_k^3\int_{\partial B_{\frac{\delta}{\varepsilon_k}}(0)}\int_{\R^2}\Big(\ln(1+|x|^2)+\ln(1+|y|^2)+\frac{1}{|x-y|} \Big)\left|\big(\tilde u_{2,k}(y)+\tilde u_{1,k}(y)\big)\eta_k(y)\right|\tilde u_{2,k}^2(x){\rm d}y{\rm d}S\\
\leqslant&\varepsilon_k^3\int_{\partial B_{\frac{\delta}{\varepsilon_k}}(0)}{\rm ln}(1+|x|^2)\tilde u_{2,k}^2(x){\rm d}S\int_{\R^2}\left|\big(\tilde u_{2,k}(y)+\tilde u_{1,k}(y)\big)\eta_k(y)\right|{\rm d}y\\
%=&\frac{\varepsilon_k^3}{2}\left|\int_{\partial B_{\frac{\delta}{\varepsilon_k}}(0)}\int_{\R^2}\Big(\ln(1+|x-y|^2)-\ln(1+\frac{1}{|x-y|^2}) \Big)\big(\tilde u_{2,k}(y)+\tilde u_{1,k}(y)\big)\eta_k(y)\tilde u_{2,k}^2(x){\rm d}y{\rm d}S\right|\\
%\leqslant&\frac{\varepsilon_k^3}{2}\int_{\partial B_{\frac{\delta}{\varepsilon_k}}(0)}\int_{\R^2}\Big(\ln(1+2|x|^2+2|y|^2)+\ln(1+\frac{1}{|x-y|^2}) \Big)\left|\big(\tilde u_{2,k}(y)+\tilde u_{1,k}(y)\big)\eta_k(y)\right|\tilde u_{2,k}^2(x){\rm d}y{\rm d}S\\
%\leqslant&\frac{\varepsilon_k^3}{2}\int_{\partial B_{\frac{\delta}{\varepsilon_k}}(0)}\int_{\R^2}\Big(\ln(1+|x|^2)^2+\ln(1+|y|^2)^2+\ln(1+\frac{1}{|x-y|})^2 \Big)\left|\big(\tilde u_{2,k}(y)+\tilde u_{1,k}(y)\big)\eta_k(y)\right|\tilde u_{2,k}^2(x){\rm d}y{\rm d}S\\
%\leqslant&\varepsilon_k^3\int_{\partial B_{\frac{\delta}{\varepsilon_k}}(0)}\int_{\R^2}\Big(\ln(1+|x|^2)+\ln(1+|y|^2)+\frac{1}{|x-y|} \Big)\left|\big(\tilde u_{2,k}(y)+\tilde u_{1,k}(y)\big)\eta_k(y)\right|\tilde u_{2,k}^2(x){\rm d}y{\rm d}S\\
&+\varepsilon_k^3\int_{\partial B_{\frac{\delta}{\varepsilon_k}}(0)}\tilde u_{2,k}^2(x){\rm d}S\int_{\R^2}{\rm ln}(1+|y|^2)\left|\big(\tilde u_{2,k}(y)+\tilde u_{1,k}(y)\big)\eta_k(y)\right|{\rm d}y\\
&+\varepsilon_k^3\int_{\partial B_{\frac{\delta}{\varepsilon_k}}(0)}\tilde u_{2,k}^2(x)\left[\int_{\R^2}\frac1{|x-y|}\big(\tilde u_{2,k}(y)+\tilde u_{1,k}(y)\big)^2{\rm d}y\right]^{\frac{1}{2}}\Big(\int_{\R^2}\frac1{|x-y|}\eta_k^2(y){\rm d}y\Big)^{\frac{1}{2}}{\rm d}S\\
\leqslant&C\rho_k^{\frac{1}{2}}\varepsilon_k^3e^{-\frac{a\delta}{\varepsilon_k}}\left\|\eta_k\right\|_2+C\rho_k^{\frac{1}{2}}\varepsilon_k^3e^{-\frac{a\delta}{\varepsilon_k}}\left(\int_{\R^2}\left[{\rm ln}(1+|y|^2)\right]^2\eta_k^2(y){\rm d}y\right)^{\frac12}\\
&+C\varepsilon_k^3e^{-\frac{a\delta}{\varepsilon_k}}\Big(\left\|\tilde u_{2,k}\right\|+\left\|\tilde u_{1,k}\right\|\Big)\left\|\eta_k\right\|\leqslant C\varepsilon_k^3e^{-\frac{a\delta}{\varepsilon_k}}\ \ \mbox{as}\ \ k\to\infty,
\end{aligned}
\end{equation}
%where \eqref{A:4} is also used. Similar to \eqref{A:12}, we also have
and
\begin{equation}\label{146}
\left|\int_{\partial B_\delta(x_{2,k})}\int_{\R^2}{\rm ln}|x-y|\bar u_{1,k}^2(y)\big(\bar u_{2,k}(x)+\bar u_{1,k}(x)\big)\bar\eta_k(x)\nu_j{\rm d}y{\rm d}S\right|\leqslant C\varepsilon_k^3e^{-\frac{a\delta}{\varepsilon_k}}\ \ \mbox{as}\ \ k\to\infty.
\end{equation}
Applying \eqref{4.8}, we deduce from \eqref{3.14} and \eqref{A.10} that
\begin{equation}\label{147}
\begin{aligned}
&\int_{B_\delta^c(x_{2,k})}\int_{\R^2}\left|\frac{x_j-y_j}{|x-y|^2}\big(\bar u_{2,k}(x)+\bar u_{1,k}(x)\big)\bar\eta_k(x)\bar u_{2,k}^2(y)\right|{\rm d}y{\rm d}x\\
\leqslant&\varepsilon_k^3\int_{B_{\delta/\varepsilon_k}^c(0)}\Big|\big( \tilde{u}_{2,k}(x)+ \tilde{u}_{1,k}(x)\big)\eta_k(x)\Big|\int_{\R^2}\frac{1}{|x-y|}\tilde{u}_{2,k}^2(y){\rm d}y{\rm d}x\\
\leqslant&C\varepsilon_k^3\|\tilde u_{2,k}\|^2\Big(\int_{B_{\delta/\varepsilon_k}^c(0)}\big|\tilde u_{2,k}+\tilde u_{1,k}\big|^2{\rm d}x\Big)^{\frac{1}{2}}\Big(\int_{\R^2}\eta_k^2{\rm d}x\Big)^{\frac{1}{2}}\leqslant C\varepsilon_k^3e^{-\frac{a\delta}{\varepsilon_k}}\ \ \mbox{as}\ \ k\to\infty,
\end{aligned}
\end{equation}
and
\begin{equation}\label{148}
\int_{ B_\delta^c(x_{2,k})}\int_{\R^2}\left|\frac{x_j-y_j}{|x-y|^2}\bar u_{1,k}^2(x)\big(\bar u_{2,k}(y)+\bar u_{1,k}(y)\big)\bar\eta_k(y)\right| {\rm d}y{\rm d}x\leqslant C\varepsilon_k^3e^{-\frac{a\delta}{\varepsilon_k}}\ \ \mbox{as}\ \ k\to\infty.
\end{equation}

To estimate the last term of $\alpha_{j,k}$ as $k\to\infty$, we first claim that
\begin{equation}\label{A.19}
\begin{aligned}
\frac{\mu_{2,k}-\mu_{1,k}}{\|\bar u_{2,k}-\bar u_{1,k}\|_{L^\infty(\R^2)}}\varepsilon_k^2=&\frac1{2\rho_k\varepsilon_k^2}\int_{\R^2}\int_{\R^2}{\rm ln}|x-y|\big(\bar u_{2,k}(x)+\bar u_{1,k}(x)\big)\bar\eta_k(x)\bar u_{2,k}^2(y){\rm d}x{\rm d}y\\
&+\frac1{2\rho_k\varepsilon_k^2}\int_{\R^2}\int_{\R^2}{\rm ln}|x-y|\bar u_{1,k}^2(x)\big(\bar u_{2,k}(y)+\bar u_{1,k}(y)\big)\bar\eta_k(y){\rm d}x{\rm d}y\\
&-\frac1{2\rho_k\varepsilon_k^2}\int_{\R^2}\big(\bar u_{2,k}^2(x)+\bar u_{1,k}^2(x)\big)\big(\bar u_{2,k}(x)+\bar u_{1,k}(x)\big)\bar\eta_k(x){\rm d}x
\end{aligned}
\end{equation}
is bounded uniformly in $k$ as $k\to\infty$.  In fact, the same argument of \eqref{A:0} yields that
%applying the exponential decay of $\tilde{u}_{i,k}$, we derive from \eqref{3.14} that
\begin{equation}\label{150}
\begin{aligned}
&\Big|\frac1{\varepsilon_k^2}\int_{\R^2}\int_{\R^2}{\rm ln}|x-y|\big(\bar u_{2,k}(x)+\bar u_{1,k}(x)\big)\bar\eta_k(x)\bar u_{2,k}^2(y){\rm d}x{\rm d}y\\
&\quad+\frac1{\varepsilon_k^2}\int_{\R^2}\int_{\R^2}{\rm ln}|x-y|\bar u_{1,k}^2(x)\big(\bar u_{2,k}(y)+\bar u_{1,k}(y)\big)\bar\eta_k(y){\rm d}x{\rm d}y\Big|\leqslant C\varepsilon_k^2\ \ \mbox{as}\ \ k\to\infty.
\end{aligned}
\end{equation}
 It follows from \eqref{4.8} that
\begin{equation}\label{152}
\begin{aligned}
&\frac1{2\varepsilon_k^2}\int_{\R^2}\left|\big(\bar u_{2,k}^2+\bar u_{1,k}^2\big)\big(\bar u_{2,k}+\bar u_{1,k}\big)\bar\eta_k\right|{\rm d}x\\
=&\frac{1}{2}\int_{\R^2}\left|\big(\tilde{u}_{2,k}^2+ \tilde{u}_{1,k}^2\big)\big(\tilde{u}_{2,k}+ \tilde{u}_{1,k}\big)\eta_k\right|{\rm d}x
\leqslant C\ \ \mbox{as}\ \ k\to\infty.
\end{aligned}
\end{equation}
We then conclude from \eqref{150} and \eqref{152} that \eqref{A.19} is bounded uniformly in $k$ as $k\to\infty$.

Moreover, by the exponential decay of $\tilde u_{1,k}$ as $k\to\infty$, it yields from \eqref{A.19} that
\begin{equation}\label{149}
\frac{\mu_{2,k}-\mu_{1,k}}{\|\bar u_{2,k}-\bar u_{1,k}\|_{L^\infty(\R^2)}}\varepsilon_k^2\int_{\partial B_\delta(x_{2,k})}\bar u_{1,k}^2\nu_j{\rm d}S
=o\big(e^{-\frac{a\delta}{\varepsilon_k}}\big)\ \ \mbox{as}\ \ k\to\infty.
\end{equation}
Applying the estimates of \eqref{143}--\eqref{148} and \eqref{149},  the proof of Lemma \ref{lem:A.6} is therefore complete.
\qed
\medskip

\begin{lemma}\label{lem:A.7}
Let $\gamma_k$ and $\delta>0$ be as in \eqref{4.31} and \eqref{4.14}, respectively. Then there exists a constant $b\in(0,a)$ such that
\begin{equation*}
\gamma_k=o\big(e^{-\frac{b\delta}{\varepsilon_k}}\big)\ \ \mbox{as}\ \ k\to\infty.
\end{equation*}
\end{lemma}
\noindent{\bf Proof.}
Since $x_{2,k}\to0$ as $k\to\infty$, we note from the first identity of \eqref{4.25} that
\begin{equation}\label{A:14}
\begin{aligned}
&\frac{\varepsilon_k^2}2\int_{B_\delta(x_{2,k})}x_{2,k}\cdot\nabla V(x)\big(\bar u_{2,k}+\bar u_{1,k}\big)\bar\eta_k{\rm d}x\\
=&\frac{\varepsilon_k^2}2\Sigma_{j=1}^2x_{2,k}^j\int_{B_\delta(x_{2,k})}\frac{\partial V(x)}{\partial x_j}\big(\bar u_{2,k}+\bar u_{1,k}\big)\bar\eta_k(x){\rm d}x=o\big(e^{-\frac{b\delta}{\varepsilon_k}}\big)\ \ \mbox{as}\ \ k\to\infty,
\end{aligned}
\end{equation}
where $x_{2,k}:=\{x_{2,k}^1,x_{2,k}^2\}\in \R^2$. By the definition of $\gamma_k$ in \eqref{4.31}, it is sufficient to prove that
\begin{equation}\label{167}
\frac{\beta_{2,k}-\beta_{1,k}}{\|\bar u_{2,k}-\bar u_{1,k}\|_{L^\infty(\R^2)}}=o\big(e^{-\frac{b\delta}{\varepsilon_k}}\big)\ \ {\rm as}\ \ k\to\infty.
\end{equation}

It follows from \eqref{4.30} that
\begin{equation*}\label{A.20}
\begin{aligned}
&\frac{\beta_{2,k}-\beta_{1,k}}{\|\bar u_{2,k}-\bar u_{1,k}\|_{L^\infty(\R^2)}}\\
=&\varepsilon_k^2\mu_{2,k}\int_{B_\delta^c(x_{2,k})}\big(\bar u_{2,k}+\bar u_{1,k}\big)\bar\eta_k{\rm d}x+\frac{\mu_{2,k}-\mu_{1,k}}{\|\bar u_{2,k}-\bar u_{1,k}\|_{L^\infty(\R^2)}}\varepsilon_k^2\int_{B_\delta^c(x_{2,k})}\bar u_{1,k}^2{\rm d}x\\
&+\frac12\varepsilon_k^2\mu_{2,k}\int_{\partial B_\delta(x_{2,k})}\big(\bar u_{2,k}+\bar u_{1,k}\big)\bar\eta_k(x-x_{2,k})\cdot\nu{\rm d}S\\
&+\frac12\frac{\mu_{2,k}-\mu_{1,k}}{\|\bar u_{2,k}-\bar u_{1,k}\|_{L^\infty(\R^2)}}\varepsilon_k^2\int_{\partial B_\delta(x_{2,k})}\bar u_{1,k}^2(x-x_{2,k})\cdot\nu{\rm d}S\\
&-\frac12\varepsilon_k^2\int_{\partial B_\delta(x_{2,k})}V(x)\big(\bar u_{2,k}+\bar u_{1,k}\big)\bar\eta_k(x-x_{2,k})\cdot\nu{\rm d}S\\
&-\frac12\int_{B_\delta^c(x_{2,k})}\int_{\R^2}\frac{(x-x_{2,k})\cdot(x-y)}{|x-y|^2}\bar u_{2,k}^2(x)\big(\bar u_{2,k}(y)+\bar u_{1,k}(y)\big)\bar\eta_k(y){\rm d}y{\rm d}x\\
&-\frac12\int_{B_\delta^c(x_{2,k})}\int_{\R^2}\frac{(x-x_{2,k})\cdot(x-y)}{|x-y|^2}\big(\bar u_{2,k}(x)+\bar u_{1,k}(x)\big)\bar\eta_k(x)\bar u_{1,k}^2(y){\rm d}y{\rm d}x\\
&-\int_{B_\delta^c(x_{2,k})}\int_{\R^2}{\rm ln}|x-y|\bar u_{2,k}^2(x)\big(\bar u_{2,k}(y)+\bar u_{1,k}(y)\big)\bar\eta_k(y){\rm d}y{\rm d}x\\
&-\int_{B_\delta^c(x_{2,k})}\int_{\R^2}{\rm ln}|x-y|\big(\bar u_{2,k}(x)+\bar{u}_{1,k}(x)\big)\bar\eta_k(x)\bar u_{1,k}^2(y){\rm d}y{\rm d}x\\
&-\frac12\int_{\partial B_\delta(x_{2,k})}\int_{\R^2}{\rm ln}|x-y|\bar u_{2,k}^2(x)\big(\bar u_{2,k}(y)+\bar u_{1,k}(y)\big)\bar\eta_k(y)(x-x_{2,k})\cdot\nu{\rm d}y{\rm d}S\\
&-\frac12\int_{\partial B_\delta(x_{2,k})}\int_{\R^2}{\rm ln}|x-y|\big(\bar u_{2,k}(x)+\bar u_{1,k}(x)\big)\bar\eta_k(x)\bar u_{1,k}^2(y)(x-x_{2,k})\cdot\nu{\rm d}y{\rm d}S\\
&+\frac12\int_{B_\delta^c(x_{2,k})}\big(\bar u_{2,k}^2+\bar u_{1,k}^2\big)\big(\bar u_{2,k}+\bar u_{1,k}\big)\bar\eta_kdx\\
&+\frac14\int_{\partial B_\delta(x_{2,k})}\big(\bar u_{2,k}^2+\bar u_{1,k}^2\big)\big(\bar u_{2,k}+\bar u_{1,k}\big)\bar\eta_k(x-x_{2,k})\cdot\nu{\rm d}S.
\end{aligned}
\end{equation*}
We next estimate the right hand side of \eqref{A.20} as follows. By the exponential decay of $\tilde{u}_{i,k}$, we obtain from \eqref{A.19} that
\begin{equation}\label{160}
\begin{aligned}
o\big(e^{-\frac{b\delta}{\varepsilon_k}}\big)=&\varepsilon_k^2\mu_{2,k}\int_{B_\delta^c(x_{2,k})}\big(\bar u_{2,k}+\bar u_{1,k}\big)\bar\eta_k{\rm d}x+\frac{\mu_{2,k}-\mu_{1,k}}{\|\bar u_{2,k}-\bar u_{1,k}\|_{L^\infty(\R^2)}}\varepsilon_k^2\int_{B_\delta^c(x_{2,k})}\bar u_{1,k}^2{\rm d}x\\
&+\frac12\varepsilon_k^2\mu_{2,k}\int_{\partial B_\delta(x_{2,k})}\big(\bar u_{2,k}+\bar u_{1,k}\big)\bar\eta_k(x-x_{2,k})\cdot\nu{\rm d}S\\
&+\frac12\frac{\mu_{2,k}-\mu_{1,k}}{\|\bar u_{2,k}-\bar u_{1,k}\|_{L^\infty(\R^2)}}\varepsilon_k^2\int_{\partial B_\delta(x_{2,k})}\bar u_{1,k}^2(x-x_{2,k})\cdot\nu{\rm d}S\\
&-\frac12\varepsilon_k^2\int_{\partial B_\delta(x_{2,k})}V(x)\big(\bar u_{2,k}+\bar u_{1,k}\big)\bar\eta_k(x-x_{2,k})\cdot\nu{\rm d}S\\
&+\frac12\int_{B_\delta^c(x_{2,k})}\big(\bar u_{2,k}^2+\bar u_{1,k}^2\big)\big(\bar u_{2,k}+\bar u_{1,k}\big)\bar\eta_kdx\\
&+\frac14\int_{\partial B_\delta(x_{2,k})}\big(\bar u_{2,k}^2+\bar u_{1,k}^2\big)\big(\bar u_{2,k}+\bar u_{1,k}\big)\bar{\eta}_k(x-x_{2,k})\cdot\nu{\rm d}S\ \ \mbox{as}\ \ k\to\infty.
\end{aligned}
\end{equation}
Similar to \eqref{147}, we obtain that there exists a constant $b\in(0,a)$ such that
\begin{equation}\label{161}
\begin{aligned}
&\int_{B_\delta^c(x_{2,k})}\int_{\R^2}\frac{(x-x_{2,k})\cdot(x-y)}{|x-y|^2}\bar u_{2,k}^2(x)\big(\bar u_{2,k}(y)+\bar u_{1,k}(y)\big)\bar\eta_k(y){\rm d}y{\rm d}x\\
%\leqslant&\varepsilon_k^4\int_{B_{\delta/\varepsilon_k}^c(0)}\int_{\R^2}\frac{|x|}{|x-y|}\tilde u_{2,k}^2(x)\big|\big(\tilde u_{2,k}(y)+\tilde u_{1,k}(y)\big)\eta_k(y)\big|{\rm d}y{\rm d}x\\
%\leqslant&\varepsilon_k^4\int_{B_{\delta/\varepsilon_k}^c(0)}|x|\tilde u_{2,k}^2(x)\int_{\R^2}\frac{1}{|x-y|}\big|\big(\tilde u_{2,k}(y)+\tilde u_{1,k}(y)\big)\eta_k(y)\big|{\rm d}y{\rm d}x\\
%\leqslant&\varepsilon_k^4\int_{B_{\delta/\varepsilon_k}^c(0)}|x|\tilde u_{2,k}^2(x)\Big(\int_{\R^2}\frac{1}{|x-y|}\big|\tilde u_{2,k}(y)+\tilde u_{1,k}(y)\big|^2{\rm d}y\Big)^{\frac{1}{2}}\Big(\int_{\R^2}\frac{1}{|x-y|}\eta_k^2(y){\rm d}y\Big)^{\frac{1}{2}}{\rm d}x\\
%\leqslant&C\varepsilon_k^4\|\tilde u_{2,k}+ \tilde u_{1,k}\|\|\eta_k\|\int_{B_{\delta/\varepsilon_k}^c(0)}|x|\tilde u_{2,k}^2(x){\rm d}x\leqslant C\varepsilon_k^4e^{-\frac{b\delta}{\varepsilon_k}}\ \ \mbox{as}\ \ k\to\infty,
%%\end{aligned}
%\end{equation}
%and
%\begin{equation}\label{162}
&\quad+\int_{B_\delta^c(x_{2,k})}\int_{\R^2}\frac{(x-x_{2,k})\cdot(x-y)}{|x-y|^2}\big(\bar u_{2,k}(x)+\bar u_{1,k}(x)\big)\bar\eta_k(x)\bar u_{1,k}^2(y){\rm d}y{\rm d}x=o(e^{-\frac{b\delta}{\varepsilon_k}})\ \ \mbox{as}\ \ k\to\infty.
\end{aligned}
\end{equation}
%Using \eqref{4.8}, we conclude from \eqref{A:4} and \eqref{A.10} that
The same argument of \eqref{A:12} yields that
\begin{equation}\label{166}
\begin{aligned}
&-\int_{B_\delta^c(x_{2,k})}\int_{\R^2}{\rm ln}|x-y|\bar u_{2,k}^2(x)\big(\bar u_{2,k}(y)+\bar u_{1,k}(y)\big)\bar\eta_k(y){\rm d}y{\rm d}x\\
&-\int_{B_\delta^c(x_{2,k})}\int_{\R^2}{\rm ln}|x-y|\big(\bar u_{2,k}(x)+\bar{u}_{1,k}(x)\big)\bar\eta_k(x)\bar u_{1,k}^2(y){\rm d}y{\rm d}x\\
&-\frac12\int_{\partial B_\delta(x_{2,k})}\int_{\R^2}{\rm ln}|x-y|\bar u_{2,k}^2(x)\big(\bar u_{2,k}(y)+\bar u_{1,k}(y)\big)\bar\eta_k(y)(x-x_{2,k})\cdot\nu{\rm d}y{\rm d}S\\
&-\frac12\int_{\partial B_\delta(x_{2,k})}\int_{\R^2}{\rm ln}|x-y|\big(\bar u_{2,k}(x)+\bar u_{1,k}(x)\big)\bar\eta_k(x)\bar u_{1,k}^2(y)(x-x_{2,k})\cdot\nu{\rm d}y{\rm d}S=o\big(e^{-\frac{b\delta}{\varepsilon_k}}\big)\ \ \mbox{as}\ \ k\to\infty.
\end{aligned}
\end{equation}
We conclude from \eqref{160}--\eqref{166} that \eqref{167} holds true, and the proof of Lemma \ref{lem:A.7} is therefore complete.
\qed
\medskip

\begin{lemma}\label{lem:A.8}
Let $\tilde u_{i,k}$ be as in \eqref{4:2} for $i=1,2$, and suppose $\eta_k$ is a solution of \eqref{4.18}. Then we have
\begin{equation*}
\int_{\R^2}\int_{\R^2}{\rm ln}|x-y|\eta_k(x)\big(\tilde u_{2,k}(x)+\tilde u_{1,k}(x)\big)\tilde u_{2,k}^2(y){\rm d}x{\rm d}y\to2\int_{\R^2}\int_{\R^2}{\rm ln}|x-y|\eta_0(x)Q(x)Q^2(y){\rm d}x{\rm d}y\ \ \mbox{as}\ \ k\to\infty,
\end{equation*}
where $\eta_0(x)$ is defined by \eqref{4.17}.
\end{lemma}
\noindent{\bf Proof.}
We note that
\begin{equation*}
\begin{aligned}
&\int_{\R^2}\int_{\R^2}{\rm ln}|x-y|\Big[\eta_k(x)\big(\tilde u_{2,k}(x)+\tilde u_{1,k}(x)\big)\tilde u_{2,k}^2(y)-2\eta_0(x)Q(x)Q^2(y)\Big]{\rm d}x{\rm d}y\\
=&\int_{\R^2}\int_{\R^2}{\rm ln}|x-y|\Big(\eta_k(x)-\eta_0(x)\Big)\big(\tilde u_{2,k}(x)+\tilde u_{1,k}(x)\big)\tilde u_{2,k}^2(y){\rm d}x{\rm d}y\\
&+\int_{\R^2}\int_{\R^2}{\rm ln}|x-y|\eta_0(x)\Big(\tilde u_{2,k}(x)+\tilde u_{1,k}(x)-2Q(x)\Big)\tilde u_{2,k}^2(y){\rm d}x{\rm d}y\\
&+2\int_{\R^2}\int_{\R^2}{\rm ln}|x-y|\eta_0(x)Q(x)\big(\tilde u_{2,k}^2(y)-Q^2(y)\big){\rm d}x{\rm d}y:=I_{1,k}+I_{2,k}+I_{3,k}.
\end{aligned}
\end{equation*}
To prove Lemma \ref{lem:A.8}, it is sufficient to prove that for $i=1,2,3$,
\begin{equation}\label{A:21}
    I_{i,k}\to0 \ \ \mbox{as}\ \ k\to\infty.
\end{equation}

By the exponential decay of $\tilde u_{i,k}$, since $\tilde u_{2,k}\to Q$ in $L^\infty(\R^2)\cap X$ as $k\to\infty$, we have for any $\epsilon>0$, there exists a large constant $R>0$ such that
\begin{equation}\label{A:15}
\begin{aligned}
    &2\rho_k\int_{B_R^c(0)}{\rm ln}\big(1+|x|^2\big)\big|\tilde u_{2,k}(x)+\tilde u_{1,k}(x)\big|{\rm d}x+2\int_{B_R^c(0)}\left|\tilde u_{2,k}(x)+\tilde u_{1,k}(x)\right|{\rm d}x\\
    &\quad\cdot\Big(\|\tilde u_{2,k}\|_*^2+C\|\tilde u_{2,k}\|^2\Big)
    < \frac{\epsilon}{2}\ \ \mbox{as}\ \ k\to\infty.
\end{aligned}
\end{equation}
For the fixed constant $R>0$ in \eqref{A:15}, since $\tilde u_{2,k}\to Q$ in $L^\infty(\R^2)\cap X$ and $\eta_k\to\eta_0$ in $C_{loc}(\R^2)$ as $k\to\infty$, we obtain from \eqref{4.8}  that
\begin{equation}\label{A:16}
\begin{aligned}
    &\|\eta_k-\eta_0\|_{C(B_R(0))}\Big(\rho_k\int_{B_R(0)}{\rm ln}(1+|x|^2)\big|\tilde u_{2,k}(x)+\tilde u_{1,k}(x)\big|{\rm d}x\\
    &\quad+\int_{B_R(0)}\big|\tilde u_{2,k}(x)
    +\tilde u_{1,k}(x)\big|{\rm d}x
   \cdot\big(\|\tilde u_{2,k}\|_*^2+C\|\tilde u_{2,k}\|^2\big)\Big)
    < \frac{\epsilon}{2}\ \ \mbox{as}\ \ k\to\infty.
\end{aligned}
\end{equation}
Similar to \eqref{A:0}, we obtain from \eqref{3.14}, \eqref{A:15} and \eqref{A:16} that for any $\epsilon>0$, there exists a large constant $R>0$ such that
%there exists a constant $R>0$ such that as $k\to\infty$,
\begin{equation}\label{A:18}
\begin{aligned}
\left|I_{1,k}\right|=&\left|\int_{\R^2}\int_{\R^2}{\rm ln}|x-y|\Big(\eta_k(x)-\eta_0(x)\Big)\big(\tilde u_{2,k}(x)+\tilde u_{1,k}(x)\big)\tilde u_{2,k}^2(y){\rm d}x{\rm d}y\right|\\
%=&\frac{1}{2}\left|\int_{\R^2}\int_{\R^2}\Big({\rm ln}(1+|x-y|^2)-{\rm ln}\Big(1+\frac{1}{|x-y|^2}\Big)\Big)\Big(\eta_k(x)-\eta_0(x)\Big)\big(\tilde u_{2,k}(x)+\tilde u_{1,k}(x)\big)\tilde u_{2,k}^2(y){\rm d}x{\rm d}y\right|\\
%\leqslant&\int_{\R^2}\int_{\R^2}\Big({\rm ln}(1+|x|^2)+{\rm ln}(1+|y|^2)+\frac{1}{|x-y|}\Big)|\eta_k(x)-\eta_0(x)||\tilde u_{2,k}(x)+\tilde u_{1,k}(x)|\tilde u_{2,k}^2(y){\rm d}x{\rm d}y\\
\leqslant&\rho_k\int_{\R^2}{\rm ln}\big(1+|x|^2\big)\left|\eta_k(x)-\eta_0(x)\right||\tilde u_{2,k}(x)+\tilde u_{1,k}(x)|{\rm d}x\\
&+\int_{\R^2}\left|\eta_k(x)-\eta_0(x)\right|\left|\tilde u_{2,k}(x)+\tilde u_{1,k}(x)\right|{\rm d}x\int_{\R^2}{\rm ln}\big(1+|y|^2\big)\tilde u_{2,k}^2(y){\rm d}y\\
&+\int_{\R^2}\left|\eta_k(x)-\eta_0(x)\right|\left|\tilde u_{2,k}(x)+\tilde u_{1,k}(x)\right|\int_{\R^2}\frac1{|x-y|}\tilde u_{2,k}^2(y){\rm d}y{\rm d}x\\
\leqslant&\rho_k\|\eta_k-\eta_0\|_{C(B_R(0))}\int_{B_R(0)}{\rm ln}(1+|x|^2)|\tilde u_{2,k}(x)+\tilde u_{1,k}(x)|{\rm d}x\\
&+2\rho_k\int_{B_R^c(0)}{\rm ln}\big(1+|x|^2\big)|\tilde u_{2,k}(x)+\tilde u_{1,k}(x)|{\rm d}x\\
&+\Big(\|\eta_k-\eta_0\|_{C(B_R(0))}\int_{B_R(0)}\left|\tilde u_{2,k}(x)+\tilde u_{1,k}(x)\right|{\rm d}x+2\int_{B_R^c(0)}\left|\tilde u_{2,k}(x)+\tilde u_{1,k}(x)\right|{\rm d}x\Big)\\
&\quad\cdot\Big(\|\tilde u_{2,k}\|_*^2+C\|\tilde u_{2,k}\|^2\Big)<\epsilon\ \ \hbox{as}\ \ k\to\infty.
\end{aligned}
\end{equation}
%where the facts that $\eta_k\to\eta_0$ in $C_{loc}(\R^2)$ and $\tilde u_{2,k}\to Q$ in $X\cap L^\infty(\R^2)$ as $k\to\infty$ are also used.
We thus have $I_{1,k}\rightarrow 0$ as $k\rightarrow\infty$.

Similarly, one can obtain from \eqref{1.6}, \eqref{3.14} and \eqref{4.8} that
\begin{equation}\label{A:19}
\begin{aligned}
\left|I_{2,k}\right|=&\left|\int_{\R^2}\int_{\R^2}{\rm ln}|x-y|\eta_0(x)\big(\tilde u_{2,k}(x)+\tilde u_{1,k}(x)-2Q(x)\big)\tilde u_{2,k}^2(y){\rm d}x{\rm d}y\right|\\
\leqslant&\rho_k\int_{\R^2}{\rm ln}(1+|x|^2)\left|\tilde u_{2,k}(x)+\tilde u_{1,k}(x)-2Q(x)\right|{\rm d}x\\
&+\int_{\R^2}\big|\tilde u_{2,k}(x)+\tilde u_{1,k}(x)-2Q(x)\big|{\rm d}x\int_{\R^2}{\rm ln}(1+|y|^2)\tilde u_{2,k}^2(y){\rm d}y\\
&+\int_{\R^2}\big|\tilde u_{2,k}(x)+\tilde u_{1,k}(x)-2Q(x)\big|\int_{\R^2}\frac1{|x-y|}\tilde u_{2,k}^2(y){\rm d}y{\rm d}x\to0\ \ \hbox{as}\ \ k\to\infty,
\end{aligned}
\end{equation}
and
\begin{equation}\label{A:20}
\begin{aligned}
|I_{3,k}|=&\left|2\int_{\R^2}\int_{\R^2}{\rm ln}|x-y|\eta_0(x)Q(x)\big(\tilde u_{2,k}^2(y)-Q^2(y)\big){\rm d}x{\rm d}y\right|\\
=&\left|\int_{\R^2}\int_{\R^2}\Big({\rm ln}(1+|x-y|^2)-{\rm ln}\Big(1+\frac{1}{|x-y|^2}\Big)\Big)\eta_0(x)Q(x)\big(\tilde u_{2,k}^2(y)-Q^2(y)\big){\rm d}x{\rm d}y\right|\\
\leqslant&2\int_{\R^2}\int_{\R^2}\Big({\rm ln}(1+|x|^2)+\ln (1+|y|^2)+\frac{1}{|x-y|}\Big)|\eta_0(x)|Q(x)\big|\tilde u_{2,k}^2(y)-Q^2(y)\big|{\rm d}x{\rm d}y\\
\leqslant&2\|\tilde u_{2,k}-Q\|_2\|\tilde u_{2,k}+Q\|_2\int_{\R^2}{\rm ln}\big(1+|x|^2\big)Q(x){\rm d}x+2\|\tilde u_{2,k}-Q\|_*\|\tilde u_{2,k}+Q\|_*\int_{\R^2}Q{\rm d}x\\
&+C\|\tilde u_{2,k}+Q\|\|\tilde u_{2,k}-Q\|\int_{\R^2}Q{\rm d}x\to0\ \ \hbox{as}\ \ k\to\infty,
\end{aligned}
\end{equation}
where we have used the fact that $\tilde u_{i,k}\to Q$ in $L^\infty(\R^2)\cap X$  as $k\to\infty$.

Following \eqref{A:18}--\eqref{A:20}, we obtain that \eqref{A:21} holds true. This completes the proof of Lemma \ref{lem:A.8}.
\qed
\medskip

% \renewcommand{\proofname}{\bf Proof.}
% \section{Nondegeneracy of \texorpdfstring{$L_+$}{L+}}

%\begin{proof}[Proof of Theorem ]

% \end{proof}

\renewcommand{\proofname}{\bf Proof.}

%%%%%%%%%%%%%%%%%%%%%%%%%%%%%%%%%%%%%%%%%%%%%%

%%%%%%%%%%%%%%%%%%%%%%%%%%%%%%%%%%%%%%%%%%%%%%

% \bibliographystyle{plain}
% \bibliography{HW}

\end{document}